\documentclass[11pt]{vortico-article-lite}

\usepackage{listings}
\usepackage{booktabs}
\usepackage{multirow}
\usepackage{tabularx}
\usepackage{enumitem}
\usepackage{xspace}
\usetikzlibrary{
  shapes.geometric,
  fit,
  backgrounds,
  decorations.pathreplacing,
  chains,
  matrix,
  shadows.blur,
  patterns,
}

\definecolor{codebg}{HTML}{F7F7F8}
\definecolor{codeframe}{HTML}{E0E0E4}
\definecolor{codekw}{HTML}{008080}
\definecolor{codestr}{HTML}{9E744F}
\definecolor{codecom}{HTML}{656775}
\definecolor{codefunc}{HTML}{E25822}

\lstdefinestyle{flamapy}{
  language=Python,
  basicstyle=\ttfamily\small,
  keywordstyle=\color{codekw}\bfseries,
  stringstyle=\color{codestr},
  commentstyle=\color{codecom}\itshape,
  emphstyle=\color{codefunc}\bfseries,
  backgroundcolor=\color{codebg},
  frame=single,
  rulecolor=\color{codeframe},
  framesep=6pt,
  xleftmargin=12pt,
  xrightmargin=12pt,
  breaklines=true,
  showstringspaces=false,
  tabsize=4,
  captionpos=b,
  numbers=left,
  numberstyle=\tiny\color{primary400},
  numbersep=8pt,
  emph={Flama,Component,Resource,CRUDResource,MLResource,
        RESTResource,HTTPEndpoint,WebSocketEndpoint,
        Router,Module,Worker,BaseModel,Schema,SchemaList,
        SchemaMetadata,
        BackgroundTask,BackgroundThreadTask,BackgroundProcessTask,
        Paginator,Config,AccessToken,RefreshToken,
        SQLAlchemyTableRepository,HTTPResourceRepository,
        WorkerComponent,ModelComponent,SchemaModule,
        ResourcesModule,ModelsModule,MCPModule,
        APIResponse,APIErrorResponse,JSONResponse,
        StreamingResponse,FileResponse,HTMLResponse,
        ServerSentEventResponse,NDJSONResponse,
        Middleware,Route,WebSocketRoute,Mount,
        MCPServer,Dialect,Shape,LLMCodec,EventBuffer,
        EngineInput,EngineDelta,StreamsBackend,
        ModelArtifact,ModelCapabilities,LLMModelCapabilities,
        MLModelCapabilities,FrameworkInfo},
  morekeywords={async,await,yield,match,case,Annotated},
}

\lstdefinestyle{flamabash}{
  language=bash,
  basicstyle=\ttfamily\small,
  keywordstyle=\color{codekw}\bfseries,
  stringstyle=\color{codestr},
  commentstyle=\color{codecom}\itshape,
  backgroundcolor=\color{codebg},
  frame=single,
  rulecolor=\color{codeframe},
  framesep=6pt,
  xleftmargin=12pt,
  xrightmargin=12pt,
  breaklines=true,
  showstringspaces=false,
  numbers=none,
  morekeywords={flama,serve,run,start,model,inspect,predict},
}

\lstdefinestyle{flamajson}{
  basicstyle=\ttfamily\small,
  stringstyle=\color{codestr},
  backgroundcolor=\color{codebg},
  frame=single,
  rulecolor=\color{codeframe},
  framesep=6pt,
  xleftmargin=12pt,
  xrightmargin=12pt,
  breaklines=true,
  showstringspaces=false,
  numbers=none,
}

\usepackage{tocloft}

\newcommand{\flama}{\textsc{Flama}\xspace}
\newcommand{\py}[1]{\lstinline[style=flamapy]{#1}}
\newcommand{\http}[1]{\texttt{#1}}
\newcommand{\flm}{\texttt{.flm}\xspace}
\newcommand{\cli}[1]{\texttt{#1}}

\newcommand{\flamapart}[2]{%
  \clearpage
  \vspace*{1.5cm}
  {\noindent\sffamily\Large\bfseries Part #1\par #2}\par
  \vspace{0.25cm}
}

\hypersetup{
  pdftitle  = {Flama: A Python Framework for Development and Deployment
               of Production-Ready APIs, Machine Learning, and LLM Services},
  pdfauthor = {José A. Perdiguero López, Miguel A. Durán-Olivencia},
  pdfsubject = {Artificial Intelligence, Software Engineering, Machine
                Learning, LLM Serving, Web APIs},
  pdfkeywords = {Python, ASGI, REST API, Artificial Intelligence, Machine
                 Learning, LLM, Model Serving, Dependency Injection,
                 Domain-Driven Design, Generative AI},
}

\vtitle{Flama: a Python framework for development and deployment
of production-ready APIs, machine learning, and LLM services}

\vauthor{José A. Perdiguero López, Miguel A. Durán-Olivencia}
\vaffiliation{Vortico Tech, Málaga, Spain}
\vemail{research@vortico.tech}
\vaddress{Vortico Tech, Málaga, 29100, Spain}
\vdate{August 2026}
\vlogo{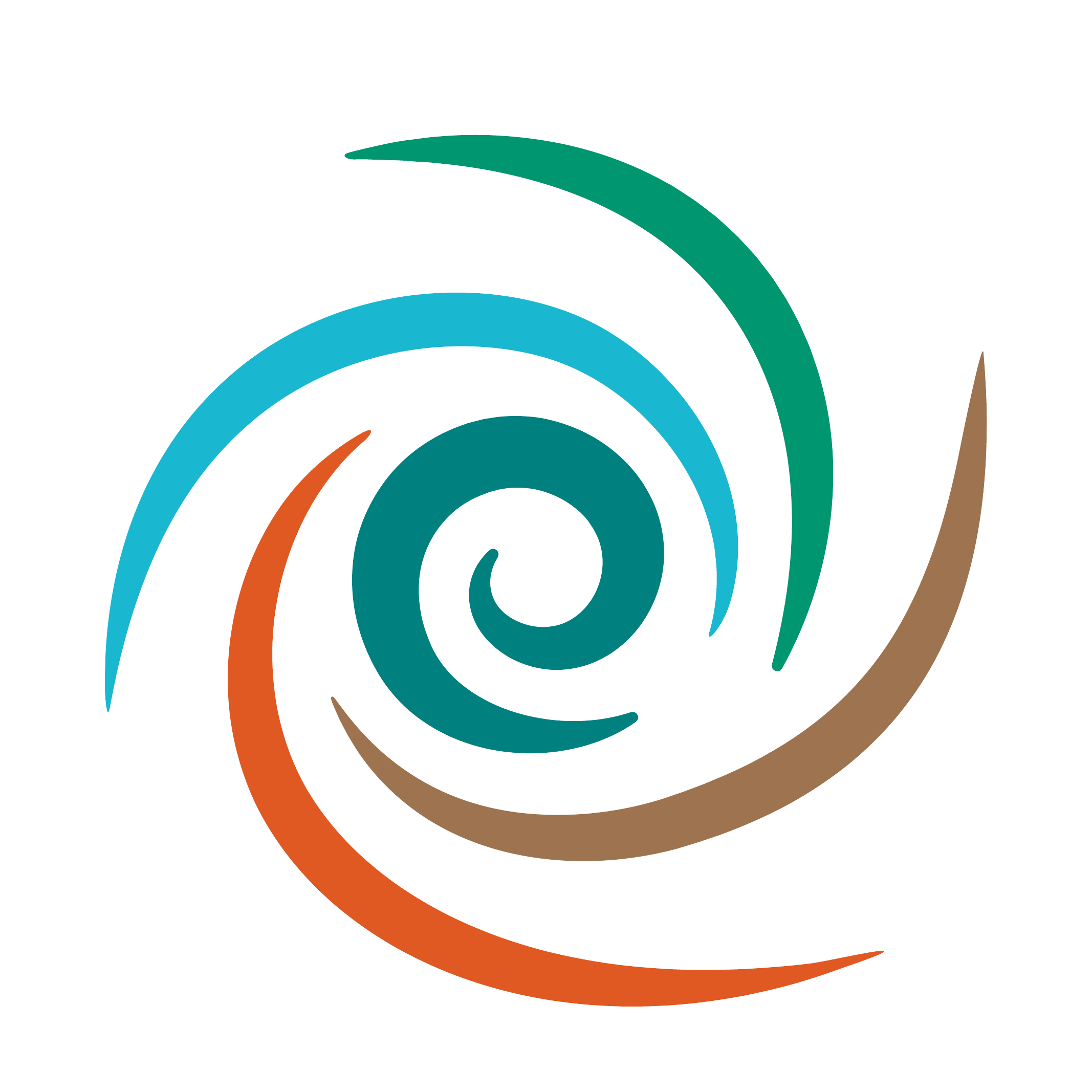}

\vabstract{%
We present \flama, an open-source Python framework designed for
the development and deployment of production-ready web APIs,
machine learning services, and large-language-model (LLM)
applications. Built on the Asynchronous Server Gateway Interface
(ASGI) standard, \flama provides a type-driven, async-first
programming model that unifies traditional REST API development,
predictive model serving, and generative AI inference within a
single, coherent architecture.

The framework is organized around seven core subsystems.
A component-based dependency injection system resolves handler
parameters from type annotations at startup, replacing ad-hoc
patterns (global singletons, request-attached state, middleware
side effects) with a uniform, testable mechanism. A pluggable
schema layer supports three validation libraries (Pydantic,
Marshmallow, and Typesystem) through an adapter that normalizes
field introspection, validation, and JSON Schema emission. An
automatic CRUD resource generator transforms a SQLAlchemy table
definition and a schema class into a complete set of REST
endpoints, backed by the Repository and Unit of Work
domain-driven design patterns for transactional consistency.
A portable binary format (\flm) packages trained models from
scikit-learn, TensorFlow, PyTorch, and Hugging Face Transformers
together with compressed metadata, hyperparameters, training
metrics, and auxiliary artifacts into self-describing files that
can be deployed with zero application code.
A multi-backend LLM serving system deploys generative models
through vLLM (Linux/CUDA) or MLX (Apple Silicon), exposing them simultaneously under four
wire-protocol dialects (OpenAI, Anthropic, Ollama, and a native
streaming protocol) with a shared codec and decoder pipeline
that normalizes reasoning channels and tool-call extraction
across heterogeneous model families.
A Rust-accelerated core, compiled via Maturin, provides
high-throughput route resolution, path and host matching, JSON
encoding, compression, and cookie and multipart parsing as
native Python extensions.
Finally, a Model Context Protocol (MCP) module enables any
\flama application to act as an MCP server, exposing tools,
resources, and prompts to AI model clients over JSON-RPC~2.0.

Built-in capabilities include JWT authentication with
permission-based middleware, two pagination strategies
(page-number and limit-offset), background task execution in
threads or processes, WebSocket endpoints with encoding
negotiation, Server-Sent Event and Newline-Delimited JSON
streaming responses, OpenAPI~3.2.0 schema generation from
handler signatures, and a command-line interface for running
applications, serving models, downloading and packaging models
from remote repositories, performing offline inspection and
inference, and migrating codebases across major versions.

We describe the architecture in detail, present the programming
model through worked examples, and compare \flama with existing
frameworks, model serving platforms, and LLM inference engines.
This document serves as both a comprehensive technical reference
and an extended tutorial for practitioners seeking to build and
deploy ML-powered and LLM-powered APIs with minimal operational
overhead.
}

\begin{document}
\makeabstract

%\tableofcontents
%\clearpage

% ============================================================
% PART I: Foundations
% ============================================================
\flamapart{I}{Foundations}
% ==========================================================================
% Begin part_1_foundations.tex
% ==========================================================================
% ============================================================
%  Part I — Foundations
%  Sections: Introduction, Design Principles, Architecture
% ============================================================

\section{Introduction}
\label{sec:introduction}

\subsection{The deployment gap in machine learning}

Fitting a model is no longer the hard part. A decade of work on
training frameworks, pre-trained model repositories, and
experiment-tracking platforms has seen to that: between
scikit-learn \citep{sklearn}, TensorFlow \citep{tensorflow},
PyTorch \citep{pytorch}, and the Hugging Face ecosystem
\citep{huggingface}, a competent practitioner can have a
working classifier in an afternoon.

Putting that model in front of users is a different matter.
\citet{sculley2015hidden} made the point with a diagram that has
been reproduced ever since: a small box marked \emph{ML code},
surrounded by much larger ones for data collection, feature
extraction, configuration, serving infrastructure, and
monitoring. Those proportions have not improved. Surveying case
studies seven years later, \citet{paleyes2022challenges} still
found deployment and integration consuming more of a project
than the modelling did.

The cause of all this is architectural. Researchers' tools and
engineers' tools grew up apart, in different communities,
assuming different things about the runtime, the interface
contract, and the operational lifecycle. What that separation
produces, in practice, is a fragmented toolchain, and a typical
production ML service might combine:

\begin{itemize}[leftmargin=2em, itemsep=4pt]
  \item A \emph{web framework} for HTTP routing, request parsing,
    input validation, authentication, and error handling (Flask
    \citep{flask}, Django \citep{django}).
  \item A \emph{model serving system} for loading trained
    artefacts, managing model versions, and running inference
    (TensorFlow Serving \citep{tfserving}, TorchServe
    \citep{torchserve}).
  \item An \emph{experiment and packaging layer} for tracking
    training runs, storing model metadata, and producing
    deployable artefacts (MLflow \citep{mlflow}, BentoML
    \citep{bentoml}).
\end{itemize}

Every one of those brings its own abstractions, configuration
formats, deployment conventions, and operational overhead. The
seams between them are where the technical debt accrues:
adapter code has to be written and then maintained to translate
the web framework's request format into the model server's
input format; authentication and authorization get duplicated,
or proxied; error handling and logging have to be reconciled
across systems that were never designed to cooperate.

\subsection{Unifying web APIs and model serving}

\flama starts from a refusal: web API development and model
serving are not distinct concerns, and do not need distinct
systems. The motivating observation behind that is a modest one,
namely that a production ML service needs what any web API needs
plus a short list of model-specific things. Routing, input
validation, authentication, pagination, database access,
background tasks, API documentation, and then, on top of all of
it, the ability to load a trained model, run inference, return
predictions, and inspect metadata. No architectural argument
requires those two sets to sit in separate processes, behind
separate middleware stacks, or on separate deployment
pipelines.

\flama is an open-source Python framework offering a single
programming model for both. Six ideas organise it, each aimed
at a specific pain point in the landscape above:

\begin{enumerate}[leftmargin=2em, itemsep=4pt]
  \item \textbf{Type-driven validation.} Request and response
    schemas are derived from Python type annotations
    \citep{pep484}. The
    framework supports three schema libraries (Pydantic
    \citep{pydantic}, Marshmallow \citep{marshmallow}, and
    Typesystem \citep{typesystem}) through a pluggable adapter
    layer, so teams are not locked into a single validation
    ecosystem.

  \item \textbf{Component-based dependency injection.} All
    handler dependencies, from database connections and parsed
    authentication tokens to validated request bodies and model
    instances, are resolved through a compile-time dependency
    graph with request-scoped caching. This replaces ad-hoc
    patterns (global singletons, request-attached state,
    middleware side effects) with a uniform, testable, and
    composable mechanism.

  \item \textbf{Resource abstraction with domain-driven design.}
    CRUD endpoints over relational tables are generated
    automatically from a SQLAlchemy \citep{sqlalchemy} table
    definition and a schema class, using the Repository and
    Unit of Work patterns \citep{evans2003ddd,
    fowler2002patterns}. Individual operations can be overridden
    or extended without abandoning the generated scaffold.

  \item \textbf{Native predictive model serving.} Trained models
    from scikit-learn, TensorFlow, PyTorch, and Hugging Face
    Transformers are packaged into a portable binary format
    (\flm), loaded as injectable components, and exposed through
    auto-generated inference endpoints. Registering one is a
    single \py{add\_model()} call, and the artifact's own
    metadata decides which backend loads it.

  \item \textbf{First-class LLM serving.} Large language models
    are deployed through a multi-backend engine (vLLM on
    Linux/CUDA, MLX on Apple Silicon) and exposed
    simultaneously under four wire-protocol dialects (OpenAI,
    Anthropic, Ollama, and a native streaming protocol) through
    a shared codec and decoder pipeline that normalizes
    reasoning channels, tool-call extraction, and multimodal
    content across heterogeneous model families. Each dialect
    mounts under its own prefix, so a single model answers at
    \texttt{/openai/v1/chat/completions} and
    \texttt{/anthropic/v1/messages} at once.

  \item \textbf{Production-ready defaults.} JWT authentication,
    pagination, background task execution in threads or
    processes, OpenAPI~3.2.0 schema generation, debug tooling,
    and a six-command CLI are included without additional
    dependencies or configuration.
\end{enumerate}

\subsection{Scope and audience}

This document does two jobs at once. It is a technical
reference, covering architecture, programming model, and
implementation in enough depth for contributors and advanced
users, and it is also an extended tutorial, walking each
subsystem through worked examples a practitioner can lift into
their own project. Those two jobs pull in slightly different
directions, and where they conflict we have favoured the
reference.

Three kinds of reader are in view:

\begin{itemize}[leftmargin=2em, itemsep=4pt]
  \item \emph{ML engineers} who need to deploy trained models as
    production services without learning a separate serving
    infrastructure.
  \item \emph{Backend engineers} who build REST APIs and want
    built-in support for data validation, CRUD generation,
    domain-driven design, and dependency injection.
  \item \emph{Framework designers} interested in the
    architectural decisions behind a unified API-and-ML
    framework, including the pluggable schema adapter, the
    component-based DI system, and the Rust-accelerated routing
    core.
\end{itemize}

\subsection{Document structure}

The document is organized into five parts:

\begin{description}[leftmargin=2em, itemsep=4pt,
  font=\sffamily\bfseries]
  \item[Part~I: Foundations]
    introduces the framework, states the design principles, and
    describes the layered architecture, the request processing
    pipeline, and the module system
    (Sections~\ref{sec:introduction}--\ref{sec:architecture}).

  \item[Part~II: The web framework]
    covers the core API-building subsystems: routing and
    endpoints, schema and data validation, dependency injection,
    resources with domain-driven design, authentication and
    authorization, pagination, background tasks, and middleware
    (Sections~\ref{sec:routing}--\ref{sec:middleware}).

  \item[Part~III: Machine learning and generative AI]
    presents the predictive model serving subsystem (the \flm
    serialization format, framework-specific model wrappers,
    three levels of serving abstraction), the generative AI
    stack (multi-backend LLM serving, wire-protocol dialects,
    the codec and decoder pipeline, streaming), and the Model
    Context Protocol (Section~\ref{sec:ml-serving}).

  \item[Part~IV: Operations and tooling]
    describes the \flama CLI in detail (including model
    acquisition, codebase migration, and LLM serving commands),
    the configuration management system, deployment patterns,
    debug mode, and the testing infrastructure
    (Sections~\ref{sec:cli}--\ref{sec:testing}).

  \item[Part~V: Ecosystem and outlook]
    positions \flama relative to existing frameworks and LLM
    inference engines through a detailed feature comparison, and
    outlines the roadmap for future development
    (Sections~\ref{sec:comparison}--\ref{sec:conclusion}).
\end{description}

The framework is released under the Apache~2.0 licence and is
available at \url{https://github.com/vortico/flama}, with
documentation at \url{https://flama.dev} and package
distribution via PyPI (\cli{pip install flama}).

\section{Design principles}
\label{sec:design-principles}

A small set of principles governs every architectural decision
in \flama. Because they constrain the design space, and because
they are what separates the framework from the alternatives
that made different trades, it is worth setting them out
explicitly rather than leaving them to be reconstructed from
the code. None of them is abstract. Each came out of a concrete
problem met in real API and ML deployment work.

\subsection{Asynchronous by default}
\label{sec:async-by-default}

\flama targets the ASGI (Asynchronous Server Gateway Interface)
specification \citep{asgispec}, which models an application as
an asynchronous callable:

\begin{lstlisting}
async def application(scope: Scope,
                      receive: Receive,
                      send: Send) -> None:
    ...
\end{lstlisting}

Three arguments arrive. The scope is a dictionary describing the
connection, giving its type, path, headers, and query string,
while the other two are channels: \py{receive} is where request
body chunks come from, and \py{send} is where response headers
and body chunks go. The whole interface is the asynchronous
answer to WSGI, which has carried the Python web ecosystem
since PEP~333 \citep{pep333} and is synchronous by construction, since
a WSGI application holds its thread for the whole of each
request.

Everything inside \flama is written against
\py{async}/\py{await}, from middleware dispatch down through
database queries to model inference. Two consequences follow,
one for performance and one for design:

\begin{enumerate}[leftmargin=2em, itemsep=4pt]
  \item \textbf{I/O multiplexing.} I/O-bound workloads
    (database queries, HTTP calls to upstream services, file
    reads, network waits) can be multiplexed on a single
    thread without blocking. A single worker process can serve
    thousands of concurrent connections, because each
    connection yields control to the event loop during I/O
    waits rather than holding a thread hostage.

  \item \textbf{Explicit concurrency boundaries.} CPU-bound
    work (model inference, data transformation, report
    generation) can be offloaded to background threads or
    processes through the framework's concurrency primitives
    (\py{BackgroundThreadTask}, \py{BackgroundProcessTask})
    without breaking the async contract. The dispatch itself is
    handled by \py{flama.concurrency.run()}, which awaits a
    coroutine directly and sends anything else to a thread pool
    with the caller's \py{contextvars} context copied across.
\end{enumerate}

None of which forces async syntax on anyone. A synchronous
handler is wrapped in \py{asyncio.to\_thread()} at resolution
time, so an ordinary function works as well as a coroutine:

\begin{lstlisting}
@app.route("/health")
def health_check() -> dict:
    return {"status": "ok"}
\end{lstlisting}

\py{health_check} is not a coroutine, the framework notices as
much, and it gets scheduled on a thread pool. The practical
payoff is migration: porting an existing synchronous codebase
does not mean rewriting every handler before it will run.

\subsection{Type annotations as the source of truth}
\label{sec:type-annotations}

Python type annotations \citep{pep484, pep526, pep585, pep604}
serve as the single source of truth for four distinct concerns
in \flama:

\begin{enumerate}[leftmargin=2em, itemsep=4pt]
  \item \textbf{Request parsing.} The names and types of handler
    parameters determine where each value is extracted from
    (path segment, query string, or request body) and how it is
    parsed.

  \item \textbf{Validation.} Schema classes referenced in type
    annotations trigger automatic validation of the
    corresponding request data. Validation errors produce
    structured 400~Bad Request responses.

  \item \textbf{Dependency resolution.} Parameters whose types
    match a registered component are resolved through the
    dependency injection system. The type annotation is the
    sole mechanism by which a handler declares its dependencies.

  \item \textbf{API documentation.} The OpenAPI 3.2.0
    specification is generated from handler signatures at
    startup: parameter types become OpenAPI parameter schemas,
    return types become response schemas, and schema classes
    become reusable component definitions.
\end{enumerate}

The four converge in a way that is easier to show than to
describe. Take this handler:

\begin{lstlisting}
import typing as t
from flama import schemas

@app.route("/users/{user_id:int}", methods=["PUT"])
async def update_user(
    user_id: int,
    data: t.Annotated[schemas.Schema, schemas.SchemaMetadata(UserUpdate)],
    connection: AsyncConnection,
    token: AccessToken,
) -> t.Annotated[schemas.Schema, schemas.SchemaMetadata(User)]:
    ...
\end{lstlisting}

From this signature alone, the framework determines that:

\begin{itemize}[leftmargin=2em, itemsep=4pt]
  \item \py{user_id} is an integer extracted from the URL path
    segment \texttt{\{user\_id:int\}}.
  \item \py{data} is validated from the JSON request body
    against the \py{UserUpdate} schema, using the configured
    schema library; the \py{SchemaMetadata} annotation is what
    tells the framework which parameter carries the request
    body and which schema to validate it against.
  \item \py{connection} is an \py{AsyncConnection} resolved by a
    registered component, typically one backed by the
    \py{SQLAlchemyModule}.
  \item \py{token} is an \py{AccessToken}, decoded by the JWT
    authentication component from the request's
    \texttt{access\_token} header or cookie.
  \item The return annotation fixes the response schema
    (\py{User}) and, with it, the OpenAPI response definition.
\end{itemize}

Nothing else is required beyond the route binding itself: no
decorators, no configuration dictionaries, no registration
calls. The signature alone is the endpoint's contract.

\subsection{Pluggable schema libraries}
\label{sec:pluggable-schemas}

Picking a validation library on a user's behalf is a decision
that ages badly. \flama declines to make it, and interposes an
adapter over three established ones instead, each available as
an optional install extra:

\begin{itemize}[leftmargin=2em, itemsep=4pt]
  \item \textbf{Pydantic}~\citep{pydantic}: the most widely
    adopted validation library in the Python ecosystem, with
    Rust-accelerated core validation.
  \item \textbf{Marshmallow} \citep{marshmallow}: a mature
    library with a rich ecosystem of plugins and a
    serialization API oriented around explicit field
    declarations.
  \item \textbf{Typesystem} \citep{typesystem}: a lightweight
    library designed for data validation and form rendering in
    web frameworks.
\end{itemize}

Whichever is chosen, the adapter translates its field and
schema representations into one internal model, the \py{Field}
and \py{Schema} structures of Section~\ref{sec:schemas}, and
everything downstream (validation, JSON Schema emission,
OpenAPI generation) reads that instead. The choice is made once,
at application initialization:

\begin{lstlisting}
app = Flama(schema_library="pydantic")
\end{lstlisting}

after which the library's own classes are used throughout the
codebase and the translation stays out of sight. Two things are
bought with this arrangement: a team already invested in one
library can adopt \flama without rewriting its schemas, and the
core validation pipeline stops being hostage to any single
library's API churn. The second is not hypothetical; Pydantic's
1.x-to-2.x transition rewrote most of its public surface, and
an adapter is where a framework absorbs that kind of change
instead of passing it on.

\subsection{Dependency injection as a first-class mechanism}
\label{sec:di-principle}

How does a handler get hold of what it needs? Python web
frameworks have accumulated several answers, none of them
uniform: module-level singletons, state hung off the request
(\py{request.state.db} and relatives), middleware that quietly
writes into the ASGI scope, explicit \py{Depends()} calls that
weld the handler to one resolution strategy.

In place of all of them \flama uses component-based dependency
injection, along the lines \citet{fowler2004injection} sets out.
A handler has no business knowing \emph{how} its dependencies
get built; what it does instead is declare, through type
annotations, \emph{what} it wants, leaving the framework to
supply the instances at call time.

Three things follow:

\begin{enumerate}[leftmargin=2em, itemsep=4pt]
  \item \textbf{Testability.} Swapping a dependency in a test
    means registering a different component, with no patching
    and no monkeypatching involved. A test wanting a fake
    database registers a component that hands back a mock
    connection, and the handler is none the wiser.

  \item \textbf{Composability.} Components can depend on other
    components, so \py{UserRepository} can ask for an
    \py{AsyncConnection} that is itself produced by a connection
    component, and the framework works out the resulting graph
    on its own.

  \item \textbf{Uniformity.} One mechanism covers database
    connections, validated request data, authentication tokens,
    model instances, pagination parameters, and anything a user
    defines. One resolution system, one caching strategy, one
    place to look when a dependency misbehaves.
\end{enumerate}

\subsection{Convention over configuration for CRUD}
\label{sec:convention-crud}

CRUD endpoints over a relational table are the most written and
least interesting code in API development. Most frameworks ask
for five to seven handler functions, each with a route binding,
input validation, error handling, and its own database
interaction. Almost all of it is boilerplate, since the handlers
take their shape from the table, and departures from the
standard pattern are rare enough to be worth handling as
exceptions rather than designing for.

So \flama generates them, from two declarations: a SQLAlchemy
\citep{sqlalchemy} table and a schema class. Create, retrieve,
update, delete, list, and bulk-drop all come out with the right
status codes, input validation, pagination, and error handling
(integrity errors become 400~Bad Request, missing resources
404~Not Found). The generated code is built on the
domain-driven design layer of Section~\ref{sec:resources-ddd},
and any single operation can be overridden or extended without
giving up the rest.

In the usual case, hundreds of lines of handler code collapse
into a class declaration of under ten.

\subsection{ML and LLM serving as native concerns}
\label{sec:ml-native}

Model serving in \flama isn't a plugin, a sidecar, or an
afterthought. It's a module, on the same footing as routing,
schema generation, and resource management. Predictive models
and large language models get the same treatment:

\begin{itemize}[leftmargin=2em, itemsep=4pt]
  \item A trained model is packaged into a \flm file (a
    compressed binary containing the model artefact, framework
    metadata, training parameters, evaluation metrics, and
    optional auxiliary artefacts). For predictive models the
    payload is the serialized weight tensor; for LLMs it is a
    tarball of the model's checkpoint directory.
  \item Registration reads only the metadata header, so it stays
    cheap. Deserialising the body is deferred to a startup
    event, which lets the server bind its port before a
    multi-gigabyte checkpoint begins loading. The result is a
    typed component, injected into handlers like any other.
  \item A predictive model gets \http{GET~/} for its metadata
    and \http{POST~/predict/} for inference, both generated. An
    LLM gets a whole multi-dialect surface instead, following
    whichever of OpenAI, Anthropic, Ollama, and the native
    protocol were asked for.
  \item The same CLI that starts a development server will also
    serve a model with no application code at all.
\end{itemize}

A team whose API happens to include a model, then, has no
second serving system to operate. The model sits in the same
process as everything else, behind the same authentication,
under the same logging and monitoring, shipped down the same
pipeline.

\subsection{Protocol-agnostic generative serving}
\label{sec:protocol-agnostic}

Generative AI has produced a thicket of mutually incompatible
wire protocols: OpenAI's Chat Completions API \citep{openai},
Anthropic's Messages API \citep{anthropic}, Ollama's local
inference protocol \citep{ollama}, and a long tail of
vendor-specific endpoints. Wire a framework's internals to any
one of them and its users inherit that choice, along with the
integration rewrite that follows every protocol revision.

The way out is to keep the model's execution and the wire
format apart. At the centre sits a canonical transport model:
a request is a sequence of typed messages carrying multimodal
content parts, a response is a stream of typed events (start,
text, tool call, trace, stop). A \emph{dialect} stands between
that model and the client, and it is three things at once, a
parser turning wire format into canonical messages, a renderer
turning canonical events into streaming wire frames, and an
assembler turning them into buffered response envelopes.

With that in place, one physical model instance serves every
dialect at once, each mounted under its own URL prefix, so that
the OpenAI SDK, the Anthropic SDK, the Ollama CLI, and a
browser pointed at the native chatbot all reach the same
application and the same weights. Normalising reasoning
channels and pulling tool calls out of whatever shape a model
emits them in belongs to the codec and decoder pipeline. That
work happens once, upstream of whichever dialect renders the
result.

\section{Architecture}
\label{sec:architecture}

Three things make up the structure of \flama, and each is taken
in turn below: the layers it is built from, the path an HTTP
request takes down through them, and the module system by which
it is extended.

\subsection{Layered design}
\label{sec:layered-design}

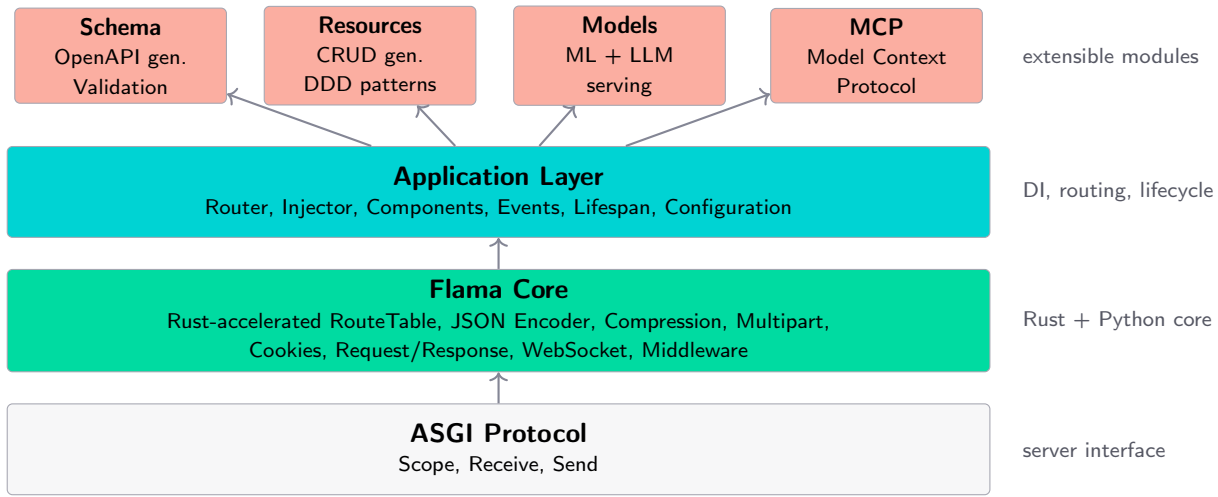
\begin{figure}[ht]
\centering
\begin{tikzpicture}[
  layer/.style={
    draw=primary300,
    fill=#1,
    minimum width=13cm,
    minimum height=1.2cm,
    font=\sffamily\small,
    rounded corners=2pt,
    align=center,
  },
  module/.style={
    draw=primary300,
    fill=#1,
    minimum width=2.8cm,
    minimum height=0.9cm,
    font=\sffamily\footnotesize,
    rounded corners=2pt,
    align=center,
  },
  annot/.style={
    font=\sffamily\scriptsize\color{primary600},
    anchor=west,
  },
]

\node[layer=primary50] (asgi) at (0, 0)
  {\textbf{ASGI Protocol}\\[-1pt]
   {\scriptsize Scope, Receive, Send}};

\node[layer=ciclon50] (core) at (0, 1.7)
  {\textbf{Flama Core}\\[-1pt]
   {\scriptsize Rust-accelerated RouteTable, JSON Encoder,
    Compression, Multipart,}\\[-1pt]
   {\scriptsize Cookies, Request/Response, WebSocket, Middleware}};

\node[layer=vortico50] (app) at (0, 3.4)
  {\textbf{Application Layer}\\[-1pt]
   {\scriptsize Router, Injector, Components, Events, Lifespan,
    Configuration}};

\node[module=flama50] (schema) at (-5.0, 5.2)
  {\textbf{Schema}\\\scriptsize OpenAPI gen.\\
   \scriptsize Validation};
\node[module=flama50] (resources) at (-1.7, 5.2)
  {\textbf{Resources}\\\scriptsize CRUD gen.\\
   \scriptsize DDD patterns};
\node[module=flama50] (models) at (1.6, 5.2)
  {\textbf{Models}\\\scriptsize ML + LLM\\
   \scriptsize serving};
\node[module=flama50] (mcp) at (5.0, 5.2)
  {\textbf{MCP}\\\scriptsize Model Context\\
   \scriptsize Protocol};

\node[annot] at (6.8, 0) {server interface};
\node[annot] at (6.8, 1.7) {Rust + Python core};
\node[annot] at (6.8, 3.4) {DI, routing, lifecycle};
\node[annot] at (6.8, 5.2) {extensible modules};

\draw[->, thick, primary400] (asgi) -- (core);
\draw[->, thick, primary400] (core) -- (app);
\draw[->, thick, primary400] (app) -- (schema);
\draw[->, thick, primary400] (app) -- (resources);
\draw[->, thick, primary400] (app) -- (models);
\draw[->, thick, primary400] (app) -- (mcp);

\end{tikzpicture}
\caption{Layered architecture of \flama. The ASGI protocol
provides the server interface. The Flama Core layer, with
Rust-compiled performance-critical paths, supplies HTTP
abstractions, route resolution, path and host matching, JSON
encoding, compression, and multipart and cookie parsing. The
Application layer adds dependency injection, component
management, and lifecycle hooks. Modules extend the
application with schema generation, CRUD resources, ML and LLM
model serving, and the Model Context Protocol.}
\label{fig:architecture}
\end{figure}

The framework is layered, with every layer handing the one above
it a set of abstractions, and every layer able to be reasoned
about on its own, which is rather the point of the arrangement.
Figure~\ref{fig:architecture} shows the whole of it.

The layers, from bottom to top, are:

\begin{description}[leftmargin=2em, itemsep=4pt,
  font=\sffamily\bfseries]
  \item[ASGI Protocol.]
    The interface between an HTTP server, Uvicorn
    \citep{uvicorn} for instance, and the application. \flama
    implements the application side of the contract and no more,
    and is not itself a server. Any compliant one will host it.

  \item[Flama Core.]
    The layer that provides the fundamental HTTP and WebSocket
    abstractions: \py{Request}, \py{Response}, \py{WebSocket},
    middleware composition, static file serving, and content
    negotiation. Performance-critical paths in this layer are
    implemented in Rust and compiled as Python extensions via
    Maturin \citep{maturin}. The Rust core comprises seven modules:
    \emph{route resolution} (iterates the registered route
    entries in a single pass, extracting path parameters as it
    matches), \emph{path and host matching} (the segment-based
    template matcher that route resolution calls, plus a
    host matcher for exact, wildcard-subdomain, and
    match-any virtual-host patterns), \emph{JSON encoding}
    (fast serialization with support for datetimes, UUIDs,
    decimals, dataclasses, and framework types),
    \emph{compression} (bz2, lzma, zlib, zstd, gzip, and brotli
    codecs with a streaming compressor), \emph{multipart parsing}
    (async form-data parsing via Tokio), \emph{cookie handling}
    (RFC-compliant parsing and Set-Cookie construction), and
    \emph{HTTP utilities} (content-type parsing). What crosses
    back into Python is always a plain object, bytes or a tuple
    or a list or a dictionary, which keeps FFI overhead down
    while the inner loops still run compiled. None of this is
    visible at install time: wheels are published for every
    supported Python (3.10 through 3.14) on Linux, macOS, and
    Windows, so no Rust toolchain is needed to \cli{pip install
    flama}.

  \item[Application Layer.]
    The \py{Flama} class, the object users actually instantiate.
    It composes a \py{Router}, an \py{Injector}, a
    \py{Components} registry, a \py{MiddlewareStack}, an
    \py{Events} manager, and a set of \py{Modules}. Orchestrating
    the request lifecycle is its job: take the ASGI connection,
    push it through the middleware stack to the router, resolve
    dependencies, run the handler, build the response.

  \item[Modules.]
    Self-contained units of related functionality. A module can
    register components, add routes, and hook startup and
    shutdown. Four ship with the framework (Schema, Resources,
    Models, MCP), and custom ones subclass \py{Module}.
\end{description}

\subsection{The application object}
\label{sec:application-object}

Everything hangs off the \py{Flama} class, which is itself the
ASGI callable, and a single line of it is already a working
application:

\begin{lstlisting}
from flama import Flama

app = Flama()
\end{lstlisting}

That already has OpenAPI schema generation, debug error pages,
and an empty route table waiting for handlers. Optional
constructor parameters take over from there:

\begin{lstlisting}
app = Flama(
    openapi={
        "info": {
            "title": "My API",
            "version": "1.0.0",
            "description": "A production API",
        }
    },
    schema_library="pydantic",
    routes=[...],
    components=[...],
    modules=[...],
    middleware=[...],
    events={"startup": [...], "shutdown": [...]},
    debug=False,
)
\end{lstlisting}

Inside, the object is an assembly of six subsystems:

\begin{center}
\small
\begin{tabularx}{\textwidth}{@{}lX@{}}
\toprule
\textbf{Subsystem} & \textbf{Responsibility} \\
\midrule
\py{Router} &
  Maps URL paths to handler callables using the
  Rust-accelerated route table. Manages route
  registration, path parameter extraction, and
  method-not-allowed responses. \\
\py{Injector} &
  Builds and caches dependency resolution trees for
  each handler at startup. At request time, walks the
  pre-compiled tree to resolve all handler parameters. \\
\py{Components} &
  A registry of all component instances available for
  injection. Includes built-in components (request data,
  validation, authentication) and user-registered
  components. \\
\py{MiddlewareStack} &
  An ordered chain of ASGI middleware that wraps the
  router. Processes requests in registration order and
  responses in reverse order. \\
\py{Modules} &
  A dictionary of named module instances (Schema,
  Resources, Models, MCP). Each module is accessible as
  an attribute of the application
  (e.g.\ \py{app.models}). \\
\py{Events} &
  Startup and shutdown event handlers, executed when the
  ASGI lifespan protocol signals that the application
  should initialize or tear down. \\
\bottomrule
\end{tabularx}
\end{center}

\subsection{Request processing pipeline}
\label{sec:request-pipeline}

\begin{figure}[ht]
\centering
\begin{tikzpicture}[
  node distance=0.35cm,
  pipestep/.style={
    draw=vortico500,
    fill=vortico50,
    minimum width=2.5cm,
    minimum height=0.75cm,
    font=\sffamily\scriptsize,
    rounded corners=2pt,
    align=center,
  },
  arr/.style={->, thick, vortico600},
]

\node[pipestep] (req) {HTTP\\Request};
\node[pipestep, right=of req] (mw1) {Middleware\\Stack};
\node[pipestep, right=of mw1] (router) {Route\\Resolution};
\node[pipestep, right=of router] (di) {Dependency\\Injection};
\node[pipestep, right=of di] (handler) {Handler\\Execution};
\node[pipestep, below=0.6cm of handler] (resp) {Response\\Building};
\node[pipestep, left=of resp] (mw2) {Middleware\\(response)};
\node[pipestep, left=of mw2] (send) {ASGI\\Send};

\draw[arr] (req) -- (mw1);
\draw[arr] (mw1) -- (router);
\draw[arr] (router) -- (di);
\draw[arr] (di) -- (handler);
\draw[arr] (handler) -- (resp);
\draw[arr] (resp) -- (mw2);
\draw[arr] (mw2) -- (send);

\end{tikzpicture}
\caption{Request processing pipeline. The request flows left to
right through the middleware stack, route resolution, dependency
injection, and handler execution. The response flows back
through the middleware stack before being sent to the client.}
\label{fig:pipeline}
\end{figure}
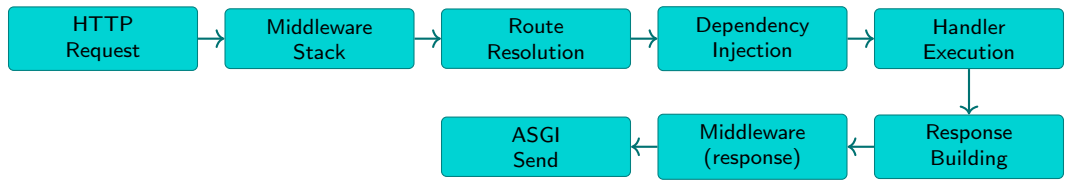

A request moves through \flama in a fixed sequence of stages,
and the sequence is worth knowing, because middleware ordering,
the timing of dependency resolution, and the point at which
errors surface all follow from it.
Figure~\ref{fig:pipeline} traces the flow.

There are eight stages, described below in the order they run.

\subsubsection{Stage 1: ASGI reception}

The server, Uvicorn or another, accepts the connection and
builds a scope dictionary out of it: method, path, headers,
query string, client address. It calls the application's
\py{__call__} with that scope and the \py{receive}/\py{send}
channels.

\subsubsection{Stage 2: Middleware processing (request phase)}

The middleware stack is an ordered sequence of ASGI callables,
each wrapping the next, and the outermost sees the request
first. Any of them can inspect or modify the scope, read the
body through \py{receive}, hand control onward, or cut the
pipeline short by answering directly, which is what an
authentication failure or a CORS preflight does.

\subsubsection{Stage 3: Route resolution}

The \py{Router} hands the path to the Rust-compiled
\py{RouteTable} and gets back one of three outcomes:

\begin{itemize}[leftmargin=2em, itemsep=2pt]
  \item \textbf{Full match}: the path matches a registered
    route pattern, and the HTTP method is allowed. The router
    extracts the path parameters (typed according to the route
    pattern) and dispatches to the matched handler.
  \item \textbf{Method not allowed}: the path matches a route,
    but the HTTP method is not supported. The router returns a
    405~Method Not Allowed response.
  \item \textbf{No match}: the path does not match any route.
    The router returns a 404~Not Found response.
\end{itemize}

\subsubsection{Stage 4: Dependency injection}

For the matched handler, the \py{Injector} walks the
pre-compiled resolution tree, built at startup as described in
Section~\ref{sec:dependency-injection}, and every parameter in
the signature resolves from one of three sources:

\begin{enumerate}[leftmargin=2em, itemsep=2pt]
  \item \textbf{Context values}: objects available directly from
    the ASGI scope or the request context (\py{Request},
    \py{Flama}, \py{Route}, \py{WebSocket}).
  \item \textbf{Components}: registered component instances
    whose \py{resolve()} method produces a value of the
    required type. Components may have their own upstream
    dependencies, which are resolved recursively.
  \item \textbf{Parameters}: primitive values extracted from
    path segments or query strings.
\end{enumerate}

Resolved values land in a per-request cache. A component
declaring \py{cacheable = True}, which is the default, resolves
at most once per request no matter how many parameters or
nested components ask for it.

\subsubsection{Stage 5: Handler execution}

With the arguments resolved, the handler runs, or, for a
class-based endpoint, the method matching the verb runs in its
place. Should that handler be synchronous rather than a
coroutine, it goes to a thread pool via
\py{asyncio.to\_thread()}.

\subsubsection{Stage 6: Response building}

Whatever the handler returned now becomes an HTTP response. A
\py{Response} object is used as it stands. Anything else is
wrapped in an \py{APIResponse}
with the route's declared response schema attached. When that
schema is present, \py{APIResponse} re-validates the value
through the schema adapter before encoding it to JSON, so a
handler whose return type is schema-annotated must return a
plain \py{dict} (or a list of dicts, for a collection response)
with the schema's field names as keys, not an instantiated
schema object.

\subsubsection{Stage 7: Middleware processing (response phase)}

The response travels back out through the same stack in reverse
order, and along the way headers get inspected and rewritten,
metrics logged, bodies compressed, and cleanup done.

\subsubsection{Stage 8: ASGI transmission}

Headers and body go out over the \py{send} channel. Any
background tasks attached to the response run once transmission
has finished.

\subsection{The module system}
\label{sec:module-system}

A module is a self-contained unit of related functionality: a
subclass of \py{Module}, implementing whichever lifecycle hooks
it needs and ignoring the rest.

\begin{lstlisting}
from flama.modules import Module

class AnalyticsModule(Module):
    name = "analytics"

    def __init__(self, dsn: str) -> None:
        super().__init__()
        self._dsn = dsn
        self.client: AnalyticsClient | None = None

    async def on_startup(self) -> None:
        self.client = AnalyticsClient(self._dsn)

    async def on_shutdown(self) -> None:
        await self.client.flush()
        self.client = None
\end{lstlisting}

A module holds its own state rather than writing it onto the
application: the instance is reachable from the application
under its \py{name} (here \py{app.analytics}), so whatever it
exposes as attributes or methods becomes the module's public
surface. Modules can also register components and routes
dynamically during startup, which makes them a natural unit for
packaging a whole subsystem (database connection management,
model lifecycle, schema generation) for reuse.

The framework ships with four modules:

\begin{description}[leftmargin=2em, itemsep=4pt,
  font=\sffamily\bfseries]
  \item[SchemaModule]
    generates the OpenAPI~3.2.0 specification
    \citep{openapi} from route metadata and type annotations.
    It walks all registered routes at startup, extracts
    parameter and body annotations, and emits a complete
    OpenAPI document. The specification is served as JSON at a
    configurable path (default \http{/schema/}), and an
    interactive Swagger UI is rendered at \http{/docs/}.

  \item[ResourcesModule]
    manages the lifecycle of REST resources
    (Section~\ref{sec:resources-ddd}), exposing
    \py{add\_resource()} to attach a resource class to the
    application under a path prefix. Database connectivity
    itself is a separate, optional module
    (\py{SQLAlchemyModule}) that a \py{CRUDResource}-based
    application registers alongside it.

  \item[ModelsModule]
    manages the lifecycle of ML models
    (Section~\ref{sec:ml-serving}): it registers
    \py{ModelComponent} instances for injection, mounts their
    endpoints, and materialises each one during startup. That
    last step runs sequentially rather than concurrently, on
    the grounds that loading several multi-gigabyte artifacts
    at once trades a slow cold start for an out-of-memory
    failure.

  \item[MCPModule]
    implements the Model Context Protocol
    (Section~\ref{sec:mcp}), enabling a \flama application to
    act as an MCP server that exposes tools, resources, and
    prompts to AI model clients over a JSON-RPC 2.0 transport.
\end{description}

Custom modules are registered at application construction time as
a list of module instances, where each subclass supplies its own
\py{name} attribute, which is how the application indexes it:

\begin{lstlisting}
app = Flama(
    modules=[AnalyticsModule(dsn=ANALYTICS_DSN)],
)
\end{lstlisting}

Once registered, a module is reachable as an attribute of the
application, so the example above answers to
\py{app.analytics}. Two modules claiming the same \py{name}
raise at construction rather than silently shadowing one
another.
% ==========================================================================
% End part_1_foundations.tex
% ==========================================================================

% ============================================================
% PART II: The web framework
% ============================================================
\flamapart{II}{The Web Framework}
% ==========================================================================
% Begin part_2_web_framework.tex
% ==========================================================================
% ============================================================
%  Part II — The Web Framework
%  Sections: Routing, Schemas, DI, Resources/DDD, Auth,
%            Pagination, Background Tasks, Middleware
% ============================================================

\section{Routing and endpoints}
\label{sec:routing}

Routing is the front door of a \flama application. It decides
which handler sees each incoming request, how path segments
become typed parameters, and what happens when nothing matches.
Since the route declaration is also where an application's
external interface gets pinned down, most of what follows in
this part depends on it, which is why it comes first: route
types, then the declaration syntax, then the compiled route
table underneath, then class-based endpoints.

\subsection{Route types}
\label{sec:route-types}

\flama distinguishes four kinds of routes, each modelling a
different relationship between a URL pattern and a handler:

\begin{description}[leftmargin=2em, itemsep=4pt,
  font=\sffamily\bfseries]
  \item[HTTP routes]
    bind a URL pattern and one or more HTTP methods to a handler
    callable. The handler may be a plain function (synchronous
    or asynchronous) or a class-based endpoint
    (Section~\ref{sec:class-endpoints}). HTTP routes are the
    most common route type and account for the majority of
    endpoints in a typical application.

  \item[WebSocket routes]
    bind a URL pattern to a WebSocket endpoint with a
    three-phase lifecycle: connection, a receive-and-respond
    loop, and disconnection. WebSocket routes support encoding
    negotiation (JSON, text, or raw bytes) for incoming
    messages and can be used for real-time features such as
    live dashboards and interactive interfaces.

  \item[Mounts]
    attach a sub-application or a \py{Router} instance under a
    path prefix. Mounts enable modular composition: a team can
    develop a self-contained set of routes in an isolated router
    and mount it into the main application at deployment time.
    Mounts also support mounting any ASGI-compliant application,
    making it possible to combine \flama with other ASGI
    frameworks in a single process.

  \item[Resource routes]
    are generated automatically from a resource class declaration
    (Section~\ref{sec:resources-ddd}). They produce a full set
    of CRUD endpoints, each backed by domain-driven design
    patterns. Resource routes are syntactic sugar: they expand
    into a mount containing standard HTTP routes.
\end{description}

\subsection{Declaring routes}
\label{sec:declaring-routes}

Routes can be declared with decorator syntax, which is the
idiomatic approach for applications where routes are defined
close to their handlers:

\begin{lstlisting}
import typing as t
from flama import Flama, schemas

app = Flama()

UserResponse = t.Annotated[
    schemas.Schema, schemas.SchemaMetadata(User)
]

@app.route("/users/{user_id:int}", methods=["GET"])
async def get_user(user_id: int) -> UserResponse:
    """Retrieve a single user by primary key."""
    ...
\end{lstlisting}

Every schema-typed value in a signature, whether it travels in
the request body or the response, is declared the same way:
\py{t.Annotated[schemas.Schema, schemas.SchemaMetadata(X)]}
\citep{pep593},
with \py{list[schemas.Schema]} in place of \py{schemas.Schema}
when the value is a collection. The annotation carries two
pieces of information the framework needs and cannot infer from
a bare class reference: that the parameter (or return value) is
a schema-validated payload at all, and which schema class
validates it. Binding these annotations to a named alias, as
with \py{UserResponse} above, keeps signatures readable when the
same schema recurs across handlers:

\begin{lstlisting}
UserRequest = t.Annotated[
    schemas.Schema, schemas.SchemaMetadata(UserInput)
]

@app.get("/users/{user_id:int}")
async def get_user(user_id: int) -> UserResponse:
    ...

@app.post("/users/")
async def create_user(data: UserRequest) -> UserResponse:
    ...

@app.put("/users/{user_id:int}")
async def update_user(
    user_id: int, data: UserRequest
) -> UserResponse:
    ...

@app.delete("/users/{user_id:int}")
async def delete_user(user_id: int) -> None:
    ...
\end{lstlisting}

A handler that returns a \py{Response} object directly, or that
has no documented payload, simply omits the return annotation;
the framework then generates no response schema for it.

For applications where routes are defined separately from their
handlers (e.g.\ in a configuration module or a factory function),
routes can be constructed programmatically:

\begin{lstlisting}
from flama.routing import Route, Router, Mount

routes = [
    Route("/users/{user_id:int}", get_user,
          methods=["GET"]),
    Route("/users/", create_user,
          methods=["POST"]),
    Mount("/admin", app=admin_router),
]

app = Flama(routes=routes)
\end{lstlisting}

\subsection{Path parameters and type converters}
\label{sec:path-parameters}

Path segments enclosed in braces are treated as parameters. Each
parameter may include a type converter, separated by a colon,
that determines how the raw string segment is parsed and
validated:

\begin{center}
\small
\begin{tabularx}{\textwidth}{@{}llX@{}}
\toprule
\textbf{Pattern} & \textbf{Type} & \textbf{Matching behaviour} \\
\midrule
\texttt{\{id\}} & \py{str} &
  Matches any single path segment (no slashes). \\
\texttt{\{id:int\}} & \py{int} &
  Matches a sequence of digits; returns an integer. \\
\texttt{\{id:float\}} & \py{float} &
  Matches a decimal number; returns a float. \\
\texttt{\{id:uuid\}} & \py{UUID} &
  Matches a UUID string (8-4-4-4-12 hex); returns a
  \py{uuid.UUID} object. \\
\texttt{\{id:decimal\}} & \py{Decimal} &
  Matches a decimal number; returns a
  \py{decimal.Decimal} object. \\
\bottomrule
\end{tabularx}
\end{center}

Type converters are applied during route resolution, before the
handler is called. If a path segment does not match the declared
type (e.g.\ \texttt{/users/abc} against \texttt{\{id:int\}}),
the route is not matched and the router continues searching for
alternative routes or returns 404.

\subsection{The Rust-accelerated route table}
\label{sec:route-table}

Route resolution runs on every request, so its cost is paid on
every request, and a router that degrades as the table grows
degrades the whole application with it. \flama delegates the
matching to a Rust-compiled \py{RouteTable}, built with Maturin
and exposed to Python as a native extension module. What that
buys is measurable: under Callgrind, a full request cycle costs
about 3.54~M estimated cycles against a table of ten routes and
3.69~M against a table of two hundred, and within that
two-hundred-route table the difference between matching the
first entry and the last is under 3\%. Table size, in other
words, has almost stopped mattering.

The \py{RouteTable} stores route patterns in a pre-compiled
data structure and resolves incoming paths by iterating over
all registered entries, performing segment-by-segment matching
for each candidate. Constant segments are compared as byte
slices, while parameterized segments are validated against their
declared type (integer, UUID, etc.). Path parameter extraction
is performed in the same pass as pattern matching, avoiding a
second traversal of the URL. The table exposes three operations:

\begin{itemize}[leftmargin=2em, itemsep=4pt]
  \item \py{add_entry(pattern, methods, index)}: registers a
    route pattern with its allowed HTTP methods and an index
    that maps back to the Python route object.
  \item \py{resolve(path, method)}: matches a request path
    against all registered patterns and returns a
    \py{ResolveResult} containing the match type (full, mount,
    or not found), the route index, the extracted path
    parameters, and the set of allowed methods.
  \item Path parameter extraction is performed in the same
    pass as pattern matching, avoiding a second traversal of
    the URL.
\end{itemize}

The Rust JSON encoder (\py{json_encoder}) serves the same
purpose for response serialization: it replaces Python's
\py{json.dumps()} with a compiled implementation that handles
the common case (flat dictionaries of strings and numbers)
significantly faster.

\subsection{Class-based endpoints}
\label{sec:class-endpoints}

For handlers that serve multiple HTTP methods on the same URL
path, or that need to manage a multi-step connection lifecycle,
\flama provides three base classes: \py{HTTPEndpoint},
\py{WebSocketEndpoint}, and \py{JSONRPCEndpoint}. The first
two are the most common and are described below; the third
is used internally by the MCP module
(Section~\ref{sec:mcp}) and follows the same pattern.

\subsubsection{HTTP endpoints}

An \py{HTTPEndpoint} groups related handlers into a single
class, with one method per HTTP verb:

\begin{lstlisting}
import typing as t
from flama import schemas
from flama.endpoints import HTTPEndpoint

UserUpdateRequest = t.Annotated[
    schemas.Schema, schemas.SchemaMetadata(UserUpdate)
]

class UserEndpoint(HTTPEndpoint):
    async def get(self, user_id: int) -> UserResponse:
        """Retrieve a user."""
        ...

    async def put(
        self, user_id: int, data: UserUpdateRequest
    ) -> UserResponse:
        """Update a user."""
        ...

    async def delete(self, user_id: int) -> None:
        """Delete a user."""
        ...
\end{lstlisting}

The framework introspects the class at registration time to
determine the set of allowed methods. Requests with unsupported
methods receive a 405~Method Not Allowed response. HEAD requests
are served automatically by any endpoint that implements GET.

Class-based endpoints are registered with the same decorator or
programmatic API as function handlers:

\begin{lstlisting}
app.route("/users/{user_id:int}")(UserEndpoint)
\end{lstlisting}

\subsubsection{WebSocket endpoints}

WebSocket connections \citep{rfc6455ws} follow a three-phase
lifecycle modelled by the \py{WebSocketEndpoint} class:

\begin{lstlisting}
from flama.endpoints import WebSocketEndpoint

class ChatEndpoint(WebSocketEndpoint):
    encoding = "json"

    async def on_connect(self, websocket):
        await websocket.accept()

    async def on_receive(self, websocket, data):
        response = process_message(data)
        await websocket.send(json=response)

    async def on_disconnect(self, websocket, websocket_code):
        cleanup(websocket)
\end{lstlisting}

The \py{encoding} attribute determines how incoming messages are
parsed (\texttt{"json"}, \texttt{"text"}, or \texttt{"bytes"};
it is \py{None} by default, which negotiates). The framework
calls \py{on_connect} when the handshake completes,
\py{on_receive} for each incoming message, and
\py{on_disconnect} when the connection closes. Sending is done
through the single \py{send()} method, which takes the payload
as one of the keyword arguments \py{data}, \py{json}, or
\py{message} rather than through per-type methods; reading
directly, outside the \py{on_receive} hook, is the mirror-image
\py{receive(data="text")}.

WebSocket routes are registered with the same decorator syntax
as HTTP routes:

\begin{lstlisting}
@app.websocket_route("/ws/updates")
class LiveUpdates(WebSocketEndpoint):
    encoding = "json"

    async def on_connect(self, websocket):
        await websocket.accept()

    async def on_receive(self, websocket, data):
        result = await compute(data)
        await websocket.send(json=result)

    async def on_disconnect(self, websocket, websocket_code):
        pass
\end{lstlisting}

The equivalent programmatic API is available through
\py{app.add\_websocket\_route(path, endpoint)}. Both approaches
produce a \py{WebSocketRoute} object in the route table.

\subsubsection{Streaming responses}

For HTTP endpoints that need to produce output incrementally
(large file downloads, real-time event streams, or progressive
result delivery), \flama provides three specialised response
classes that stream their content to the client without buffering
the entire body in memory.

\paragraph{General streaming.}
The \py{StreamingResponse} class accepts a synchronous or
asynchronous generator and emits each yielded chunk directly to
the client:

\begin{lstlisting}
from flama import Flama
from flama.http.responses import StreamingResponse

app = Flama()

async def generate_report():
    for chunk in compute_report_chunks():
        yield chunk

@app.get("/report")
async def stream_report():
    return StreamingResponse(
        generate_report(),
        media_type="text/csv",
    )
\end{lstlisting}

\paragraph{Server-Sent Events (SSE).}
The \py{ServerSentEventResponse} implements the WHATWG
Server-Sent Events living standard \citep{sse}, setting the
\http{Content-Type} to \texttt{text/event-stream},
\http{Cache-Control} to \texttt{no-cache}, and
\http{Connection} to \texttt{keep-alive}. Each yielded item is
either a plain string (sent as a bare \texttt{data:} line) or a
\py{ServerSentEvent} object, serialised as an SSE frame with
optional \texttt{id}, \texttt{event}, \texttt{data}, and
\texttt{retry} fields. A \py{ServerSentEvent} with only
\texttt{comment} set is emitted as a comment-only heartbeat
line, keeping the connection alive across idle intervals
without dispatching a client-side event:

\begin{lstlisting}
from flama.http.responses import ServerSentEvent, ServerSentEventResponse

async def token_stream():
    for token in model.generate_tokens(prompt):
        yield ServerSentEvent(event="token", data=token,
                              id=str(seq))

@app.get("/stream")
async def sse_endpoint():
    return ServerSentEventResponse(token_stream())
\end{lstlisting}

SSE is the primary streaming format for the LLM serving
subsystem's OpenAI, Anthropic, and native dialects. Client
libraries can resume interrupted streams via the
\texttt{Last-Event-ID} header, and the native dialect leverages
this for transparent reconnection with sequence-based replay.

\paragraph{Newline-Delimited JSON (NDJSON).}
The \py{NDJSONResponse} \citep{ndjson} emits one compact JSON object per line
with the \http{Content-Type} set to
\texttt{application/x-ndjson}. Each yielded item is encoded as
a self-contained JSON line followed by a newline character:

\begin{lstlisting}
from flama.http.responses import NDJSONResponse

async def generate_events():
    async for event in engine.stream():
        yield event.to_dict()

@app.get("/events")
async def ndjson_endpoint():
    return NDJSONResponse(generate_events())
\end{lstlisting}

NDJSON is the streaming format used by the Ollama dialect.
Unlike SSE, NDJSON does not support named event types or client
reconnection semantics, but its simplicity makes it well-suited
for log-style streaming where each line is a self-contained
record.

\medskip

Combined with WebSocket endpoints, these three streaming
response types give \flama full support for real-time
communication patterns: WebSocket for bidirectional messaging,
SSE for structured event streams with reconnection, NDJSON for
line-oriented progressive delivery, and \py{StreamingResponse}
for arbitrary binary or text streaming.

\subsection{OpenAPI schema generation}
\label{sec:openapi}

Every registered route contributes to an OpenAPI~3.2.0
specification \citep{openapi} that is built at startup by the
\py{SchemaModule}. The generation process is automatic and
requires no additional annotations beyond the handler's type
signature:

\begin{itemize}[leftmargin=2em, itemsep=4pt]
  \item Path parameters become OpenAPI path parameter objects,
    with types inferred from the route pattern's type converters.
  \item Query parameters are detected from handler parameters
    that are not path parameters and not schema objects.
  \item Request bodies are generated from handler parameters
    whose type is a registered schema class.
  \item Response schemas are generated from the handler's return
    type annotation.
  \item Schema classes are converted to reusable
    \texttt{components/schemas} definitions via the schema
    adapter's JSON Schema emission.
\end{itemize}

Handlers can supply additional metadata (tags, summaries,
descriptions, and response codes) as YAML in their docstrings.
The docstring is split on \texttt{---} and its last segment is
parsed with \py{yaml.safe\_load()}, then merged into the
generated OpenAPI operation object. The separator is therefore
only needed when the docstring opens with human-readable prose
that should not be parsed as YAML; a docstring that is entirely
YAML, as below, needs no separator:

\begin{lstlisting}
@app.get("/users/{user_id:int}")
async def get_user(user_id: int) -> UserResponse:
    """
    tags:
        - Users
    summary:
        Retrieve a user by ID
    responses:
        200:
            description: The requested user
        404:
            description: User not found
    """
    ...
\end{lstlisting}

The generated specification is served as JSON at \http{/schema/}
and rendered as an interactive Swagger UI at \http{/docs/}.

\section{Schema and data validation}
\label{sec:schemas}

Validation is the business of checking that incoming data has
the structure and types it claims before any of it reaches
application logic. For a web API that means path parameters,
query strings, request bodies, and, if the developer wants it,
response bodies too. When something fails the check the client
deserves an error it can act on, which in \flama's case is a
400 carrying a per-field \py{detail} object rather than a bare
status line. Skip the check and malformed input travels
inward instead, surfacing later as a cryptic runtime error, as
corrupted data, or as a security problem.

Four things make up the machinery: the adapter that keeps the
framework independent of any one schema library, the internal
schema representation everything else is written against, the
validation pipeline itself, and response serialization.

\subsection{The adapter pattern}
\label{sec:schema-adapter}

Python's ecosystem offers several mature libraries for data
validation and serialization. Each library has a distinct API,
a different approach to field declaration, and its own trade-offs
between performance, flexibility, and ease of use. Coupling a
web framework to a single validation library forces users to
adopt that library's conventions for their entire codebase, even
if another library is better suited to their domain or their
team's existing investment.

\flama avoids this coupling by interposing an adapter layer
between the framework's validation pipeline and the schema
library. The adapter normalizes three operations:

\begin{enumerate}[leftmargin=2em, itemsep=4pt]
  \item \textbf{Field introspection}: given a schema class,
    extract the list of fields, their types, their nullability,
    their default values, and whether they are required.
  \item \textbf{Validation}: given a dictionary of raw data and
    a schema class, validate the data and return either a
    validated object or a list of per-field errors.
  \item \textbf{JSON Schema emission}: given a schema class,
    produce a JSON Schema object suitable for inclusion in an
    OpenAPI specification.
\end{enumerate}

Three adapters are provided:

\begin{center}
\small
\begin{tabularx}{\textwidth}{@{}lX@{}}
\toprule
\textbf{Library} & \textbf{Characteristics} \\
\midrule
Pydantic \citep{pydantic} &
  The most widely adopted validation library.
  Rust-accelerated core validation. Declarative field
  definitions with Python type annotations.
  First-class JSON Schema support. \\
Marshmallow \citep{marshmallow} &
  A mature library with explicit field declarations,
  a rich plugin ecosystem, and a serialization API
  oriented around \py{load()}/\py{dump()} semantics. \\
Typesystem \citep{typesystem} &
  A lightweight library focused on data validation
  and form rendering. Suitable for small projects
  where Pydantic or Marshmallow would be
  over-specified. \\
\bottomrule
\end{tabularx}
\end{center}

The adapter is selected at application construction time
and applies globally:

\begin{lstlisting}
app = Flama(schema_library="pydantic")
\end{lstlisting}

Users define schemas using their chosen library's native API
(e.g.\ Pydantic \py{BaseModel} subclasses, Marshmallow
\py{Schema} subclasses) and reference them in handler
signatures. The framework translates between the library's
types and its own internal model transparently.

\subsection{Internal schema representation}
\label{sec:internal-schema}

The framework's core validation pipeline operates on an internal
schema representation that is independent of any external
library. This representation consists of two frozen data classes:

\begin{description}[leftmargin=2em, itemsep=4pt,
  font=\sffamily\bfseries]
  \item[Field]
    represents a single typed field. It records the field's name,
    Python type, nullability, whether the field is required,
    its default value (if any), whether it accepts multiple
    values (i.e.\ a list), and its JSON Schema equivalent. Fields
    are the atomic unit of validation: each field is validated
    independently, and per-field error messages are collected
    into the validation error response.

  \item[Schema]
    is a named collection of \py{Field} objects. A schema can be
    constructed from a type annotation
    (\py{Schema.from_type(UserUpdate)}) or from an explicit
    field list (\py{Schema.build("UserUpdate", fields=[...])}),
    enabling programmatic schema construction for generated
    resources.
\end{description}

This internal representation serves as the lingua franca
between the adapter layer and the rest of the framework. The
validation pipeline, the OpenAPI generator, and the response
serializer all consume \py{Field} and \py{Schema} objects
rather than library-specific types.

\subsection{Request validation pipeline}
\label{sec:validation-pipeline}

Validation in \flama is not a standalone function that the
handler calls manually. It is implemented as a sequence of
dependency-injection components
(Section~\ref{sec:dependency-injection}), each responsible for
validating one layer of the incoming request. The components
execute automatically during the dependency resolution phase
(Stage~4 of the request pipeline), before the handler is called.

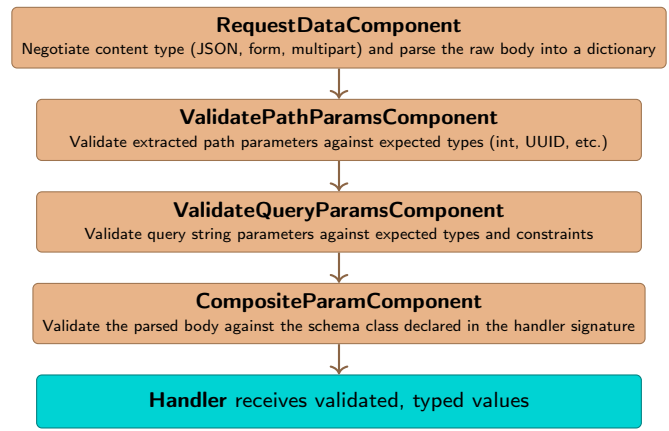
\begin{figure}[ht]
\centering
\begin{tikzpicture}[
  node distance=0.4cm,
  comp/.style={
    draw=bosque500,
    fill=bosque50,
    minimum width=8cm,
    minimum height=0.7cm,
    font=\sffamily\scriptsize,
    rounded corners=2pt,
    align=center,
  },
  arr/.style={->, thick, bosque600},
]

\node[comp] (codec) {\textbf{RequestDataComponent}\\[-1pt]
  {\tiny Negotiate content type (JSON, form, multipart)
   and parse the raw body into a dictionary}};
\node[comp, below=of codec] (path)
  {\textbf{ValidatePathParamsComponent}\\[-1pt]
  {\tiny Validate extracted path parameters against
   expected types (int, UUID, etc.)}};
\node[comp, below=of path] (query)
  {\textbf{ValidateQueryParamsComponent}\\[-1pt]
  {\tiny Validate query string parameters against
   expected types and constraints}};
\node[comp, below=of query] (body)
  {\textbf{CompositeParamComponent}\\[-1pt]
  {\tiny Validate the parsed body against the schema
   class declared in the handler signature}};
\node[comp, below=of body, fill=vortico50, draw=vortico500]
  (handler)
  {\textbf{Handler} receives validated, typed values};

\draw[arr] (codec) -- (path);
\draw[arr] (path) -- (query);
\draw[arr] (query) -- (body);
\draw[arr] (body) -- (handler);

\end{tikzpicture}
\caption{Request validation pipeline. Each component resolves
one aspect of the incoming data and passes validated, typed
values to downstream components or the handler.}
\label{fig:validation-pipeline}
\end{figure}

Four components carry a typical body-validating request through
the pipeline, described below. Two siblings not shown in the
figure, \py{PrimitiveParamComponent} and \py{FileParamComponent},
resolve primitive-typed path and query parameters and
\py{UploadFile} parameters respectively, using the same
mechanism.

\begin{enumerate}[leftmargin=2em, itemsep=4pt]
  \item \textbf{RequestDataComponent.} This component reads the
    \texttt{Content-Type} header and selects the appropriate
    codec: \py{JSONDataCodec} for \texttt{application/json},
    \py{URLEncodedCodec} for
    \texttt{application/\allowbreak x-\allowbreak www-\allowbreak form-\allowbreak urlencoded}, and
    \py{MultiPartCodec} for \texttt{multipart/\allowbreak form-\allowbreak data}. The
    codec parses the raw request body into \py{types.RequestData}.
    If no codec matches the content type, a
    \py{NoCodecAvailable} error is raised.

  \item \textbf{ValidatePathParamsComponent.} This component
    receives the path parameters extracted by the route table
    and validates them against the types declared in the route
    pattern. For example, if the route pattern is
    \texttt{/users/\{user\_id:int\}}, this component verifies
    that the extracted value is a valid integer.

  \item \textbf{ValidateQueryParamsComponent.} This component
    validates query string parameters against the types
    declared in the handler signature. Parameters that are not
    path parameters and not schema objects are assumed to be
    query parameters.

  \item \textbf{CompositeParamComponent.} This is the component
    that actually binds a value to a handler parameter annotated
    \py{t.Annotated[schemas.Schema, schemas.SchemaMetadata(X)]}
    (or \py{list[schemas.Schema]} for a collection). It reads the
    parsed body produced by \py{RequestDataComponent} and
    validates it against \py{X} through the schema adapter's
    \py{validate()} method, raising a
    \py{SchemaValidationError} on failure. A separate
    \py{ValidateRequestDataComponent} performs the same
    validation and is available for a handler that wants the
    raw validated dictionary directly, by declaring a parameter
    of type \py{ValidatedRequestData}, without binding it to a
    specific schema-annotated parameter.
\end{enumerate}

If any validation component encounters an error, the pipeline
short-circuits and the framework returns a 400~Bad Request
response with a structured JSON body. The \py{detail} object is
keyed by field name; the shape of each field's error entry
comes directly from the configured schema library's own error
representation. With the Pydantic adapter, for example:

\begin{lstlisting}[style=flamajson]
{
  "status_code": 400,
  "detail": {
    "age": {
      "type": "int_parsing",
      "loc": ["age"],
      "msg": "Input should be a valid integer, unable to
              parse string as an integer",
      "input": "not-an-int"
    }
  },
  "error": "ValidationError"
}
\end{lstlisting}

\subsection{Response serialization}
\label{sec:response-serialization}

Return values from handlers are serialized through the
\py{APIResponse} class. When the route declares a response
schema, \py{APIResponse} re-validates the returned value through
the schema adapter, which reconstructs the schema object from
the value's keys and dumps it back to a plain dictionary; the
returned value must therefore already be a dictionary (or a
list of dictionaries) keyed by the schema's field names, not an
instantiated schema class. The resulting dictionary is then
encoded to JSON using the Rust-accelerated JSON encoder.

The response type and its JSON Schema are inferred from the
handler's return type annotation and contribute to the OpenAPI
specification. This means that the return type annotation has
two purposes: it controls the runtime serialization of the
response, and it generates the documentation that clients use
to understand the response format.

\section{Dependency injection}
\label{sec:dependency-injection}

Dependency injection is how a handler's parameters get resolved
without the handler having to build or find its own dependencies
\citep{fowler2004injection}. In \flama it is not an optional
convenience layered over something simpler. It is the mechanism
the framework itself uses to hand over request data, validated
inputs, database connections, authentication tokens, and model
instances, so every request makes at least one pass through the
injector whether the developer engages with it or not.

What follows covers the component model, the resolution tree
built at startup, the per-request cache, and the context types
available without registering anything.

\subsection{The component model}
\label{sec:component-model}

A \textbf{component} is a class that knows how to produce a
value of a given type. It is the fundamental building block of
\flama's dependency injection system. Every component implements
two methods:

\begin{enumerate}[leftmargin=2em, itemsep=4pt]
  \item \py{can_handle_parameter(parameter)}: inspects a
    parameter's type annotation and returns \py{True} if the
    component can produce a value for that parameter. The
    default implementation checks whether the parameter's type
    matches the return type of the component's \py{resolve()}
    method.

  \item \py{resolve(**kwargs)}: produces the value. The method's
    return type annotation indicates the type of the produced
    value, and the method's parameter annotations indicate the
    component's own upstream dependencies. These dependencies
    are resolved recursively using the same injection
    mechanism.
\end{enumerate}

The following example defines a component that produces an
asynchronous database connection:

\begin{lstlisting}
from flama import Component

class DatabaseConnectionComponent(Component):
    async def resolve(self, app: Flama) -> AsyncConnection:
        return await app.sqlalchemy.open_connection()
\end{lstlisting}

The component declares that it requires the \py{Flama}
application instance (a context value, always available) and
produces an \py{AsyncConnection}, which it obtains from the
\py{SQLAlchemyModule} registered on the application and reached,
like any module, under its name. When a handler declares a
parameter of type \py{AsyncConnection}, the injector identifies
this component as the provider and calls its \py{resolve()}
method with the application instance.

Components can depend on other components, forming a dependency
graph of arbitrary depth:

\begin{lstlisting}
class UserRepositoryComponent(Component):
    def resolve(self,
                connection: AsyncConnection) -> UserRepository:
        return UserRepository(connection)
\end{lstlisting}

Here, \py{UserRepositoryComponent} depends on
\py{AsyncConnection}, which is itself resolved by
\py{DatabaseConnectionComponent}. The injector resolves the full
chain automatically.

\subsection{Resolution trees}
\label{sec:resolution-trees}

At startup, the injector examines every registered handler and
builds a \textbf{resolution tree} for each of its parameters.
This tree is a directed acyclic graph (DAG) where each node
represents a value that must be produced, and each edge
represents a dependency relationship.

Each node in the tree is one of three types:

\begin{description}[leftmargin=2em, itemsep=4pt,
  font=\sffamily\bfseries]
  \item[ContextNode]
    the value is available directly from the ASGI scope or the
    request context. Context values include the \py{Request}
    object, the \py{Flama} application, the matched \py{Route},
    and the \py{WebSocket} connection (for WebSocket handlers).

  \item[ComponentNode]
    the value is produced by a registered component. The node
    stores a reference to the component instance and has child
    nodes for the component's own dependencies.

  \item[ParameterNode]
    the value is a primitive (string, integer, float, boolean,
    UUID) extracted from a path segment or query string
    parameter.
\end{description}

\begin{figure}[ht]
\centering
\begin{tikzpicture}[
  nodebase/.style={
    text width=2.75cm,
    minimum height=0.85cm,
    inner sep=3pt,
    font=\sffamily\scriptsize,
    rounded corners=2pt,
    align=center,
  },
  treenode/.style={nodebase, draw=vortico500, fill=vortico50},
  ctxnode/.style={nodebase, draw=bruma500,   fill=bruma50},
  prmnode/.style={nodebase, draw=bosque500,  fill=bosque50},
  swatch/.style={draw, minimum width=0.30cm, minimum height=0.30cm,
                 rounded corners=1pt, inner sep=0pt},
  arr/.style={->, thick, primary500},
]

\node[treenode, text width=3.2cm] (handler) at (0, 0)
  {\textbf{update\_user()}};

\node[prmnode]  (uid)     at (-5.1, -1.9)
  {user\_id: \py{int}\\(path parameter)};
\node[treenode] (data)    at (-1.7, -1.9)
  {data: UserUpdate\\(schema validation)};
\node[treenode] (session) at ( 1.7, -1.9)
  {connection:\\AsyncConnection\\(component)};
\node[treenode] (token)   at ( 5.1, -1.9)
  {token: AccessToken\\(component)};

\node[ctxnode] (request) at (-1.7, -4.0) {request: Request\\(context)};
\node[ctxnode] (app)     at ( 1.7, -4.0) {app: Flama\\(context)};
\node[ctxnode] (scope)   at ( 5.1, -4.0) {scope: Scope\\(context)};

\draw[arr] (handler) -- (uid);
\draw[arr] (handler) -- (data);
\draw[arr] (handler) -- (session);
\draw[arr] (handler) -- (token);
\draw[arr] (data)    -- (request);
\draw[arr] (session) -- (app);
\draw[arr] (token)   -- (scope);

\node[font=\sffamily\tiny, text=primary600, anchor=west]
  (legkey) at (-3.50, -5.45) {Node type:};
\node[swatch, draw=bosque500, fill=bosque50, right=0.22cm of legkey] (sw1) {};
\node[font=\sffamily\tiny, anchor=west, right=0.10cm of sw1] (lb1) {Parameter};
\node[swatch, draw=vortico500, fill=vortico50, right=0.45cm of lb1] (sw2) {};
\node[font=\sffamily\tiny, anchor=west, right=0.10cm of sw2] (lb2) {Component};
\node[swatch, draw=bruma500, fill=bruma50, right=0.45cm of lb2] (sw3) {};
\node[font=\sffamily\tiny, anchor=west, right=0.10cm of sw3] (lb3) {Context};

\end{tikzpicture}
\caption{Resolution tree for a handler with four parameters.
The injector traverses the tree bottom-up: context values are
read from the ASGI scope, component values are produced by
calling \py{resolve()}, and primitive parameters are extracted
from the URL path. The tree is compiled at startup and
flattened into an ordered sequence of resolution steps for
efficient request-time execution. Node fill indicates where a value comes from, as given in the key.}
\label{fig:resolution-tree}
\end{figure}
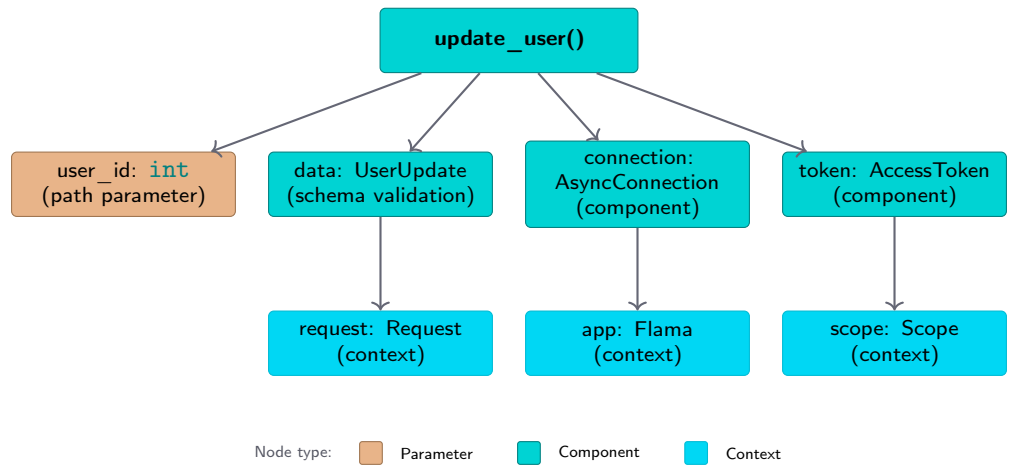

Figure~\ref{fig:resolution-tree} shows the resolution tree for
a handler with four parameters: a path parameter, a schema
validation result, a database connection (component), and an
authentication token (component). The injector compiles this
tree at startup and flattens it into an ordered sequence of
\textbf{resolution steps} that can be executed without graph
traversal at request time. Circular dependencies are detected
during compilation and raise a descriptive error.

\subsection{Caching}
\label{sec:di-caching}

Components may declare \py{cacheable = True} (the default), in
which case their resolved value is stored in a per-request LRU
cache keyed by the component's identity string. The cache
ensures that a database connection, an authentication token, or
any other expensive-to-produce value is resolved at most once
per request, even if multiple handler parameters or nested
components depend on it.

The cache is scoped to a single request and is discarded after
the response has been sent. This prevents stale values from
leaking across requests and ensures that each request sees a
fresh set of dependencies.

\subsection{Context types}
\label{sec:context-types}

The following types are available as context values without
requiring a component. Any handler or component can request them
by declaring parameters with these type annotations:

\begin{center}
\small
\begin{tabularx}{\textwidth}{@{}lX@{}}
\toprule
\textbf{Type} & \textbf{Value} \\
\midrule
\py{Scope} &
  The ASGI scope dictionary, containing the request
  method, path, headers, query string, and client
  address. \\
\py{Receive} &
  The ASGI receive channel, from which request body
  chunks are read. \\
\py{Send} &
  The ASGI send channel, through which response
  headers and body are emitted. \\
\py{Request} &
  The HTTP request object, with properties for method,
  path, query parameters, headers, cookies, and
  methods for reading the body. \\
\py{Response} &
  The HTTP response object. Available in middleware
  and post-processing hooks. \\
\py{WebSocket} &
  The WebSocket connection object. Available only in
  WebSocket handlers. \\
\py{App} &
  The application instance (\py{Flama}). Provides
  access to the application's state, configuration,
  modules, and registered routes. \\
\py{Route} &
  The matched route object. Contains the route pattern,
  allowed methods, and handler reference. \\
\py{Exception} &
  The current exception, if the handler is being
  invoked in an error-handling context. \\
\bottomrule
\end{tabularx}
\end{center}

\section{Resources and domain-driven design}
\label{sec:resources-ddd}

A SQLAlchemy table and a schema class are enough, between them,
to generate nine REST endpoints with correct status codes,
pagination, and error handling, backed by the domain-driven
design patterns that keep the transactions honest. This is the
part of \flama that saves the most typing, and it is also the
part that most needs an escape hatch, since a generated CRUD
API is only useful for as long as it does what the domain
wants. Overriding a single operation is therefore possible
without giving up the other eight.

\subsection{The resource abstraction}
\label{sec:resource-abstraction}

A \textbf{resource} in \flama is a class that declares the data
model, the validation schema, and the set of operations that an
API exposes over a domain entity. The framework provides two
resource base classes:

\begin{description}[leftmargin=2em, itemsep=4pt,
  font=\sffamily\bfseries]
  \item[Resource]
    a generic resource for virtual entities that do not
    correspond to a database table: health checks, calculators,
    proxies to external APIs, or aggregation endpoints. The
    developer defines the operations explicitly.

  \item[CRUDResource]
    a database-backed resource that derives its operations from
    a SQLAlchemy \citep{sqlalchemy} table definition. The
    framework's metaclass inspects the table columns, generates
    input and output schemas, creates a repository class, and
    produces handler methods for all standard CRUD operations.
\end{description}

\subsection{CRUD resource declaration}
\label{sec:crud-declaration}

A \py{CRUDResource} is declared by providing three pieces of
information: the SQLAlchemy table, the schema class for
output (and optionally separate schemas for input), and a
human-readable name:

\begin{lstlisting}
import sqlalchemy
from flama.resources.crud import CRUDResource

metadata = sqlalchemy.MetaData()

users = sqlalchemy.Table(
    "users",
    metadata,
    sqlalchemy.Column("id", sqlalchemy.Integer,
                      primary_key=True,
                      autoincrement=True),
    sqlalchemy.Column("name", sqlalchemy.String(100),
                      nullable=False),
    sqlalchemy.Column("email", sqlalchemy.String(200),
                      nullable=False, unique=True),
    sqlalchemy.Column("created_at", sqlalchemy.DateTime,
                      server_default=sqlalchemy.func.now()),
)

class UserResource(CRUDResource):
    name = "user"
    verbose_name = "User"
    model = users
    schema = UserSchema
    input_schema = UserInputSchema
    output_schema = UserOutputSchema
\end{lstlisting}

The class declaration only defines the resource; it still has to
be attached to the application, which mounts its generated
routes under the given path prefix:

\begin{lstlisting}
app.resources.add_resource("/users/", UserResource)
\end{lstlisting}

\py{CRUDResource} operates on an async SQLAlchemy connection, so
the application also needs the \py{SQLAlchemyModule}
(Section~\ref{sec:ddd-patterns}) registered to actually have a
database engine to connect to; without it, the generated
handlers have nothing to run their queries against.

From this declaration, the metaclass generates the following
endpoints:

\begin{center}
\small
\begin{tabularx}{\textwidth}{@{}lllX@{}}
\toprule
\textbf{Method} & \textbf{Path} &
\textbf{Status} & \textbf{Behaviour} \\
\midrule
\http{POST} & \texttt{/users/} & 201 &
  Validate input, insert row, return created resource. \\
\http{GET} & \texttt{/users/\{resource\_id\}/} & 200 &
  Retrieve single resource by primary key, or 404. \\
\http{PUT} & \texttt{/users/\{resource\_id\}/} & 200 &
  Full replacement of a single resource. \\
\http{PATCH} & \texttt{/users/\{resource\_id\}/} & 200 &
  Partial update of a single resource. \\
\http{DELETE} & \texttt{/users/\{resource\_id\}/} & 204 &
  Delete single resource, or 404. \\
\http{GET} & \texttt{/users/} & 200 &
  List resources with pagination. \\
\http{PUT} & \texttt{/users/} & 200 &
  Bulk replace all resources. \\
\http{PATCH} & \texttt{/users/} & 200 &
  Bulk partial replace of all resources. \\
\http{DELETE} & \texttt{/users/} & 204 &
  Bulk delete all resources (drop), reporting the number
  deleted. \\
\bottomrule
\end{tabularx}
\end{center}

The single-resource routes are keyed on \texttt{resource\_id}
rather than the column's own name, which is also the keyword
that \py{app.resolve\_url()} expects when building a URL for
one of them (e.g.\
\py{app.resolve\_url("user:retrieve", resource\_id=1)}).

Each operation includes error handling: integrity errors (e.g.\
a duplicate email violating a unique constraint) are caught as
\py{IntegrityError} and produce 400~Bad Request responses, and
missing resources produce 404~Not Found responses. Custom
methods can be added using the \py{@ResourceRoute.method()}
decorator, and any generated method can be overridden by
defining it in the subclass.

\subsection{Domain-driven design patterns}
\label{sec:ddd-patterns}

The generated CRUD endpoints do not interact with the database
directly. Instead, they delegate data access and transactional
management to two domain-driven design patterns
\citep{evans2003ddd, fowler2002patterns}: the Repository and
the Unit of Work (called Worker in \flama).

\subsubsection{The Repository pattern}

A \textbf{repository} encapsulates all interaction with a data
source behind a collection-like interface. The calling code
does not know whether the data comes from a relational database,
an in-memory cache, or a remote HTTP service. This separation
has three benefits: it makes the data access logic testable in
isolation, it allows the data source to be changed without
modifying the business logic, and it provides a clear boundary
for caching and query optimization.

\flama provides two repository families:

\begin{description}[leftmargin=2em, itemsep=4pt,
  font=\sffamily\bfseries]
  \item[SQLAlchemyTableRepository]
    operates on a SQLAlchemy table using async connections. Every
    method other than \py{create} locates rows through
    SQLAlchemy clauses and exact-match filters passed as
    \py{*clauses}/\py{**filters}, the same vocabulary as a
    SQLAlchemy \py{select()}, rather than a bare primary key.
    It provides six methods:
    \py{create(*data)} is variadic, inserting one row per
    positional dict and always returning a list of the created
    rows;
    \py{retrieve(*clauses, **filters)} fetches a single row,
    raising if none or more than one match;
    \py{update(data, *clauses, **filters)} updates the matching
    rows with \py{data} and returns them;
    \py{delete(*clauses, **filters)} removes the matching rows;
    \py{list(*clauses, order_by=None, order_direction="asc",
    **filters)} returns an async iterator over the matching rows
    (compatible with the pagination system);
    \py{drop(*clauses, **filters)} removes the matching rows and
    returns how many were dropped.

  \item[HTTPResourceRepository]
    proxies CRUD operations over HTTP to a remote service, so
    one \flama application can consume resources exposed by
    another while its calling code still speaks in repository
    terms. A subclass declares the remote resource's path
    (\py{\_resource = "/user"}) and is constructed with a
    \py{Client} pointing at the service. Because it addresses a
    REST resource rather than a table, it is keyed on identity
    rather than on query clauses: \py{create(data)},
    \py{retrieve(id)}, \py{update(id, data)},
    \py{partial\_update(id, data)}, and \py{delete(id)} act on a
    single record, while \py{list()}, \py{replace(data)},
    \py{partial\_replace(data)}, and \py{drop()} act on the
    collection. \py{list()} follows the remote endpoint's
    pagination transparently, yielding records across pages as
    an async iterable.
\end{description}

The two families therefore share a vocabulary rather than a
literal signature: the SQLAlchemy repository selects rows with
clauses and filters, while the HTTP repository addresses
resources by identifier, because that is what each underlying
data source exposes.

\subsubsection{The Unit of Work pattern (Worker)}

A \textbf{Worker} is the \flama implementation of the Unit of
Work pattern. The transactional boundary logic (connection
acquisition, begin, commit, and rollback) is defined by the
abstract base class \py{AbstractWorker}, which implements the
async context manager protocol (\py{__aenter__}/\py{__aexit__}).
The concrete \py{Worker} subclass provides default (no-op)
implementations of \py{set\_up()}, \py{tear\_down()},
\py{commit()}, and \py{rollback()}, suitable for applications
that override these methods with database-specific behaviour.
All repository operations within a Worker's context execute
atomically: either all operations commit, or all are rolled
back.

In practice, a \py{Worker} for a SQLAlchemy-backed application
subclasses \py{SQLAlchemyWorker} rather than the bare
\py{Worker}, which supplies the \py{set\_up()}/\py{tear\_down()}/
\py{commit()}/\py{rollback()} implementations that talk to a
SQLAlchemy connection:

\begin{lstlisting}
from flama.ddd.workers.sqlalchemy import SQLAlchemyWorker

class AppWorker(SQLAlchemyWorker):
    users: UserRepository
    posts: PostRepository
\end{lstlisting}

The Worker's type annotations declare which repositories it
manages; repository instances are created lazily from these
annotations when the Worker is used as an async context manager.
A \py{WorkerComponent} makes a specific Worker instance available
for dependency injection, so a handler that declares a parameter
of that Worker's type receives a fully configured instance; the
component still has to be registered explicitly, alongside the
\py{SQLAlchemyModule} that gives it a database engine to connect
to:

\begin{lstlisting}
from flama import Flama
from flama.ddd import WorkerComponent
from flama.sqlalchemy import SQLAlchemyModule

app = Flama(
    modules=[SQLAlchemyModule(DATABASE_URL)],
    components=[WorkerComponent(worker=AppWorker())],
)

@app.post("/users/")
async def create_user(worker: AppWorker) -> None:
    async with worker:
        # create() is variadic and always returns a list, one
        # created row per positional dict passed in.
        [user] = await worker.users.create(
            {"name": "Ada", "email": "ada@example.com"}
        )
        await worker.posts.create(
            {"author_id": user["id"],
             "title": "First post",
             "body": "Hello, world."}
        )
        # Commits on successful exit.
        # Rolls back on exception.
\end{lstlisting}

The Worker manages four lifecycle operations: connection
acquisition, transaction begin, commit (on successful exit of
the \py{async with} block), and rollback (on exception).

\begin{figure}[ht]
\centering
\begin{tikzpicture}[
  node distance=0.6cm and 1.5cm,
  box/.style={
    draw=#1,
    fill=#1!8!white,
    minimum width=3.5cm,
    minimum height=0.9cm,
    font=\sffamily\scriptsize,
    rounded corners=2pt,
    align=center,
  },
  arr/.style={->, thick, primary500},
]

\node[box=vortico500] (handler)
  {\textbf{Handler}\\(receives Worker via DI)};
\node[box=flama500, below=of handler]
  (worker) {\textbf{Worker}\\(transactional boundary)};
\node[box=bosque500, below left=0.7cm and 0.3cm of worker]
  (repo1) {\textbf{UserRepository}\\(CRUD operations)};
\node[box=bosque500, below right=0.7cm and 0.3cm of worker]
  (repo2) {\textbf{PostRepository}\\(CRUD operations)};
\node[box=bruma500, below=2.6cm of worker]
  (db) {\textbf{Database}\\(async connection, transaction)};

\draw[arr] (handler) -- node[right, font=\tiny\sffamily,
  text=primary600] {inject} (worker);
\draw[arr] (worker) -- (repo1);
\draw[arr] (worker) -- (repo2);
\draw[arr] (repo1) -- (db);
\draw[arr] (repo2) -- (db);

\end{tikzpicture}
\caption{Domain-driven design integration. The handler receives
a Worker via dependency injection. The Worker manages
repositories within a transactional boundary. All operations
within the \py{async with} block either commit together or roll
back together.}
\label{fig:ddd}
\end{figure}
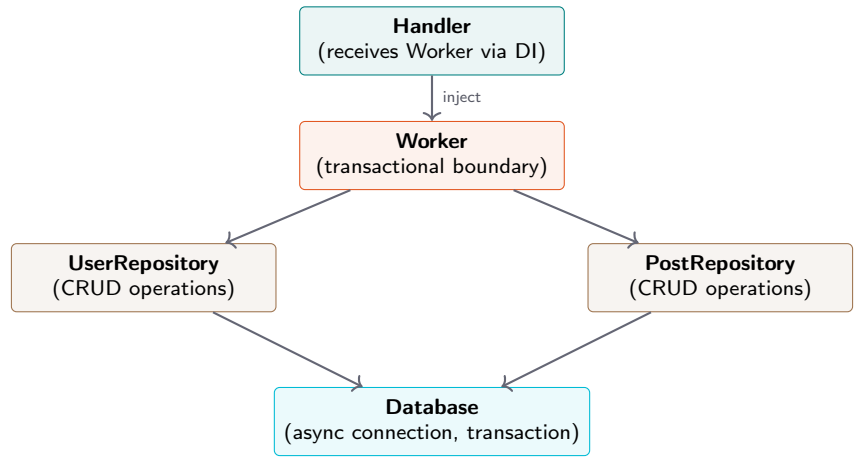

\section{Authentication and authorization}
\label{sec:authentication}

\flama ships its own JWT \citep{rfc7519jwt} implementation
rather than reaching for a third-party library, and the reason
is integration rather than distrust of the alternatives. A
token that the framework decodes itself can be handed to a
handler as a typed parameter by the same injector that supplies
everything else, checked by middleware that already knows how
routes are tagged, and described in the generated OpenAPI
document without a separate registration step. Three pieces
follow: the JWT implementation, the components that make a
decoded token available to handlers, and the middleware that
turns its claims into route-level access control.

\subsection{JWT implementation}
\label{sec:jwt-implementation}

The framework provides a complete JWT implementation covering
the three aspects of token-based authentication:

\begin{description}[leftmargin=2em, itemsep=4pt,
  font=\sffamily\bfseries]
  \item[Token encoding and decoding.]
    The \py{JWT} class encodes a payload dictionary into a
    signed token string and decodes a token string back into
    a payload, verifying the signature in the process.
    Standard claims (\texttt{iss}, \texttt{sub}, \texttt{aud},
    \texttt{exp}, \texttt{iat}, \texttt{nbf}) are validated
    automatically on decoding. Expired tokens and tokens with
    future \texttt{nbf} claims are rejected.

  \item[Signing algorithms.]
    HMAC-SHA256 (HS256) is the default signing algorithm.
    HS384 and HS512 are also implemented, for applications that
    require a longer MAC. All three are symmetric: the signing
    key is shared between the token issuer and the token
    consumer. Asymmetric algorithms (RSA, ECDSA) are not
    implemented, so issuer and consumer must currently share the
    same secret key.

  \item[JSON Web Signature (JWS).]
    The compact serialization format is implemented following
    the {JSON} Web Signature specification \citep{rfc7515jws}.
    A token consists of three Base64-URL-encoded
    segments separated by dots: the header (algorithm and token
    type), the payload (claims), and the signature.
\end{description}

\subsection{Token components}
\label{sec:token-components}

Authentication integrates with the dependency injection system
through two components, each of which resolves a typed token
object from the incoming request:

\begin{description}[leftmargin=2em, itemsep=4pt,
  font=\sffamily\bfseries]
  \item[AccessTokenComponent]
    extracts the JWT from an \texttt{access\_token} header or
    an \texttt{access\_token} cookie (header checked first), by
    default expecting the header value in the form
    \texttt{Bearer~<token>}. It decodes the token against the
    secret it was constructed with and returns an
    \py{AccessToken} object containing the decoded header and
    payload.

  \item[RefreshTokenComponent]
    performs the same extraction and decoding for refresh
    tokens, reading a \texttt{refresh\_token} header or cookie
    instead. Refresh tokens are used in token rotation flows
    where the client exchanges an expired access token and a
    valid refresh token for a new pair of tokens.
\end{description}

Neither component is registered by default: both take the
signing secret as a required constructor argument, so the
application declares the ones it needs explicitly, alongside
the middleware that enforces permissions on top of them
(Section~\ref{sec:permission-middleware}):

\begin{lstlisting}
from flama import Flama
from flama.authentication import AccessTokenComponent, AuthenticationMiddleware

app = Flama(
    components=[AccessTokenComponent(secret=SECRET_KEY)],
    middleware=[AuthenticationMiddleware()],
)
\end{lstlisting}

Any handler that needs to know the identity of the authenticated
user simply declares a parameter of type \py{AccessToken}:

\begin{lstlisting}
import typing as t
from flama import schemas
from flama.authentication import AccessToken

ProfileResponse = t.Annotated[
    schemas.Schema, schemas.SchemaMetadata(Profile)
]

@app.get("/profile")
async def get_profile(token: AccessToken) -> ProfileResponse:
    user_id = token.payload.sub
    return await load_profile(user_id)
\end{lstlisting}

The framework resolves the token transparently. If the request
does not contain a valid token, the component raises an
exception that the middleware translates into a
401~Unauthorized response. Standard claims
(\py{iss}/\py{sub}/\py{aud}/\py{exp}/\py{nbf}/\py{iat}/\py{jti})
are attributes of \py{token.payload}; anything else, such as
application-specific user data, is namespaced under
\py{token.payload.data}, a plain dictionary, since it falls
outside the JWT standard and is not validated on decoding.

\subsection{Permission-based middleware}
\label{sec:permission-middleware}

The \py{AuthenticationMiddleware} enforces route-level access
control based on JWT permission claims. Routes are tagged with
required permissions at registration time:

\begin{lstlisting}
@app.route("/admin/users", methods=["DELETE"],
           tags={"permissions": ["admin", "delete"]})
async def delete_all_users() -> None:
    ...
\end{lstlisting}

When a request arrives at a route whose tags declare required
permissions, the middleware performs the following steps:

\begin{enumerate}[leftmargin=2em, itemsep=4pt]
  \item Resolve an \py{AccessToken} for the request, the same
    way a handler parameter of that type would be resolved.
  \item Read the user's permissions from
    \py{token.payload.data["permissions"]} (a list of strings),
    unioned with every permission listed under each role in
    \py{token.payload.data["roles"]} (a mapping of role name to
    its list of permissions), so a client can be granted access
    directly or through a role.
  \item Verify that the user's permissions are a superset of
    the route's required permissions.
  \item If the check passes, forward the request to the handler.
    If the token is missing or invalid, return
    401~Unauthorized. If the token is valid but the permissions
    are insufficient, return 403~Forbidden.
\end{enumerate}

The tag key the middleware reads (\texttt{permissions} by
default) and a list of URL regex patterns to exempt from
authentication entirely are both configurable on
\py{AuthenticationMiddleware}'s constructor.

This design separates authentication (verifying identity) from
authorization (verifying permissions): the token components
handle authentication, and the middleware handles authorization.
The two concerns can be used independently: a handler can
request an \py{AccessToken} without the middleware being active,
and the middleware can enforce permissions without the handler
needing to inspect the token.

\section{Pagination}
\label{sec:pagination}

A list endpoint over a table of any size eventually has to
answer the question of what to leave out. Serialising a
million rows into one JSON response wastes bandwidth, delays
the first byte, and may simply exhaust the client, so what is
wanted instead is a way for the caller to ask for a slice and a
standard envelope telling it which slice arrived. \flama offers
two strategies for this. Both work by wrapping the list handler,
adding their query parameters to its signature without the
handler's code being aware of them, and folding the returned
collection into the envelope on the way out.

\subsection{Pagination strategies}

\begin{description}[leftmargin=2em, itemsep=4pt,
  font=\sffamily\bfseries]
  \item[Page-number pagination.]
    The client specifies a \texttt{page} (1-indexed) and a
    \texttt{page\_size}. This strategy is natural for tabular
    UIs where the user navigates between numbered pages.

  \item[Limit-offset pagination.]
    The client specifies a \texttt{limit} (maximum number of
    items) and an \texttt{offset} (number of items to skip).
    This strategy is suited to infinite-scroll interfaces and
    cursor-based navigation.
\end{description}

Both strategies support an optional \texttt{count} query
parameter (boolean, default \py{False}) that controls whether
the total number of items is reported. It is off by default
because counting is the expensive part of paginating a large
table; when omitted, the envelope still carries the pagination
parameters but reports \py{null} for the count.

\subsection{Usage}

Pagination is applied declaratively at the route level:

\begin{lstlisting}
@app.route("/items/", pagination="page_number")
async def list_items(**kwargs) -> list[Item]:
    return await repository.list()
\end{lstlisting}

The \py{**kwargs} is not decoration. The paginator rewrites the
handler's signature to add its own query parameters, and it
refuses to wrap a handler that has nowhere to put them, raising
\py{TypeError} at registration time if the parameter is absent.

The paginator intercepts the handler's return value, slices the
collection according to the pagination parameters, and wraps the
result in a response envelope:

A request to \texttt{/items/?page=2\&page\_size=20\&count=true}
then produces:

\begin{lstlisting}[style=flamajson]
{
  "meta": {
    "page": 2,
    "page_size": 20,
    "count": 148
  },
  "data": [
    {"id": 21, "name": "Widget A"},
    {"id": 22, "name": "Widget B"}
  ]
}
\end{lstlisting}

The \texttt{meta} object provides the pagination parameters used
for the current request and the total count (\py{null} unless
\texttt{count=true} was requested). For limit-offset pagination,
the \texttt{meta} object contains \texttt{limit},
\texttt{offset}, and \texttt{count} fields instead. The default
page size is~10.

The paginator is implemented as a handler wrapper. It replaces
the handler's \py{__signature__} with one in which \py{**kwargs}
has been substituted by the strategy's own parameters, which is
why the framework can resolve and document them without the
handler ever naming them. It then intercepts the raw return
value and delegates slicing and counting to the appropriate
strategy class. Because the substitution happens on the
signature the rest of the framework reads, the pagination
parameters also appear in the auto-generated OpenAPI
specification.

\section{Background tasks}
\label{sec:background-tasks}

Sending a confirmation email, generating a report, writing an
audit log, kicking off a retraining run: none of these should
hold up the HTTP response, and all of them have to happen
somewhere. A synchronous framework leaves two options, neither
good, which are to make the caller wait or to stand up an
external queue such as Celery for what may be a single line of
work. \flama runs them itself, after the response body has gone
out, with no additional infrastructure involved.

\subsection{Concurrency models}

Two execution strategies are available, corresponding to the
two dominant forms of concurrency in Python:

\begin{description}[leftmargin=2em, itemsep=4pt,
  font=\sffamily\bfseries]
  \item[BackgroundThreadTask]
    executes the callable in a thread via
    \py{asyncio.to_thread()}. This model is appropriate for
    I/O-bound work: sending HTTP requests, writing to a
    message queue, or performing database operations.

  \item[BackgroundProcessTask]
    executes the callable in a separate process via
    \py{multiprocessing.Process}. This model is appropriate for
    CPU-bound work: model retraining, image resizing, PDF
    generation, or any computation that would block the event
    loop if run in a thread.
\end{description}

\subsection{Usage}

Tasks are created and attached to responses:

\begin{lstlisting}
import typing as t
from flama import BackgroundThreadTask, schemas
from flama.http import APIResponse

OrderRequest = t.Annotated[
    schemas.Schema, schemas.SchemaMetadata(OrderInput)
]

@app.post("/orders/")
async def create_order(data: OrderRequest):
    order = await save_order(data)
    task = BackgroundThreadTask(
        send_confirmation_email,
        recipient=order["email"],
        order_id=order["id"],
    )
    return APIResponse(order, status_code=201,
                       background=task)
\end{lstlisting}

The response is sent immediately. The email is sent in a
background thread after the response body has been fully
transmitted.

For handlers that need to schedule multiple tasks, the
\py{BackgroundTasks} class aggregates them into a single
container:

\begin{lstlisting}
from flama import BackgroundTasks

tasks = BackgroundTasks()
tasks.add_task("thread", send_email, order["email"])
tasks.add_task("process", generate_invoice, order["id"])

return APIResponse(order, background=tasks)
\end{lstlisting}

Tasks in a \py{BackgroundTasks} container are executed
sequentially in the order they were added.

\section{Middleware}
\label{sec:middleware}

Logging, error handling, authentication, compression, CORS:
concerns that belong to every request and to no handler in
particular. Middleware is where they live. In \flama it is an
ordered stack of ASGI callables each wrapping the next, on the
onion model that ASGI and WSGI frameworks have converged on,
where a request enters at the outermost layer, works inward
through each wrapper to the handler, and the response comes
back out through the same layers in reverse. Any layer may
inspect or modify what passes through it in either direction,
or answer immediately and let nothing further in.

\subsection{Middleware stack}

Construction is bottom-up, so the middleware registered last
ends up outermost and sees each request first. Worth keeping in
mind when order matters: authentication that must run before
compression has to be registered after it.

Depth costs less than one might expect. Measured under
Callgrind, a request through an application with no middleware
at all costs 3.52~M estimated cycles, with five 3.53~M, and with
ten 3.54~M, so each additional layer adds on the order of
0.1\%. The stack is cheap enough that the decision about what
to put in it can be made on architectural grounds rather than
on a per-request budget.

\subsection{Built-in middleware}

\flama ships with the following middleware:

\begin{center}
\small
\begin{tabularx}{\textwidth}{@{}lX@{}}
\toprule
\textbf{Middleware} & \textbf{Purpose} \\
\midrule
BaseHTTPMiddleware &
  A convenience class for writing HTTP middleware
  using a simple \py{dispatch(request, call_next)}
  interface instead of raw ASGI callables. \\
ExceptionMiddleware &
  Maps Python exceptions to HTTP responses. Supports
  custom handlers per exception type. In debug mode,
  renders interactive HTML error pages with full
  tracebacks, source code context, and local variable
  values. \\
ServerErrorMiddleware &
  Catches any exception that escapes the
  ExceptionMiddleware. Produces a generic
  500~Internal Server Error response and logs the
  traceback. \\
CORSMiddleware &
  Adds Cross-Origin Resource Sharing headers.
  Configurable allowed origins, methods, and
  headers. Handles preflight OPTIONS requests
  automatically. \\
CompressionMiddleware &
  Negotiates a compression codec from the client's
  \texttt{Accept-Encoding} header and compresses the
  response body. Brotli and gzip are tried by default,
  in that order; a configurable minimum response size
  gates when compression is attempted. \\
HTTPSRedirectMiddleware &
  Redirects HTTP requests to HTTPS with a 301
  permanent redirect. \\
TrustedHostMiddleware &
  Validates the \texttt{Host} header against a
  whitelist of allowed hosts. Returns 400 for
  requests with untrusted hosts. \\
SessionMiddleware &
  Provides cookie-based HTTP sessions. Session data is
  serialized to JSON and signed as a JWS token with
  HMAC-SHA256, with expiry enforced from the token's
  issued-at claim. \\
CorrelationIdMiddleware &
  Assigns a correlation ID to every request, taken
  from an incoming header or generated as a UUID4,
  and echoes it back on the response for downstream
  log correlation. \\
AuthenticationMiddleware &
  Enforces JWT permission-based access control
  (Section~\ref{sec:authentication}). Defined in
  the \py{flama.authentication} module rather
  than the core middleware module. \\
\bottomrule
\end{tabularx}
\end{center}

\subsection{Custom middleware}

Custom middleware subclasses \py{Middleware} and implements the
ASGI \py{__call__}. The downstream application is not passed to
the constructor; \py{MiddlewareStack} injects it as
\py{self.app} when it assembles the chain, which leaves
\py{__init__} free to take the middleware's own configuration:

\begin{lstlisting}
import time
import logging

from flama.middleware import Middleware

logger = logging.getLogger(__name__)

class TimingMiddleware(Middleware):
    async def __call__(self, scope, receive, send):
        if scope["type"] != "http":
            await self.app(scope, receive, send)
            return

        start = time.monotonic()
        await self.app(scope, receive, send)
        duration = time.monotonic() - start
        path = scope.get("path", "/")
        logger.info("%s completed in %.3fs",
                    path, duration)
\end{lstlisting}

Middleware is registered at application construction time:

\begin{lstlisting}
from flama import Flama
from flama.middleware import CORSMiddleware

app = Flama(
    middleware=[
        TimingMiddleware(),
        CORSMiddleware(allow_origins=["*"]),
    ],
)
\end{lstlisting}

What the stack receives are middleware \emph{instances}, already
configured, and it wires each one to its downstream neighbour
during startup. Earlier versions took the class and its
arguments separately, wrapped in a \py{Middleware(...)} call;
that form was removed in 2.0 and now raises \py{TypeError}.
% ==========================================================================
% End part_2_web_framework.tex
% ==========================================================================

% ============================================================
% PART III: Machine learning and generative AI
% ============================================================
\flamapart{III}{Machine Learning and Generative AI}
% ==========================================================================
% Begin part_3_machine_learning.tex
% ==========================================================================
% ============================================================
%  Part III — Machine Learning
%  Sections: ML Serving, FLM Format, Framework Support, MCP
% ============================================================

\section{Machine-learning model serving}
\label{sec:ml-serving}

The central premise of \flama is that the gap between training a
machine-learning model and deploying it behind a production API
should be as small as the gap between writing a function and
exposing it as an HTTP endpoint. In practice, this gap is often
the dominant source of friction in ML projects: the model is
trained in a notebook or a training script, exported as a pickle
file or a checkpoint directory, and then handed to a backend
engineer who must write serialization code, input validation,
error handling, health checks, and deployment scripts from
scratch. Closing that gap takes four things, and they make up
the rest of this part: a portable binary format for the
serialized model, an integration with dependency injection so
the loaded model behaves like any other dependency, a resource
abstraction that turns one class declaration into a working
endpoint, and serializers for the four frameworks that produce
most models in practice.

The order is bottom-up. The binary format that holds a model on
disk comes first, then the framework-specific serializers that
read and write it, then the mechanics of getting a serialized
model into a running application, and last the three levels of
integration on offer, which run from a single predict endpoint
to a full resource with inspection and streaming.

\subsection{Design goals}
\label{sec:ml-design-goals}

Four requirements shaped the design of the model serving
subsystem:

\begin{enumerate}[leftmargin=2em, itemsep=4pt]
  \item \textbf{Framework agnosticism.}
    The deployment system must support models trained with
    scikit-learn, TensorFlow/Keras, PyTorch, and HuggingFace
    Transformers. Adding a new framework should require only a
    serializer and a model wrapper, not changes to the serving
    infrastructure.

  \item \textbf{Self-describing artifacts.}
    A deployed model must carry its own metadata: the
    framework and version that produced it, the model's
    class name, its hyperparameters, its training metrics,
    and any auxiliary artifacts (tokenizer files, vocabulary
    files, label maps). This metadata must be accessible
    without loading the model weights, enabling fast
    inspection and cataloguing.

  \item \textbf{Efficient transport.}
    Model files are frequently large, hundreds of megabytes for
    a convolutional network and gigabytes for a language model,
    so the format has to compress without making the load path
    expensive. The asymmetry is deliberate and shows up in
    measurement: for a scikit-learn artifact under protocol~2,
    a dump costs around 43.7~M estimated cycles against 6.4~M
    for the corresponding load. Packaging happens once and
    loading happens on every cold start, so that is the right
    way round.

  \item \textbf{Zero-configuration serving.}
    Given a model file, it should be possible to expose it as
    a REST endpoint with a single command and no application
    code. The framework should generate input/output schemas,
    OpenAPI documentation, and error handling automatically.
\end{enumerate}

\section{The FLM binary format}
\label{sec:flm-format}

The \texttt{.flm} format (for \textbf{F}lama \textbf{L}earned
\textbf{M}odel) packages a serialized model, its metadata, and
whatever auxiliary artifacts it needs into one self-describing
file. Two properties drove the design. It is versioned, because
the storage requirements of a scikit-learn pickle and a
multi-file LLM checkpoint have nothing in common and a single
layout would have had to be stretched to cover both. And the
metadata sits at a known offset ahead of the weights, so
reading it costs a header parse rather than a full
decompression, which is what lets the framework decide how to
load a multi-gigabyte artifact before committing to loading
it.

\subsection{File structure}
\label{sec:flm-structure}

An FLM file consists of a 16-byte outer header followed by a
body whose layout is determined by the protocol version declared
in the header. Two protocol versions are defined: version~1 for
traditional ML models serialized as opaque binary blobs, and
version~2 for models that may be stored as directory bundles
(such as LLM checkpoints) and that benefit from per-section
compression control.

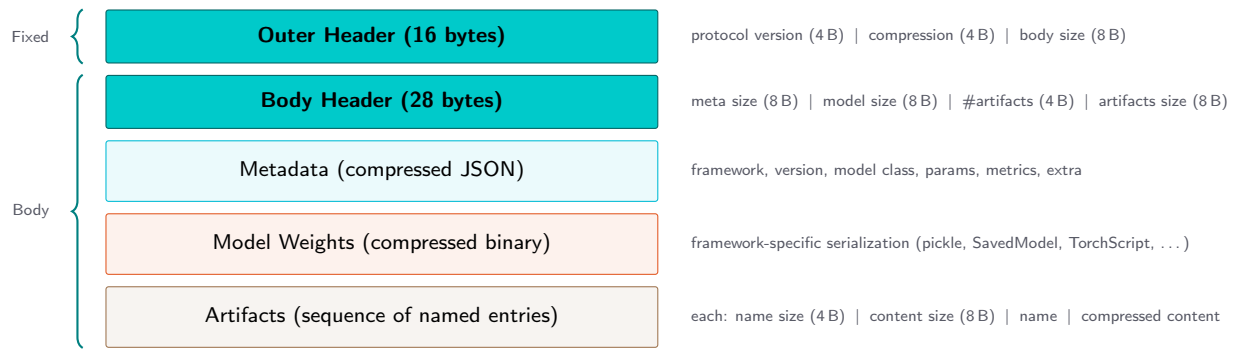
\begin{figure}[ht]
\centering
\begin{tikzpicture}[
  block/.style={
    draw=#1,
    fill=#1!8!white,
    minimum width=7.3cm,
    minimum height=0.8cm,
    font=\sffamily\scriptsize,
    rounded corners=1pt,
    align=center,
  },
  hdr/.style={
    draw=vortico700,
    fill=vortico100,
    minimum width=7.3cm,
    minimum height=0.7cm,
    font=\sffamily\scriptsize\bfseries,
    rounded corners=1pt,
    align=center,
  },
  lbl/.style={
    font=\sffamily\tiny,
    text=primary600,
  },
  node distance=0.15cm,
]

\node[hdr] (oh) {Outer Header (16 bytes)};
\node[lbl, right=0.3cm of oh.east, anchor=west]
  {protocol version (4\,B) $\,|\,$ compression (4\,B)
   $\,|\,$ body size (8\,B)};

\node[hdr, below=of oh] (bh)
  {Body Header (28 bytes)};
\node[lbl, right=0.3cm of bh.east, anchor=west]
  {meta size (8\,B) $\,|\,$ model size (8\,B)
   $\,|\,$ \#artifacts (4\,B) $\,|\,$ artifacts size (8\,B)};

\node[block=bruma500, below=of bh] (meta)
  {Metadata (compressed JSON)};
\node[lbl, right=0.3cm of meta.east, anchor=west]
  {framework, version, model class, params, metrics, extra};

\node[block=flama500, below=of meta] (model)
  {Model Weights (compressed binary)};
\node[lbl, right=0.3cm of model.east, anchor=west]
  {framework-specific serialization (pickle, SavedModel, TorchScript, \ldots)};

\node[block=bosque500, below=of model] (art)
  {Artifacts (sequence of named entries)};
\node[lbl, right=0.3cm of art.east, anchor=west]
  {each: name size (4\,B) $\,|\,$ content size (8\,B) $\,|\,$ name $\,|\,$ compressed content};

\draw[decorate, decoration={brace, amplitude=4pt, mirror},
  thick, vortico500]
  ([xshift=-0.3cm]oh.north west) --
  ([xshift=-0.3cm]oh.south west)
  node[midway, left=0.3cm, lbl] {Fixed};

\draw[decorate, decoration={brace, amplitude=4pt, mirror},
  thick, vortico500]
  ([xshift=-0.3cm]bh.north west) --
  ([xshift=-0.3cm]art.south west)
  node[midway, left=0.3cm, lbl] {Body};

\end{tikzpicture}
\caption{Structure of the FLM binary format (protocol
version~1). The outer header identifies the format version
and compression algorithm. The body contains three sections:
compressed JSON metadata, compressed model weights, and a
sequence of compressed named artifact entries. Each section
can be decompressed independently.}
\label{fig:flm-structure}
\end{figure}

\subsubsection{Outer header}

The outer header uses network byte order
(\texttt{big-endian}) and consists of three fields packed
according to the \texttt{struct} format \texttt{!I I Q}:

\begin{center}
\small
\begin{tabularx}{\textwidth}{@{}rrlX@{}}
\toprule
\textbf{Offset} & \textbf{Size} & \textbf{Field} &
\textbf{Description} \\
\midrule
0 & 4\,B & Protocol version &
  Unsigned 32-bit integer, \texttt{1} or \texttt{2}.
  \texttt{2} is the default written by \py{dump()}. \\
4 & 4\,B & Compression format &
  Enum value identifying the compression algorithm. \\
8 & 8\,B & Body size &
  Unsigned 64-bit integer. The total size of the body
  in bytes. \\
\bottomrule
\end{tabularx}
\end{center}

\subsubsection{Body header (protocol version 1)}

The body begins with a 28-byte header packed as
\texttt{!Q Q I Q}:

\begin{center}
\small
\begin{tabularx}{\textwidth}{@{}rrlX@{}}
\toprule
\textbf{Offset} & \textbf{Size} & \textbf{Field} &
\textbf{Description} \\
\midrule
0 & 8\,B & Meta size &
  Size of the compressed metadata section. \\
8 & 8\,B & Model size &
  Size of the compressed model weights section. \\
16 & 4\,B & Artifacts count &
  Number of artifact entries. \\
20 & 8\,B & Artifacts size &
  Total size of all compressed artifact entries. \\
\bottomrule
\end{tabularx}
\end{center}

\subsubsection{Metadata section}

The metadata section is a JSON document compressed with the
algorithm specified in the outer header. It is stored at a
fixed offset immediately after the body header, which means
it can be decompressed and inspected without reading the model
weights.

The metadata is represented internally by the
\py{ModelArtifact} data structure, which comprises four nested
frozen dataclasses:

\begin{lstlisting}
@dataclass(frozen=True)
class FrameworkInfo:
    lib: str      # "sklearn", "tensorflow",
                  # "torch", "transformers"
    version: str  # e.g. "1.7.2", "2.20.0"

@dataclass(frozen=True)
class ModelInfo:
    obj: str              # class name
    info: dict | None     # JSON Schema (framework-specific)
    params: dict | None   # hyperparameters
    metrics: dict | None  # training metrics

@dataclass(frozen=True)
class Metadata:
    id: str | UUID        # unique model identifier
    timestamp: datetime   # serialization timestamp
    framework: FrameworkInfo
    model: ModelInfo
    extra: dict | None    # user-defined metadata

@dataclass(frozen=True)
class ModelArtifact:
    meta: Metadata
    model: Any            # the deserialized model object
    artifacts: dict[str, Path] | None
\end{lstlisting}

\subsubsection{Model weights section}

The model weights section contains the serialized model,
compressed with the same algorithm as the metadata. The
serialization format is framework-specific and is described in
Section~\ref{sec:framework-serializers}.

\subsubsection{Artifacts section}

The artifacts section is a sequence of named entries. Each entry
consists of a 12-byte header (\texttt{!I Q}: 4 bytes for the
name length, 8 bytes for the compressed content length),
followed by the UTF-8 name string and the compressed content.
Artifacts are used for auxiliary files that the model needs at
inference time but that are not part of the weight tensors:
tokenizer vocabularies, label maps, preprocessing
configurations, and custom post-processing scripts.

On deserialization, artifacts are extracted into a temporary
directory that is cleaned up automatically when the
\py{ModelArtifact} object is garbage-collected (via
\py{weakref.finalize}).

\subsubsection{Protocol version~2}

Protocol version~2 restructures the body layout to accommodate
two advances that emerged from the LLM serving subsystem: the
need to store model weights as directory bundles (tarballs of
Hugging Face checkpoint trees rather than opaque byte blobs) and
the desire for per-section compression control so that metadata
can remain uncompressed for fast random-access inspection while
large model payloads are compressed independently.

The v2 body header is 24~bytes, packed as \texttt{!Q Q Q}:

\begin{center}
\small
\begin{tabularx}{\textwidth}{@{}rrlX@{}}
\toprule
\textbf{Offset} & \textbf{Size} & \textbf{Field} &
\textbf{Description} \\
\midrule
0 & 8\,B & Meta size &
  Size of the metadata section (including its
  discriminator byte). \\
8 & 8\,B & Artifacts size &
  Total size of the artifacts section (including its
  discriminator byte). \\
16 & 8\,B & Model size &
  Size of the model section (including its
  discriminator and kind bytes). \\
\bottomrule
\end{tabularx}
\end{center}

Each section begins with a one-byte \emph{compression
discriminator} that declares how that section is compressed:

\begin{center}
\small
\begin{tabularx}{\textwidth}{@{}rlX@{}}
\toprule
\textbf{Byte} & \textbf{Name} & \textbf{Meaning} \\
\midrule
\texttt{0x00} & inherit &
  Use the file-level compression declared in the
  outer header. \\
\texttt{0x01} & bz2 & Override with bz2. \\
\texttt{0x02} & lzma & Override with lzma. \\
\texttt{0x03} & zlib & Override with zlib. \\
\texttt{0x04} & zstd & Override with zstd. \\
\texttt{0xFF} & none & Passthrough (no compression). \\
\bottomrule
\end{tabularx}
\end{center}

The model section carries an additional one-byte
\emph{kind discriminator} immediately after its compression
byte:

\begin{center}
\small
\begin{tabularx}{\textwidth}{@{}rlX@{}}
\toprule
\textbf{Byte} & \textbf{Kind} & \textbf{Payload format} \\
\midrule
\texttt{0x00} & binary &
  Opaque serialized bytes (pickle, Keras, torch.export).
  Used for traditional ML models. \\
\texttt{0x01} & bundle &
  A tar stream of the model directory. Used for
  Transformers checkpoints and LLM artifacts,
  which consist of multiple files (weight shards,
  tokenizer, configuration). \\
\bottomrule
\end{tabularx}
\end{center}

The combination of per-section compression and model kinds means
that an LLM checkpoint can be stored as an uncompressed tar
bundle (fast extraction at load time, since model files are
already stored in efficient formats like safetensors) while the
metadata remains independently accessible and the artifacts
section uses a different compression level. The \texttt{inherit}
discriminator allows sections that do not need special treatment
to fall back to the file-level default, maintaining backward
compatibility with the compression semantics of protocol
version~1.

\subsubsection{Model capabilities}

Protocol version~2 also introduces a \py{capabilities} field in
the metadata section that declares what a packaged model can
ingest and produce. Capabilities are represented by the
\py{ModelCapabilities} hierarchy:

\begin{lstlisting}
@dataclass(frozen=True)
class LLMModelCapabilities(ModelCapabilities):
    kind: ClassVar[ModelFamily] = "llm"

    text: bool = True
    image: bool = False
    audio: bool = False
    video: bool = False
    tools: bool = False
    reasoning: bool = False
\end{lstlisting}

Capabilities are detected at serialization time by each
framework-specific serializer and persisted in the manifest.
Consumers---backend dispatch, input validation, serving-layer
capability advertisement---read this single source of truth
rather than re-probing the model at load time.

\subsubsection{Artifact families}

The metadata's \py{FrameworkInfo} dataclass carries a
\py{family} discriminator (\texttt{"ml"} or \texttt{"llm"})
that records the artifact's intent at serve time. The family is
chosen by the producer at dump time (e.g.\ via
\cli{flama get --family llm}) and is never inferred from the
library at load time. LLM artifacts always record
\texttt{"transformers"} as their library because the on-disk
format is a Hugging Face checkpoint tarball; the runtime that
actually serves them (vLLM or MLX) is selected at load time by
an import probe and is not persisted in the manifest.

\subsection{Compression}
\label{sec:compression}

The FLM format supports four compression algorithms:

\begin{center}
\small
\begin{tabularx}{\textwidth}{@{}llX@{}}
\toprule
\textbf{Enum value} & \textbf{Algorithm} &
\textbf{Notes} \\
\midrule
1 & bz2 &
  Moderate compression ratio, slow decompression. \\
2 & lzma &
  High compression ratio, very slow
  decompression. \\
3 & zlib &
  Balanced compression and speed.
  Available in all Python installations. \\
4 & zstd (default) &
  Best trade-off: near-lzma compression ratios
  at near-zlib decompression speeds. Uses
  \texttt{python-zstd} or the native
  \texttt{compression.zstd} module available
  in Python~3.14+. \\
\bottomrule
\end{tabularx}
\end{center}

The default is zstd \citep{zstd}, which provides compression
ratios comparable to lzma while decompressing at speeds
comparable to zlib. For neural network weights (which consist
largely of floating-point tensors with limited redundancy), zstd
achieves typical compression ratios of 1.5--3$\times$ on
serialized model files.

\subsection{Serialization and deserialization API}
\label{sec:serialization-api}

The top-level API consists of two functions:

\begin{lstlisting}
from flama.serialize import dump, load

# Serialize a trained model to disk
dump(
    model,
    path="classifier.flm",
    family="ml",
    compression="zstd",
    model_id="classifier-v2",
    params={"n_estimators": 100,
            "max_depth": 8},
    metrics={"accuracy": 0.947,
             "f1": 0.932},
    extra={"dataset": "prod-2024-q3",
           "author": "team-ml"},
)

# Load a serialized model from disk
artifact = load(path="classifier.flm")

print(artifact.meta.framework.lib)    # "sklearn"
print(artifact.meta.model.obj)        # "RandomForestClassifier"
print(artifact.meta.model.metrics)    # {"accuracy": 0.947, ...}
prediction = artifact.model.predict([[5.1, 3.5, 1.4, 0.2]])
\end{lstlisting}

Both functions accept either a path (string or \py{Path}) or a
binary file object, making them usable with local files, S3
objects, HTTP responses, or any other binary stream.
\py{dump()} additionally requires a \py{family} keyword
(\texttt{"ml"} or \texttt{"llm"}); it is never inferred from the
model object, since the same on-disk \texttt{transformers}
library backs both a predictive pipeline and an LLM checkpoint.

On loading, the framework version stored in the metadata is
compared with the installed version. If they differ, a
\py{FrameworkVersionWarning} is emitted to alert the user that
the model was trained with a different version of the framework
and that predictions may not be reproducible.

\section{Framework-specific serializers}
\label{sec:framework-serializers}

Each supported machine learning framework has a dedicated
serializer that implements three operations: \py{dump()} (model
to bytes), \py{load()} (bytes to model), and \py{info()} (model
to JSON Schema describing the model's architecture or
parameters). All serializers inherit from a common abstract base
class:

\begin{lstlisting}
class BaseModelSerializer(ABC):
    lib: ClassVar[str]

    @abstractmethod
    def dump(self, obj, /, **kwargs) -> bytes:
        """Serialize a model to bytes."""
        ...

    @abstractmethod
    def load(self, model: bytes, /, **kwargs) -> Any:
        """Deserialize a model from bytes."""
        ...

    @abstractmethod
    def info(self, model, /) -> dict | None:
        """Extract structural information as JSON."""
        ...

    @abstractmethod
    def version(self) -> str:
        """Return the installed framework version."""
        ...
\end{lstlisting}

A factory class (\py{ModelSerializer}) selects the appropriate
serializer by inspecting the model object's module hierarchy.
A model whose class resides under \py{sklearn.*} dispatches to
the scikit-learn serializer, one under \py{torch.*} to the
PyTorch serializer, and so on, so a live model object never
needs its framework declared. The one case that does is a
directory path, which carries no class to inspect: passing one
to \py{dump()} without a \py{lib} argument raises
\py{ValueError}.

\subsection{Scikit-learn}
\label{sec:serializer-sklearn}

\begin{center}
\small
\begin{tabularx}{\textwidth}{@{}lX@{}}
\toprule
\textbf{Operation} & \textbf{Implementation} \\
\midrule
\py{dump()} &
  Serializes the model with \py{pickle.dumps()} and
  encodes the result as Base64. Pickle is the standard
  serialization format for scikit-learn \citep{sklearn}
  and supports the full range of estimator types,
  including pipelines and custom transformers. \\
\py{load()} &
  Decodes the Base64 string and deserializes with
  \py{pickle.loads()}. \\
\py{info()} &
  Calls \py{model.get\_params()} and recursively sanitizes
  the result into a JSON-compatible dictionary (non-finite
  floats become \py{null}). Nested estimators (e.g.\ the
  base estimator in a \py{BaggingClassifier}) are
  serialized as their class names with parameters. \\
\bottomrule
\end{tabularx}
\end{center}

\subsection{TensorFlow and Keras}
\label{sec:serializer-tensorflow}

\begin{center}
\small
\begin{tabularx}{\textwidth}{@{}lX@{}}
\toprule
\textbf{Operation} & \textbf{Implementation} \\
\midrule
\py{dump()} &
  Saves the model to a temporary file in the native
  \texttt{.keras} format using the standalone Keras~3
  package's \py{keras.models.save\_model()}, reads
  the file into memory, and encodes as Base64.
  The \texttt{.keras} format \citep{kerasformat}
  captures the model architecture, weights, optimizer
  state, and compilation configuration in a single
  file. \\
\py{load()} &
  Writes the Base64-decoded bytes to a temporary file
  and loads with
  \py{keras.models.load\_model(path)}. \\
\py{info()} &
  Calls \py{model.to\_json()} and parses the JSON
  string into a dictionary. The result is a complete
  description of the model's layer structure, including
  layer types, output shapes, activation functions,
  and connections. \\
\bottomrule
\end{tabularx}
\end{center}

\subsection{PyTorch}
\label{sec:serializer-pytorch}

\begin{center}
\small
\begin{tabularx}{\textwidth}{@{}lX@{}}
\toprule
\textbf{Operation} & \textbf{Implementation} \\
\midrule
\py{dump()} &
  Exports the model to an \py{ExportedProgram}
  \citep{torchexport} via
  \py{torch.export.export(obj, example\_inputs,
  dynamic\_shapes)}, then serializes the exported
  graph to a byte buffer with \py{torch.export.save()}.
  The result is Base64-encoded. The export API
  captures the model's computation graph with
  explicit dynamic-shape annotations, producing
  a portable, optimizable artifact that is
  independent of the Python class definition. If
  \py{example\_inputs} is not provided, the
  serializer infers a suitable shape by inspecting
  the first \py{torch.nn.Linear} layer in the
  module. \\
\py{load()} &
  Decodes the Base64 string and loads the exported
  program with \py{torch.export.load(BytesIO(data))}.
  The loaded module can run inference without the
  original class definition or training code. \\
\py{info()} &
  Extracts the module list (as string representations),
  named parameters (as string representations, which
  include shape and dtype), and the full state dictionary
  with every tensor converted to a nested list. The result
  is self-contained, at the cost of size for large models,
  and requires no forward pass execution. \\
\bottomrule
\end{tabularx}
\end{center}

The \py{torch.export} API replaces the earlier TorchScript
approach and offers several advantages for deployment:
the exported graph preserves dynamic shapes (e.g.\ variable
batch sizes) through explicit \py{Dim} annotations rather than
trace-time heuristics, it supports the full Python operator set
without the restrictions imposed by TorchScript's subset
compiler, and it integrates with PyTorch's ahead-of-time
compilation pipeline for further optimization at load time.
The dynamic shapes are declared via a dictionary mapping input
names to dimension constraints:

\begin{lstlisting}
from flama.serialize import dump

dump(
    model,
    path="classifier.flm",
    example_inputs=(torch.randn(2, 768),),
    dynamic_shapes={"x": {0: torch.export.Dim("batch",
                                              min=1)}},
)
\end{lstlisting}

\subsection{HuggingFace Transformers}
\label{sec:serializer-transformers}

The Transformers serializer follows a fundamentally different
strategy from the other three: rather than a flat binary blob,
it packages a whole directory, matching protocol version~2's
\texttt{bundle} model kind (Section~\ref{sec:flm-format}).

\begin{center}
\small
\begin{tabularx}{\textwidth}{@{}lX@{}}
\toprule
\textbf{Operation} & \textbf{Implementation} \\
\midrule
\py{dump()} &
  Accepts either a directory of pretrained model files or
  a live \py{transformers.Pipeline} (whose weights are
  first written out with \py{save\_pretrained()} to a
  temporary directory). Either way, the directory is
  packed into an uncompressed tar archive in memory using
  the framework's own Rust-accelerated tar routine, since
  the on-disk files (safetensors shards, tokenizer,
  configuration) are already in efficient formats and gain
  little from re-compression. \\
\py{load()} &
  Takes the path to the bundle already extracted to disk
  (raw bytes are rejected; Transformers reads a snapshot
  directory, not a byte stream) and calls
  \py{transformers.pipeline(task=task, model=str(path),
  **kwargs)}, returning a ready-to-use
  \py{transformers.Pipeline}. \\
\py{info()} &
  Reads back \py{pipeline.model.config.to\_dict()} (the
  model architecture and hyperparameters),
  \py{pipeline.task}, and
  \py{pipeline.model.name\_or\_path}. \\
\bottomrule
\end{tabularx}
\end{center}

Capability detection is correspondingly more involved than for
the other three frameworks, which always report an empty
capability set. The Transformers serializer inspects the bundle
on disk: \texttt{vision\_config}/\texttt{audio\_config} blocks
in \texttt{config.json} (corroborated against the actual tensor
names in the safetensors header, since a checkpoint can ship a
multimodal config without the corresponding tower weights) drive
image and audio support, and tool and reasoning support are
detected by rendering the tokenizer's chat template against
sentinel inputs and checking whether the rendered output reflects
them.

\section{Model wrappers}
\label{sec:model-wrappers}

Between the serialized model on disk and the HTTP endpoint that
serves predictions, there are two cooperating layers: a
\textbf{model wrapper}, engine-agnostic and shared by every
framework, and a \textbf{backend}, one concrete class per
framework, that the wrapper delegates the actual inference call
to. The split exists because the two concerns change for
different reasons: the wrapper owns lazy deserialization,
metadata access, and the DI/HTTP-facing surface, none of which
differ across frameworks, while inference semantics do differ
across frameworks (a scikit-learn estimator expects a NumPy
array; a PyTorch module expects a \py{torch.Tensor}; a
Transformers pipeline tokenizes internally) and are confined to
the backend.

The wrapper base class defines lazy access to the model's
metadata, bundled artifacts, and backend, plus \py{inspect()}:

\begin{lstlisting}
class BaseModel(Generic[B]):
    def __init__(self, backend: B | None = None,
                 meta: Metadata | None = None,
                 artifacts: Artifacts | None = None, *,
                 name: str | None = None,
                 path: Path | None = None,
                 autoload: bool = False):
        ...

    def inspect(self) -> dict:
        return {"meta": self.meta.to_dict(),
                "manifest": list(self.manifest)}
\end{lstlisting}

\py{manifest} is the list of bundled artifact \emph{names},
read cheaply from the FLM header; the artifacts themselves are
only extracted to disk once the model is actually loaded.
\py{MLModel(BaseModel[MLBackend])} is the single concrete
wrapper for every predictive framework; it adds
\py{predict(x)} and \py{stream(x)}, both implemented once and
delegated straight to \py{self.backend.predict(x)}. There is no
per-framework subclass of the wrapper: the framework-specific
code lives entirely in the backend that
\py{MLBackend.from\_model\_artifact()} selects at load time,
based on \py{Metadata.framework.lib}.

\subsection{Framework-specific backends}

Each framework provides a concrete \py{MLBackend} subclass. Its
\py{predict(x)} method translates the framework-neutral input (a
list of lists, received as JSON from the client) into the
framework's native input format, runs inference, and converts
the output back to a JSON-serializable list:

\begin{center}
\small
\renewcommand{\arraystretch}{1.3}
\begin{tabularx}{\textwidth}{@{}lX@{}}
\toprule
\textbf{Framework} &
\textbf{\py{predict(x)} implementation} \\
\midrule
scikit-learn &
  \py{self.model.predict(x).tolist()} \\
TensorFlow &
  \py{self.model.predict(np.array(x)).tolist()} \\
PyTorch &
  \py{self.model(torch.Tensor(x)).tolist()} \\
Transformers &
  \py{self.model(x)}, delegating entirely to a
  \py{transformers.Pipeline}'s own tokenize/infer/decode
  cycle. \\
\bottomrule
\end{tabularx}
\end{center}

In every case, \py{self.model} is the object the framework's
serializer produced at load time (an estimator, a Keras model,
an exported graph module, or a \py{transformers.Pipeline}), and
each backend raises \py{FrameworkNotInstalled} up front if the
underlying library is not importable, rather than failing with
an opaque \py{ImportError} mid-request.

\section{Three levels of model integration}
\label{sec:integration-levels}

\flama provides three progressively more automated levels for
integrating a machine-learning model into a web application.
Each level builds on the one below it, and the developer
chooses the level that matches the degree of customization
required. At one extreme, the developer writes the endpoint
code and merely uses the framework for model loading and
lifecycle management. At the other extreme, the developer
provides only a model file path and a name, and the framework
generates the complete endpoint infrastructure automatically.

\subsection{Level 1: Model components}
\label{sec:model-components}

At the lowest level, a model is loaded as a dependency-injection
component. The \py{ModelComponentBuilder} factory reads the
metadata header from an FLM file to determine the artifact
family, then produces a \py{ModelComponent} that makes the model
available for injection into any handler:

\begin{lstlisting}
from flama.models.components import ModelComponentBuilder

# Build a lazy component from a .flm file
component = ModelComponentBuilder.build("classifier.flm")
\end{lstlisting}

The builder performs the following steps:

\begin{enumerate}[leftmargin=2em, itemsep=4pt]
  \item Read the FLM file's metadata header (a cheap operation
    that does not deserialize the model body).
  \item Determine the artifact family (\texttt{"ml"} or
    \texttt{"llm"}) from the metadata.
  \item For ML artifacts, create a fresh, per-instance subclass
    of \py{MLModel}; for LLM artifacts, of \py{LLMModel}. The
    subclass carries no extra behaviour of its own: its only
    purpose is to give this particular registered model its own
    type.
  \item Wrap it in a \py{ModelComponent} subclass whose
    \py{resolve()} method returns that instance, with its return
    type annotation set to the freshly created subclass. Because
    the injector keys components by the return annotation of
    \py{resolve()}, this is what lets two different registered
    models, each with its own dynamically created type, resolve
    to two different handler parameters without colliding.
\end{enumerate}

The actual model body is not deserialized at build time. Heavy
loading (backend initialization, weight deserialization) is
deferred to \py{component.startup()}, so the server port binds
before model loading begins.

A handler can then receive the model via dependency injection.
Since the component's type is generated at build time rather
than imported, the handler is typed against
\py{component.get\_model\_type()} and the component is
registered explicitly, both on the application and on its
startup event:

\begin{lstlisting}
import typing as t
from flama import Flama, schemas

component = ModelComponentBuilder.build("classifier.flm")
Model = component.get_model_type()

app = Flama(events={"startup": [component.startup]})
app.add_component(component)

@app.post("/predict")
async def predict(
    model: Model,
    data: t.Annotated[schemas.Schema, schemas.SchemaMetadata(PredictInput)],
) -> t.Annotated[schemas.Schema, schemas.SchemaMetadata(PredictOutput)]:
    result = model.predict(data["input"])
    return {"output": result}
\end{lstlisting}

This level is appropriate when the developer needs full control
over the endpoint's URL, methods, request/response schemas,
and error handling, but still wants the model loading and
lifecycle managed by the framework.

\subsection{Level 2: The \py{add\_model()} API}
\label{sec:add-model}

The \py{ModelsModule} provides a higher-level API that creates a
complete model resource from a single method call:

\begin{lstlisting}
app = Flama()

app.models.add_model(
    path="/classifier",
    model="classifier.flm",
    name="classifier",
)
\end{lstlisting}

This call generates two endpoints:

\begin{center}
\small
\begin{tabularx}{\textwidth}{@{}llX@{}}
\toprule
\textbf{Method} & \textbf{Path} &
\textbf{Behaviour} \\
\midrule
\http{GET} & \texttt{/classifier/} &
  Returns the model's metadata (framework, version,
  class, hyperparameters, metrics, artifact list). \\
\http{POST} & \texttt{/classifier/predict/} &
  Accepts \py{PredictInput} (\py{\{"input": [...]\}}),
  runs inference, and returns
  \py{PredictOutput} (\py{\{"output": [...]\}}). \\
\bottomrule
\end{tabularx}
\end{center}

\subsection{Level 3: Model resources}
\label{sec:model-resources}

The most declarative level uses a class-based resource with a
metaclass that generates all the infrastructure automatically:

\begin{lstlisting}
from flama.models import MLResource

class SentimentClassifier(MLResource):
    name = "sentiment"
    verbose_name = "Sentiment Classifier"
    model_path = "models/sentiment.flm"
\end{lstlisting}

Like a \py{CRUDResource}, the class declaration alone does not
attach it to the application; it still has to be registered
through the models module:

\begin{lstlisting}
app.models.add_model_resource(path="/sentiment", resource=SentimentClassifier)
\end{lstlisting}

The \py{MLResourceType} metaclass performs the following
operations when the class is defined:

\begin{enumerate}[leftmargin=2em, itemsep=4pt]
  \item Loads the model component from \py{model\_path} using
    the same builder as Level~1.
  \item Stores the component, the model wrapper, and the model
    type in the class's internal namespace.
  \item Calls three mixins (\py{InspectMixin},
    \py{PredictMixin}, \py{StreamMixin}) to generate endpoint
    methods. Each mixin's \py{\_add\_*()} method creates a
    handler function on the class with the correct signature
    and type annotations.
\end{enumerate}

This level is appropriate for applications that deploy multiple
models and want each model to be a self-contained, reusable
unit with its own configuration.

\section{Large language model serving}
\label{sec:llm-serving}

Everything so far has concerned \emph{predictive} models: the
binary format, the serializers, and the three levels of
integration that put a trained classifier behind a REST
endpoint. The \emph{generative} case reuses that architecture
but asks more of it, since a language model has to stream, has
to speak several wire protocols at once, and has to run on
whatever accelerator the host happens to have.

Those are the three requirements that separate it from the
predictive case:

\begin{enumerate}[leftmargin=2em, itemsep=4pt]
  \item \textbf{Streaming output.} Language model inference
    produces tokens incrementally. Users expect to see output
    appear as it is generated, not after the entire sequence
    has been computed. The serving layer must therefore deliver
    partial results via Server-Sent Events or
    Newline-Delimited JSON as the model generates.

  \item \textbf{Multi-protocol compatibility.} The generative AI
    ecosystem has converged on several incompatible wire
    protocols (OpenAI, Anthropic, Ollama). Practitioners
    integrate with these protocols using existing client SDKs
    and tooling. The serving layer must expose the same physical
    model through all major protocols without duplicating the
    inference pipeline.

  \item \textbf{Hardware-aware backends.} LLM inference is
    compute-bound and benefits from hardware-specific
    optimization (PagedAttention on CUDA, Metal acceleration on
    Apple Silicon). The serving layer must select the
    highest-performance backend available at runtime without
    requiring the user to write backend-specific code.
\end{enumerate}

\subsection{Architecture overview}
\label{sec:llm-architecture}

The LLM serving subsystem is organized as a five-layer pipeline:

\begin{description}[leftmargin=2em, itemsep=4pt,
  font=\sffamily\bfseries]
  \item[Wire dialect]
    parses incoming request JSON into canonical transport
    objects (messages, tools) and renders outgoing events into
    protocol-specific streaming frames or buffered envelopes.
    Four dialects are provided: OpenAI, Anthropic, Ollama, and
    Native.

  \item[Transport]
    defines the canonical request and response model,
    independent of any wire protocol. A request is a sequence of
    typed messages with multimodal content parts; a response is a
    stream of typed events (start, text, tool call, trace, stop).

  \item[Engine]
    bridges the transport layer and the backend. It converts
    canonical messages into tokenized \py{EngineInput} (token IDs
    plus decoded media), invokes the backend's generation
    routine, and produces a stream of \py{EngineDelta} objects
    carrying incremental text, token counts, and finish reasons.

  \item[Codec/Decoder]
    transforms the raw \py{EngineDelta} stream into canonical
    output events. The \py{LLMCodec} implements a finite-state
    machine that recognizes reasoning channels (think tags,
    channel markers) and tool-call bodies in the generated text,
    emitting structured \py{TextEvent}, \py{ToolEvent}, and
    \py{TraceEvent} objects.

  \item[Backend]
    is the hardware-bound inference engine. Two backends are
    supported: vLLM (Linux/CUDA) and MLX (Apple Silicon). The
    backend is selected at model load time by probing which
    library is importable.
\end{description}

A typical end-to-end inference flow proceeds as follows:

\begin{enumerate}[leftmargin=2em, itemsep=2pt]
  \item An HTTP request arrives at a dialect-specific endpoint
    (e.g.\ \texttt{/v1/chat/completions}).
  \item The dialect's \emph{parser} converts the wire-format
    JSON into canonical \py{Message} and \py{Tool} objects.
  \item A \py{Shape} (raw, chat, or conversation) renders the
    messages into \py{EngineInput} by applying the backend's
    chat template and tokenizer.
  \item The backend generates tokens, yielding
    \py{EngineDelta} objects incrementally.
  \item The \py{LLMCodec} decodes deltas into canonical
    \py{Event} objects.
  \item An \py{EventBuffer} feeds events into the dialect's
    \emph{renderer}, which produces SSE or NDJSON frames.
  \item For buffered (non-streaming) requests, the dialect's
    \emph{assembler} coalesces all events into a single response
    envelope.
\end{enumerate}

\subsection{Backends}
\label{sec:llm-backends}

A backend is responsible for loading a model into memory,
applying a chat template to format messages, tokenizing input,
and generating tokens. The backend abstraction is minimal: any
object that exposes \py{encode()}, \py{chat\_template()},
\py{prepare\_input()}, and \py{generate()} can serve as a
backend. Two implementations are provided:

\begin{description}[leftmargin=2em, itemsep=4pt,
  font=\sffamily\bfseries]
  \item[vLLM \citep{vllm}]
    is a high-throughput inference engine for Linux systems with
    NVIDIA GPUs. It implements PagedAttention for efficient
    KV-cache management, continuous batching for maximizing GPU
    utilization, and tensor parallelism for multi-GPU
    deployments. vLLM is the preferred backend for production
    deployments where throughput and latency are critical.

  \item[MLX \citep{mlx}]
    is Apple's framework for machine learning on Apple Silicon.
    It provides Metal-accelerated tensor operations with a
    NumPy-compatible API and unified memory that eliminates
    data transfers between CPU and GPU. The MLX backend uses
    \texttt{mlx-lm} for text generation and \texttt{mlx-vlm}
    for vision-language models. It is the preferred backend on
    macOS systems.
\end{description}

Both backends share the \py{TransformerLLMBackend} abstract base
class, which delegates chat-template rendering to the Hugging
Face \texttt{tokenizer} and \texttt{AutoProcessor} stack. This
design means that any model checkpoint compatible with the
Hugging Face model format can be loaded by either backend; the
runtime engine is what differs, not the model loading or
template application logic.

Backend selection is automatic: at model load time, the
framework probes which libraries are importable and selects the
first available option. On a Linux system with CUDA, vLLM is
used; on macOS with Apple Silicon, MLX is used. If neither is
available, the framework raises a clear error indicating that one
of the two runtimes must be installed.
The selection is transparent to the application: the same
\py{add\_model()} call works identically regardless of the
underlying backend.

\subsection{Transport layer}
\label{sec:llm-transport}

The transport layer defines the canonical data model that sits
between the wire protocols and the engine. It comprises three
sub-layers: input shapes, messages, and output events.

\subsubsection{Input shapes}

A \py{Shape} determines how a client's messages are formatted
before they reach the tokenizer. Three shapes are supported:

\begin{description}[leftmargin=2em, itemsep=4pt,
  font=\sffamily\bfseries]
  \item[Raw]
    sends the prompt text directly to the tokenizer without any
    formatting. This is suitable for completions-style
    inference where the client provides the exact token sequence.

  \item[Chat]
    wraps the input into a two-message sequence (optional system
    message plus user message) and applies the model's chat
    template. This is the default shape for single-turn
    interactions.

  \item[Conversation]
    passes the full message history (system, user, assistant,
    tool results) through the chat template. This enables
    multi-turn dialogues where the model has access to the
    entire conversation context.
\end{description}

\subsubsection{Messages and content parts}

The \py{Message} hierarchy represents a single conversational
turn. Each message has a role (\texttt{system},
\texttt{user}, \texttt{assistant}, or \texttt{tool}) and a
sequence of typed content parts:

\begin{itemize}[leftmargin=2em, itemsep=2pt]
  \item \textbf{Text}: plain text content.
  \item \textbf{Image}: a base64-encoded image or a URL,
    with MIME type and optional detail level.
  \item \textbf{Audio}: a base64-encoded audio segment with
    sample rate and format metadata.
  \item \textbf{Tool call}: a function invocation request from
    the assistant, with a call ID, function name, and JSON
    arguments.
  \item \textbf{Tool result}: the return value of a tool call,
    associated with the original call ID.
\end{itemize}

This multimodal message representation is the canonical form
into which every wire protocol's request is parsed. The dialect
parsers handle the translation from protocol-specific formats
(OpenAI's content-part arrays, Anthropic's top-level system
field, Ollama's sibling \texttt{images} array) into this uniform
representation.

\subsubsection{Output events}

The response from an LLM is represented as a stream of typed
events:

\begin{description}[leftmargin=2em, itemsep=4pt,
  font=\sffamily\bfseries]
  \item[StartEvent]
    signals the beginning of a generation. Carries the model
    name and the generation configuration.

  \item[TextEvent]
    delivers a fragment of generated text. In streaming mode,
    one \py{TextEvent} is emitted per decoded token or
    token group.

  \item[ToolEvent]
    delivers a complete tool-call request extracted from the
    generated text. Carries the function name, JSON arguments,
    and an optional call ID. Tool calls are emitted atomically
    (the full call body is accumulated before emission) to
    ensure well-formed JSON.

  \item[TraceEvent]
    delivers content from a secondary output channel (e.g.\
    reasoning traces, thinking tokens, analysis steps). Trace
    events are emitted when the codec detects channel markers
    in the generated text.

  \item[StopEvent]
    signals the end of generation. Carries the stop reason
    (end-of-sequence, length limit, tool call, or content
    filter), token usage statistics, and optional metadata.
\end{description}

\subsection{Codec and decoder pipeline}
\label{sec:llm-codec}

The \py{LLMCodec} is the bridge between the backend's raw token
stream and the canonical event model. Its core responsibility is
\emph{structured output recognition}: detecting reasoning
channels and tool calls in the generated text and emitting
appropriate events rather than treating all output as
undifferentiated text.

\subsubsection{Decoder detection}

Different model families use different conventions for
structured output. Some models wrap reasoning in
\texttt{<think>...</think>} tags; others use
\texttt{<|begin\_of\_thought|>} markers; still others emit tool
calls as JSON objects preceded by special tokens. The
\py{Decoder} is responsible for detecting which conventions a
particular model uses.

Detection proceeds in three stages:

\begin{enumerate}[leftmargin=2em, itemsep=2pt]
  \item \textbf{Pinned configuration}: if the user explicitly
    specifies a channel scanner, tool scanner, or tool parser,
    those are used directly without auto-detection.
  \item \textbf{Chat template analysis}: the decoder examines
    the model's chat template for known marker patterns and
    selects scanners accordingly.
  \item \textbf{Preflight probing}: if template analysis is
    inconclusive, a short preflight generation is performed and
    the output is analysed for structural patterns.
\end{enumerate}

The decoder comprises three pluggable components:

\begin{description}[leftmargin=2em, itemsep=4pt,
  font=\sffamily\bfseries]
  \item[Channel scanner]
    recognizes the start and end of secondary output channels.
    Built-in scanners support \texttt{think} tags, generic
    \texttt{channel} markers, and passthrough (no channels).

  \item[Tool scanner]
    recognizes the start of a tool-call body in the generated
    text. It detects special tokens (e.g.\
    \texttt{<tool\_call>}) or JSON-object openings that signal
    a function invocation.

  \item[Tool parser]
    extracts the function name and arguments from the scanned
    tool-call body. Built-in parsers support JSON objects, JSON
    arrays, named JSON sequences, and Python-style call
    notation.
\end{description}

\subsubsection{Finite-state decoding}

The codec maintains a three-state finite-state machine that
processes each \py{EngineDelta}:

\begin{itemize}[leftmargin=2em, itemsep=2pt]
  \item \textbf{Outside}: the default state. Text is emitted as
    \py{TextEvent}. If the channel scanner detects a channel
    opening, the state transitions to \emph{channel}. If the
    tool scanner detects a tool-call start, the state
    transitions to \emph{tool}.
  \item \textbf{Channel}: text is accumulated as trace content
    and emitted as \py{TraceEvent}. When the channel scanner
    detects the closing marker, the state returns to
    \emph{outside}.
  \item \textbf{Tool}: text is accumulated until the tool body
    is complete (balanced braces or explicit end marker). The
    tool parser then extracts the function name and arguments,
    and a \py{ToolEvent} is emitted atomically. The state
    returns to \emph{outside}.
\end{itemize}

This architecture ensures that the downstream dialect renderers
receive clean, typed events regardless of how the model formats
its output. A model that emits tool calls as raw JSON in a
\texttt{<tool\_call>} block and a model that uses
Python-style \texttt{function(args)} notation both produce
identical \py{ToolEvent} objects.

\subsection{Wire dialects}
\label{sec:llm-dialects}

A \py{Dialect} is the complete adapter between a wire protocol
and the canonical transport model. It comprises three components:

\begin{description}[leftmargin=2em, itemsep=4pt,
  font=\sffamily\bfseries]
  \item[Parser]
    converts wire-format request JSON into canonical
    \py{Message} and \py{Tool} objects. Each protocol has
    different conventions for representing messages (OpenAI uses
    content-part arrays; Anthropic uses a top-level system
    field; Ollama uses sibling image arrays), and the parser
    normalizes these into the common representation.

  \item[Renderer]
    converts canonical output events into streaming wire frames
    (SSE or NDJSON). The renderer is invoked incrementally as
    events arrive and produces one or more protocol-specific
    frames per event.

  \item[Assembler]
    converts a complete sequence of canonical events into a
    buffered response envelope. The assembler is used for
    non-streaming requests where the client expects a single
    JSON response rather than a stream.
\end{description}

Four dialects are provided:

\subsubsection{OpenAI dialect}

The OpenAI dialect implements the Chat Completions, Completions,
Responses, and Models APIs. It is mounted at
\texttt{/v1/chat/completions}, \texttt{/v1/completions},
\texttt{/v1/responses}, and \texttt{/v1/models}. Streaming
responses use SSE with \texttt{chat.completion.chunk} or
\texttt{text\_completion} events. Buffered responses return
\texttt{chat.completion} or \texttt{response} envelopes.

This dialect enables any client library that targets the OpenAI
API (the OpenAI Python SDK, LangChain, LlamaIndex, or any
HTTP client) to interact with a \flama-served model without
modification.

\subsubsection{Anthropic dialect}

The Anthropic dialect implements the Messages and Models APIs,
mounted at \texttt{/v1/messages} and \texttt{/v1/models}.
Streaming uses SSE with Anthropic-specific event types
(\texttt{message\_start}, \texttt{content\_\allowbreak block\_\allowbreak start},
\texttt{content\_\allowbreak block\_\allowbreak delta}, \texttt{message\_delta},
\texttt{message\_stop}). The parser handles Anthropic's
conventions: top-level \texttt{system} field, \texttt{thinking}
content blocks for reasoning traces, and \texttt{tool\_use}
blocks with explicit IDs.

\subsubsection{Ollama dialect}

The Ollama dialect implements the Chat, Generate, Tags, Show,
and Version APIs, mounted at \texttt{/api/chat},
\texttt{/api/generate}, \texttt{/api/tags}, \texttt{/api/show},
and \texttt{/api/version}. Unlike the other dialects, Ollama
uses Newline-Delimited JSON (NDJSON) for streaming rather than
SSE. The parser normalizes Ollama's sibling \texttt{images}
arrays into structured multimodal content parts.

\subsubsection{Native dialect}

The native dialect is \flama's own protocol, designed for
maximum fidelity to the internal event model. Unlike the other
three, it takes no URL prefix of its own, so its routes sit
directly under the model's mount point: \texttt{/} configures
the model's generation parameters, \texttt{/query/} returns a
buffered response, \texttt{/stream/} opens a live stream,
\texttt{/stream/\{stream\_id\}/} replays a past one, and
\texttt{/chat/} serves the chatbot interface.

The streaming format uses SSE with sequential event IDs that
enable client reconnection via the \texttt{Last-Event-ID}
header. The buffered \texttt{/query/} response is not assembled
by the dialect the way the other three assemble theirs: the
native \py{Assembler} raises \py{NotImplementedError}, because
the wire format has no buffered envelope of its own. Instead the
handler coalesces the event stream inline and returns the
canonical event model directly, as an envelope carrying
\texttt{id}, \texttt{created}, \texttt{stop\_reason}, token
counts, and the channel-tagged \texttt{blocks} (one per
contiguous run on the same channel).

The native dialect also mounts a built-in chatbot web interface
at \texttt{/chat/} (Section~\ref{sec:chatbot-ui}).

\subsection{Serving configuration}
\label{sec:llm-serving-config}

An LLM model is added to an application with the same
\py{add\_model()} API used for predictive models, with the
addition of a \py{serving} parameter that declares which
dialects to expose:

\begin{lstlisting}
from flama import Flama

app = Flama()

app.models.add_model(
    path="/llm/",
    model="assistant.flm",
    name="assistant",
    serving=("native", "openai", "anthropic", "ollama"),
    params={"temperature": 0.7, "max_tokens": 512},
)
\end{lstlisting}

This single call generates the complete set of endpoints for all
four dialects, mounted under the specified path prefix. The
resulting URL structure is:

\begin{center}
\small
\begin{tabularx}{\textwidth}{@{}llX@{}}
\toprule
\textbf{Dialect} & \textbf{Endpoint} &
\textbf{Format} \\
\midrule
Native &
  \texttt{/llm/query/} &
  JSON (buffered blocks) \\
Native &
  \texttt{/llm/stream/} &
  SSE (canonical events) \\
Native &
  \texttt{/llm/chat/} &
  SSE + chatbot UI \\
OpenAI &
  \texttt{/llm/openai/v1/chat/completions} &
  SSE / JSON \\
Anthropic &
  \texttt{/llm/anthropic/v1/messages} &
  SSE / JSON \\
Ollama &
  \texttt{/llm/ollama/api/chat} &
  NDJSON / JSON \\
\bottomrule
\end{tabularx}
\end{center}

Generation parameters (\py{temperature}, \py{top\_p},
\py{max\_tokens}, etc.) can be set at the serving level as
defaults and overridden per-request by the client through
protocol-specific fields.

\subsection{Stream persistence and replay}
\label{sec:stream-persistence}

The native dialect supports stream persistence, enabling clients
to replay past generations and resume interrupted streams. The
\py{StreamsBackend} abstraction provides durable event storage,
keyed throughout by a \texttt{(model, stream~id)} pair:
\py{append()} adds one event to a stream, \py{read()} returns a
half-open range of it, \py{pop()} reads a range and drops it,
\py{discard()} removes a stream outright, and \py{length()}
reports how many events each stream currently holds. Backends
are opened and closed with the application through
\py{aopen()}/\py{aclose()}.

Two implementations ship with the framework. The default
\py{FileStreamsBackend} writes each stream to
\texttt{<root>/<model>/<stream\_id>.jsonl}, one self-contained
JSON event per line, and deliberately leaves those logs in place
after shutdown so they remain available for inspection;
\py{InMemoryStreamsBackend} keeps the same interface without
touching disk. Range reads are what make replay cheap: a client
reconnecting with a \texttt{Last-Event-ID} header is served the
stored events from that sequence number onward, followed by live
generation for whatever has not been produced yet. A
\py{CleanupTask} bounds the cost of retention, evicting streams
by age and by aggregate disk usage.

\subsection{Chatbot template}
\label{sec:chatbot-ui}

The native dialect includes a built-in chatbot web interface that
provides a complete conversational UI without any frontend code.
The interface is a self-contained HTML page rendered from a
compiled template and served at the \texttt{/chat/} endpoint.

The chatbot UI provides:

\begin{itemize}[leftmargin=2em, itemsep=2pt]
  \item Real-time token streaming over SSE with automatic
    reconnection.
  \item Markdown rendering for structured responses.
  \item \LaTeX\ mathematics rendering via KaTeX for models that
    produce mathematical content.
  \item Mermaid diagram rendering for models that produce
    structured diagrams.
  \item Syntax highlighting for fenced code blocks.
  \item Conversation history management with multi-turn
    context.
\end{itemize}

The page is served by the native dialect's \texttt{/chat/}
handler, which renders the packaged \texttt{chatbot/chat.html}
template with the model's stream URL as its only context
variable. The template ships with the framework rather than
being a configuration point: an application that wants a
different interface replaces the route with its own handler and
consumes the same \texttt{/stream/} endpoint, which is the
public contract the bundled page itself is written against.

\section{The Model Context Protocol}
\label{sec:mcp}

The Model Context Protocol (MCP) \citep{mcp} is an open
standard for connecting AI models to external tools, data
sources, and programmatic capabilities. As AI applications
increasingly
require interaction with external systems (databases, APIs,
file systems, code execution environments), a standardized
protocol for discovering and invoking these capabilities becomes
necessary. \flama includes a first-class MCP module that allows
any \flama application to act as an MCP server, exposing tools,
resources, and prompts over a JSON-RPC 2.0 transport.

\subsection{MCP server}
\label{sec:mcp-server}

An MCP server in \flama is a registry of three kinds of
capabilities:

\begin{description}[leftmargin=2em, itemsep=4pt,
  font=\sffamily\bfseries]
  \item[Tools]
    are callable functions that perform actions: querying a
    database, calling an external API, running a computation,
    or controlling a device. Each tool has a name, a
    description, and an input schema generated from the
    handler's function signature.

  \item[Resources]
    are read-only data sources identified by URIs. A resource
    handler returns the content of the resource (text, JSON, or
    binary) when requested by the client. Resources are used
    to expose static or dynamic data to the model.

  \item[Prompts]
    are reusable prompt templates with parameters. The MCP
    client can list available prompts, retrieve a specific
    prompt with arguments filled in, and use the result as
    input to the language model.
\end{description}

\begin{lstlisting}
from flama import Flama
from flama.mcp import MCPServer

mcp = MCPServer(
    name="data-tools",
    version="1.0.0",
    instructions="Tools for querying the data warehouse.",
)

@mcp.tool(description="Run a SQL query")
def query_db(sql: str, limit: int = 100) -> str:
    result = execute_query(sql, limit=limit)
    return format_as_table(result)

@mcp.resource(uri="schema://tables",
              description="List all database tables")
def list_tables() -> str:
    return "\n".join(get_table_names())

@mcp.prompt(description="Generate a data analysis prompt")
def analyze(table: str, question: str) -> list:
    return [
        {"role": "system",
         "content": "You are a data analyst."},
        {"role": "user",
         "content": f"Table: {table}\n{question}"},
    ]

app = Flama()
app.mcp.add_server("/mcp", name="data-tools",
                   server=mcp)
\end{lstlisting}

\subsection{Stateless JSON-RPC transport}
\label{sec:jsonrpc}

The MCP module communicates over HTTP using the JSON-RPC~2.0
protocol \citep{jsonrpc}. Each MCP server is mounted as a POST-only HTTP
endpoint (hidden from the OpenAPI schema) that accepts JSON-RPC
requests and returns JSON-RPC responses.

\flama implements the current stateless revision of the MCP
specification, in which every request is self-contained: there
is no \texttt{initialize}/\texttt{initialized} session
handshake. Instead, the client's identity, capabilities, and
requested protocol version travel in the \py{\_meta} field of
each JSON-RPC request. The server validates the protocol
version, extracts client capabilities, and responds accordingly.

Three routing headers accompany every request to enable
efficient HTTP-level dispatch:

\begin{itemize}[leftmargin=2em, itemsep=2pt]
  \item \texttt{Mcp-Method}: the JSON-RPC method name (e.g.\
    \texttt{tools/call}).
  \item \texttt{Mcp-Name}: the target name (tool name, resource
    URI, or prompt name), extracted from the body's
    \texttt{params} field.
  \item \texttt{MCP-Protocol-Version}: the protocol revision
    requested by the client.
\end{itemize}

The server validates that these headers are consistent with the
request body before dispatching the call. Every response carries
the \texttt{MCP-Protocol-Version} header to confirm the
negotiated version.

The endpoint handles the following methods:

\begin{center}
\small
\begin{tabularx}{\textwidth}{@{}lX@{}}
\toprule
\textbf{JSON-RPC method} &
\textbf{Behaviour} \\
\midrule
\texttt{server/discover} &
  Returns the server's capabilities (tools,
  resources, prompts), supported extensions, server
  info, and optional instructions. Replaces the
  earlier session-based \texttt{initialize}
  handshake. \\
\texttt{ping} &
  Returns an empty object. Used for health checks. \\
\texttt{tools/list} &
  Returns the list of registered tools with names,
  descriptions, input schemas (JSON Schema 2020-12),
  and output schemas. \\
\texttt{tools/call} &
  Invokes a named tool with arguments and returns
  the result as a content block. Supports both
  immediate and task-based (asynchronous) execution. \\
\texttt{resources/list} &
  Returns registered resources with URIs and MIME
  types. \\
\texttt{resources/read} &
  Returns the content of a resource identified by
  URI. \\
\texttt{prompts/list} &
  Returns registered prompts with names and
  descriptions. \\
\texttt{prompts/get} &
  Returns a prompt's messages with arguments filled
  in. \\
\texttt{tasks/get} &
  Retrieves the current status of an asynchronous
  task. \\
\texttt{tasks/cancel} &
  Cancels a running task. \\
\bottomrule
\end{tabularx}
\end{center}

Every JSON-RPC response includes a \py{\_meta.dev.flama} field
with \flama's branding information, allowing MCP clients to
identify the server framework.

\subsubsection{Tool schemas}

Tool input and output schemas are emitted as self-contained
JSON Schema 2020-12 documents with local \texttt{\$defs}
references. Input schemas are generated automatically from the
tool handler's function signature (parameter names, types, and
defaults). Output schemas are generated from the return type
annotation when present; tools without a return annotation
advertise no output schema.

\begin{lstlisting}
@mcp.tool(description="Add two numbers")
def add(a: float, b: float) -> float:
    return a + b
\end{lstlisting}

This tool produces an input schema with two required
\texttt{number} properties and an output schema declaring a
\texttt{number} result, both as standalone JSON Schema documents
that clients can validate against independently.

\subsubsection{Trace context}

The MCP module extracts W3C Trace Context \citep{w3ctrace}
fields from the request's \py{\_meta}: \texttt{traceparent},
\texttt{tracestate}, and \texttt{baggage}. These are exposed as
an injectable \py{TraceContext} component that tool handlers can
use to propagate distributed traces to downstream services.

\subsection{Tasks extension}
\label{sec:mcp-tasks}

Some tool invocations are inherently long-running: training a
model, executing a complex query, or waiting for an external
approval. The MCP Tasks extension allows such tools to return
immediately with a task identifier that the client can poll for
completion.

A tool opts into task-based execution by declaring
\py{task=True} at registration time:

\begin{lstlisting}
@mcp.tool(description="Train a model", task=True)
async def train(dataset: str, epochs: int = 10) -> str:
    result = await run_training(dataset, epochs)
    return f"Model trained: accuracy {result.accuracy}"
\end{lstlisting}

When a client whose capabilities include the Tasks extension
invokes this tool, the server returns a task object immediately
with status \texttt{running}. The framework executes the handler
in the background and updates the task's status (\texttt{running}
$\to$ \texttt{completed} or \texttt{failed}) as the handler
progresses. The client retrieves the result via
\texttt{tasks/get}.

The task store persists task state across requests, and clients
can cancel running tasks via \texttt{tasks/cancel}.

\subsection{Elicitation extension}
\label{sec:mcp-elicitation}

Elicitation enables a tool handler to request additional input
from the user mid-execution. This is useful for tools that need
confirmation, disambiguation, or iterative refinement before
completing their work.

A tool signals that it needs more input by returning an
\py{Elicit} object:

\begin{lstlisting}
from flama.mcp import Elicit, Elicitation

@mcp.tool(description="Delete records")
def delete_records(table: str, where: str,
                   elicitation: Elicitation) -> str:
    if "confirm" not in elicitation:
        return Elicit.require(
            message=f"Delete from {table} where {where}?",
            schema={"confirm": {"type": "boolean"}},
        )

    if elicitation.get("confirm"):
        count = execute_delete(table, where)
        return f"Deleted {count} records."
    return "Cancelled."
\end{lstlisting}

When the handler returns an \py{Elicit}, the server responds
with \texttt{resultType: inputRequired} and a
\texttt{requestState} token that encodes the continuation
context. The client presents the elicitation prompt to the user,
collects the response, and sends a follow-up \texttt{tools/call}
with the user's answers and the \texttt{requestState} token. The
framework deserializes the state, injects the collected answers
as the \py{Elicitation} parameter, and re-executes the handler.

This mechanism is fully stateless: the continuation state is
serialized into the response and returned by the client on the
next request, with no server-side session required.

\subsection{Application templates extension}
\label{sec:mcp-apps}

The MCP Apps extension allows servers to expose prefetchable UI
templates that clients can render inline. An application template
is a named HTML or Markdown document that the client can
retrieve and display as part of a tool's result.

\begin{lstlisting}
@mcp.app_template(
    "ui://chart",
    name="chart",
    description="Render a data chart",
    mime_type="text/html",
)
def chart_template(data: str) -> str:
    return render_chart_html(data)
\end{lstlisting}

Templates are discovered through
\texttt{resources/templates/list} and read through
\texttt{resources/read}. The server advertises the Apps
extension in its capabilities when at least one template is
registered.

\subsection{Multiple servers}
\label{sec:multiple-mcp}

A single \flama application can host multiple MCP servers, each
mounted at a different path and exposing a different set of
capabilities:

\begin{lstlisting}
app.mcp.add_server("/mcp/data", name="data-tools",
                   server=data_mcp)
app.mcp.add_server("/mcp/admin", name="admin-tools",
                   server=admin_mcp)
\end{lstlisting}

Tools, resources, and prompts can also be registered directly
on the application using the \py{MCPModule}'s decorator API,
targeting a specific server by name:

\begin{lstlisting}
@app.mcp.tool(description="Health check",
              mcp="admin-tools")
def health() -> str:
    return "OK"
\end{lstlisting}

If only one MCP server is registered, the \py{mcp} parameter
can be omitted and the tool is added to the sole server
automatically.
% ==========================================================================
% End part_3_machine_learning.tex
% ==========================================================================

% ============================================================
% PART IV: Operations and tooling
% ============================================================
\flamapart{IV}{Operations and Tooling}
% ==========================================================================
% Begin part_4_operations.tex
% ==========================================================================
% ============================================================
%  Part IV — Operations
%  Sections: CLI, Configuration, Deployment, Testing
% ============================================================

\section{Command-line interface}
\label{sec:cli}

Six commands, built on Click~\citep{click}, cover the
operational lifecycle: running an application from an import
path, serving a model file with no application code at all,
driving multi-model deployments from a configuration file,
fetching and packaging models from a remote repository, working
with a model offline from the terminal, and migrating a codebase
across a major version. Two audiences use them, and the second
matters more for the design than the first. Developers reach
for the CLI interactively, but deployment scripts, CI pipelines,
and container orchestrators reach for it unattended, which is
why every option can also be set from a \texttt{FLAMA\_*}
environment variable. Each command is set out below with its
options and a worked example.

\subsection{Overview}
\label{sec:cli-overview}

The CLI is invoked through the \verb|flama| entry point and
organizes its functionality into six top-level commands:

\begin{lstlisting}[style=flamabash]
$ flama --help
Usage: flama [OPTIONS] COMMAND [ARGS]...

  Fire up your models with Flama

Options:
  --version  Check the version of your locally installed Flama
  --help     Get help about how to use Flama CLI

Commands:
  get      Download and package a model as .flm.
  model    Interact with an ML model without server.
  run      Run a Flama Application based on a route.
  serve    Serve an ML model file within a Flama Application.
  start    Start a Flama Application based on a config file.
  upgrade  Upgrade a Flama codebase to a newer major version.
\end{lstlisting}

All options support environment variable binding through the
\texttt{FLAMA\_*} prefix (e.g.\
\texttt{FLAMA\_APP=\allowbreak mymodule:\allowbreak app flama run}). This makes the CLI
fully compatible with containerized deployments where
configuration is injected through environment variables rather
than command-line arguments.

\subsection{\texttt{flama run}}
\label{sec:cli-run}

The \verb|run| command starts a Uvicorn \citep{uvicorn} server
hosting a user-defined \flama application identified by its
Python import path:

\begin{lstlisting}[style=flamabash]
$ flama run myapp:app --server-host 0.0.0.0 --server-port 8000
\end{lstlisting}

The argument \verb|myapp:app| follows the standard ASGI
convention: the module path on the left of the colon and the
application variable name on the right. The command accepts the
full set of Uvicorn server options, prefixed with
\verb|--server-|, including:

\begin{center}
\small
\begin{tabularx}{\textwidth}{@{}lllX@{}}
\toprule
\textbf{Option} & \textbf{Default} & \textbf{Type} &
\textbf{Purpose} \\
\midrule
\verb|--server-host| & \texttt{127.0.0.1} & str &
  Bind address \\
\verb|--server-port| & \texttt{8000} & int &
  Bind port \\
\verb|--server-reload| & off & flag &
  Auto-reload on code changes \\
\verb|--server-workers| & 1 & int &
  Number of worker processes \\
\verb|--server-loop| & \texttt{auto} & choice &
  Event loop implementation \\
\verb|--server-http| & \texttt{auto} & choice &
  HTTP protocol implementation \\
\verb|--server-ws| & \texttt{auto} & choice &
  WebSocket implementation \\
\verb|--server-log-level| & \texttt{info} & choice &
  Logging verbosity \\
\verb|--server-ssl-certfile| & --- & path &
  SSL certificate file \\
\verb|--server-ssl-keyfile| & --- & path &
  SSL private key file \\
\bottomrule
\end{tabularx}
\end{center}

The \verb|run| command is the appropriate choice when the
developer has written a complete \flama application with custom
routes, middleware, and configuration.

\subsection{\texttt{flama serve}}
\label{sec:cli-serve}

The \verb|serve| command is a one-liner for deploying one or more
models behind a REST API with no application code. It accepts
one or more \verb|--model| specifications and generates a \flama
application on the fly:

\begin{lstlisting}[style=flamabash]
$ flama serve --model classifier.flm
\end{lstlisting}

This command:
\begin{enumerate}[leftmargin=2em, itemsep=2pt]
  \item Reads the model file and detects the artifact family from
    its metadata.
  \item Generates a temporary Python module from a Jinja2
    template (\texttt{app.py.j2}) that creates a \flama
    application and registers each model via
    \verb|app.models.add_model()|.
  \item Starts a Uvicorn server hosting the generated
    application.
\end{enumerate}

The generated application exposes each model at its specified
URL prefix, plus an OpenAPI schema at \texttt{/schema/} and a
Swagger UI at \texttt{/docs/}.

Application-level options customize the generated application:

\begin{center}
\small
\begin{tabularx}{\textwidth}{@{}lllX@{}}
\toprule
\textbf{Option} & \textbf{Default} & \textbf{Env var} &
\textbf{Purpose} \\
\midrule
\verb|--app-title| & \texttt{Flama} & \texttt{APP\_TITLE} &
  Application name \\
\verb|--app-version| & \texttt{0.1.0} & \texttt{APP\_VERSION} &
  Application version \\
\verb|--app-description| & (default) & \texttt{APP\_DESCRIPTION} &
  OpenAPI description \\
\verb|--app-debug| & off & \texttt{APP\_DEBUG} &
  Enable debug mode \\
\verb|--app-schema| & \texttt{/schema/} & \texttt{APP\_SCHEMA} &
  OpenAPI schema route \\
\verb|--app-docs| & \texttt{/docs/} & \texttt{APP\_DOCS} &
  Swagger UI route \\
\bottomrule
\end{tabularx}
\end{center}

The \verb|--model| option accepts three forms. The simplest is a
bare path; the full form uses comma-separated key=value pairs;
and the file form loads a complete specification from JSON, YAML,
or TOML:

\begin{lstlisting}[style=flamabash]
# Bare path (defaults: url=/, name=model)
$ flama serve --model classifier.flm

# Full form with explicit keys
$ flama serve --model file=classifier.flm,url=/sentiment,name=v2

# Load spec from file
$ flama serve --model @spec.json
\end{lstlisting}

The model specification supports the following keys:

\begin{center}
\small
\begin{tabularx}{\textwidth}{@{}lllX@{}}
\toprule
\textbf{Key} & \textbf{Default} & \textbf{Applies to} &
\textbf{Purpose} \\
\midrule
\texttt{file} & (required) & all &
  Path to the \flm file \\
\texttt{url} & \texttt{/} & all &
  URL prefix for the model's endpoints \\
\texttt{name} & \texttt{model} & all &
  Model name \\
\texttt{serving} & (auto) & LLM &
  Colon-separated dialect list
  (native, openai, anthropic, ollama) \\
\texttt{params} & (none) & LLM &
  Colon-separated key=value generation
  parameters \\
\texttt{channel\_scanner} & (auto) & LLM &
  Override channel detection \\
\texttt{tool\_scanner} & (auto) & LLM &
  Override tool-call detection \\
\texttt{tool\_parser} & (auto) & LLM &
  Override tool-call parsing \\
\bottomrule
\end{tabularx}
\end{center}

A complete LLM deployment with multiple dialects:

\begin{lstlisting}[style=flamabash]
$ flama serve \
    --model file=assistant.flm,url=/llm,name=assistant,\
serving=native:openai:anthropic,\
params=temperature=0.7:max_tokens=4096 \
    --server-host 0.0.0.0 \
    --server-port 8000
\end{lstlisting}

This produces a single application with the native chatbot UI at
\texttt{/llm/chat/}, OpenAI-compatible endpoints at
\texttt{/llm/\allowbreak openai/\allowbreak v1/\allowbreak chat/\allowbreak completions}, and Anthropic-compatible
endpoints at \texttt{/llm/\allowbreak anthropic/\allowbreak v1/\allowbreak messages}. Multiple
\verb|--model| flags can be passed to serve several models in the
same application.

\subsection{\texttt{flama start}}
\label{sec:cli-start}

The \verb|start| command is designed for multi-model
deployments managed through a JSON configuration file. It reads
a configuration file (default: \verb|flama.json|) that
specifies the application metadata, the list of models to serve,
and the server options:

\begin{lstlisting}[style=flamajson]
{
  "app": {
    "title": "ML Platform",
    "version": "2.0.0",
    "description": "Production model serving",
    "debug": false,
    "schema": "/schema/",
    "docs": "/docs/",
    "models": [
      {
        "url": "/sentiment",
        "path": "models/sentiment.flm",
        "name": "sentiment"
      },
      {
        "url": "/toxicity",
        "path": "models/toxicity.flm",
        "name": "toxicity"
      },
      {
        "url": "/assistant",
        "path": "models/assistant.flm",
        "name": "assistant",
        "serving": ["native", "openai"]
      }
    ]
  },
  "server": {
    "host": "0.0.0.0",
    "port": 8000,
    "workers": 4,
    "log_level": "info"
  }
}
\end{lstlisting}

The command accepts a \verb|--create-config| option that
generates a template configuration file:

\begin{lstlisting}[style=flamabash]
# Generate a minimal config with host and port only
$ flama start --create-config simple

# Generate a full config with all server options
$ flama start --create-config full
\end{lstlisting}

This is particularly useful for teams that manage model
deployments through infrastructure-as-code workflows: the
configuration file can be version-controlled, reviewed, and
deployed through CI/CD pipelines.

\subsection{\texttt{flama get}}
\label{sec:cli-get}

The \verb|get| command downloads a model from a remote source and
packages it into the \flm format, ready for serving with
\verb|flama serve| or offline interaction with
\verb|flama model|. This command bridges the gap between model
repositories and the \flama deployment pipeline.

\begin{lstlisting}[style=flamabash]
$ flama get --source huggingface --family llm \
    Qwen/Qwen3-8B
\end{lstlisting}

The command downloads all model files concurrently (with
configurable parallelism), packages them into a \flm file using
protocol version~2, and records the artifact family in the
manifest. The \verb|--family| flag is required and determines
how the model is dispatched at load time:

\begin{center}
\small
\begin{tabularx}{\textwidth}{@{}lllX@{}}
\toprule
\textbf{Option} & \textbf{Default} & \textbf{Type} &
\textbf{Purpose} \\
\midrule
\verb|--source| & (required) & choice &
  Model source provider (currently
  \texttt{huggingface}) \\
\verb|--family| & (required) & choice &
  Artifact family (\texttt{ml} or \texttt{llm}).
  Drives runtime dispatch \\
\verb|-o, --output| & (auto) & path &
  Output \flm path (default:
  \texttt{<model-name>.flm}) \\
\verb|--max-concurrent| & \texttt{8} & int &
  Maximum parallel file downloads \\
\bottomrule
\end{tabularx}
\end{center}

Downloads employ bounded exponential backoff with jitter for
transient failures (HTTP~429, 5xx, network errors), and each
file is streamed to a temporary path and atomically renamed on
success to prevent partial downloads from corrupting the output.

\begin{lstlisting}[style=flamabash]
# Download a traditional ML model
$ flama get --source huggingface --family ml \
    scikit-learn/Fish-Weight

# Download an LLM with custom output path
$ flama get --source huggingface --family llm \
    -o assistant.flm \
    google/gemma-4-E2B-it
\end{lstlisting}

\subsection{\texttt{flama model}}
\label{sec:cli-model}

The \verb|model| command group provides offline interaction
with serialized models, without starting a server. It accepts a
model path as a positional argument and exposes three
subcommands: \verb|inspect| (metadata display), \verb|run|
(one-shot inference), and \verb|stream| (streaming inference).
Both \verb|run| and \verb|stream| work for ML and LLM models
alike; for LLM models, additional options control the transport
shape, system instructions, generation parameters, and output
channels.

The group-level options \verb|--channel-scanner|,
\verb|--tool-scanner|, and \verb|--tool-parser| apply to LLM
artifacts only and override the automatic decoder detection
described in Section~\ref{sec:llm-codec}.

\subsubsection{\texttt{flama model inspect}}

The \verb|inspect| subcommand displays the metadata stored in
the FLM file:

\begin{lstlisting}[style=flamabash]
$ flama model classifier.flm inspect --pretty
\end{lstlisting}

\begin{lstlisting}[style=flamajson]
{
  "meta": {
    "id": "classifier-v2",
    "timestamp": "2025-01-15T10:30:00",
    "framework": {
      "lib": "sklearn",
      "version": "1.7.2"
    },
    "model": {
      "obj": "RandomForestClassifier",
      "info": {
        "n_estimators": 100,
        "max_depth": 8,
        "criterion": "gini"
      },
      "params": {
        "n_estimators": 100,
        "max_depth": 8
      },
      "metrics": {
        "accuracy": 0.947,
        "f1": 0.932
      }
    },
    "extra": {
      "dataset": "prod-2024-q3",
      "author": "team-ml"
    }
  },
  "artifacts": {}
}
\end{lstlisting}

This subcommand is useful for verifying that a model was
serialized correctly, checking its training metrics before
deployment, and inventorying models in a model registry.

\subsubsection{\texttt{flama model run}}

The \verb|run| subcommand performs one-shot inference without a
server. For ML models, the input is a JSON array of feature
vectors and the output is a JSON array of predictions. For LLM
models, the input is a prompt and the output is the generated
response:

\begin{lstlisting}[style=flamabash]
# ML model: batch predictions from stdin
$ echo '[[5.1,3.5,1.4,0.2],[6.7,3.0,5.2,2.3]]' \
    | flama model classifier.flm run --pretty

# ML model: from file to file
$ flama model classifier.flm run \
    -i input.json -o output.json

# LLM model: single-turn generation
$ echo "Explain quantum entanglement briefly" \
    | flama model assistant.flm run --transport chat

# LLM model: with system instruction and params
$ echo "What is Python?" \
    | flama model assistant.flm run \
        --system "Be concise." \
        --param temperature=0.7
\end{lstlisting}

For LLM models, the \verb|--transport| flag determines how the
input is formatted before tokenization: \texttt{raw} sends the
text directly, \texttt{chat} wraps it as a single user message
with the model's chat template, and \texttt{conversation}
expects a JSON array of messages. Generation parameters are
passed as repeatable \verb|--param key=value| flags.

The \verb|--channel| flag includes secondary output channels
(reasoning traces, analysis steps) in the output, which are
otherwise suppressed by default. With one channel the output is
plain text; with multiple channels each block is emitted as a
JSON object with \texttt{channel} and \texttt{text} fields.

This subcommand is valuable in three scenarios:
\begin{itemize}[leftmargin=2em, itemsep=2pt]
  \item \textbf{CI/CD validation}: after model training, a
    pipeline step can run \verb|flama model run| on a
    reference dataset and compare outputs to expected values.
  \item \textbf{Batch scoring}: for workloads where
    request-by-request API calls add unnecessary overhead,
    predictions can be computed in a single pass over a JSON
    file.
  \item \textbf{Debugging}: during development, a practitioner
    can test a model's behaviour on specific inputs
    without starting a server.
\end{itemize}

\subsubsection{\texttt{flama model stream}}

The \verb|stream| subcommand performs streaming generation,
printing tokens as they are produced. It supports both ML and
LLM models, though streaming is most useful for LLMs where
generation takes appreciable time:

\begin{lstlisting}[style=flamabash]
$ echo "Write a haiku about Rust" \
    | flama model assistant.flm stream \
        --transport chat --channel all
\end{lstlisting}

The options mirror those of \verb|run| (\verb|--transport|,
\verb|--system|, \verb|--param|, \verb|--channel|), with the
addition of \verb|--buffer| which accumulates all output and
writes it at once rather than printing incrementally.

This is particularly useful for interactive experimentation
with LLMs during development, as it provides immediate visual
feedback on generation quality and speed without requiring a
running server.

\subsection{\texttt{flama upgrade}}
\label{sec:cli-upgrade}

The \verb|upgrade| command assists with codebase migration
across major \flama versions. It rewrites import statements and
renamed symbols across all Python files in the specified paths,
applying automated transformations that cover the majority of
breaking changes between versions.

\begin{lstlisting}[style=flamabash]
# Preview changes as a unified diff (default)
$ flama upgrade src/ tests/

# Apply changes in place
$ flama upgrade --write src/ tests/
\end{lstlisting}

The command operates in two modes: \verb|--diff| (default)
previews the changes as a unified diff and exits with code~1 if
any changes would be made, making it suitable for CI
integration; \verb|--write| applies the changes in place.

\begin{center}
\small
\begin{tabularx}{\textwidth}{@{}lllX@{}}
\toprule
\textbf{Option} & \textbf{Default} & \textbf{Type} &
\textbf{Purpose} \\
\midrule
\verb|--to| & (latest) & version &
  Target version \\
\verb|--from| & (detect) & version &
  Source version to migrate from \\
\verb|--diff/--write| & diff & flag &
  Preview or apply changes \\
\verb|--select| & (all) & csv &
  Run only specified operations \\
\verb|--skip| & (none) & csv &
  Skip specified operations \\
\bottomrule
\end{tabularx}
\end{center}

Symbols that have no automatic replacement are flagged with a
\texttt{\# flama-upgrade} marker and listed as manual
follow-ups, ensuring that no breaking change passes silently.

\section{Configuration and deployment}
\label{sec:deployment}

One application has to run in development, in staging, and in
production, at different addresses, with different worker
counts, different SSL material, and different model paths.
Configuration is what absorbs that variation, and \flama splits
it in two: what the application is (title, version, the models
it serves) is kept apart from how it is served (host, port,
workers). An application can be built from a Python import
string, from a dictionary, or from a JSON file, and the three
paths converge on the same objects. What follows covers those
objects, the settings reader an application uses for its own
configuration, and the deployment patterns the framework
supports.

\subsection{Configuration architecture}
\label{sec:config-architecture}

The configuration system consists of three dataclasses that
separate application concerns from server concerns:

\begin{description}[leftmargin=2em, itemsep=4pt,
  font=\sffamily\bfseries]
  \item[App]
    encapsulates the application identity (title, version,
    description), feature configuration (schema route, docs
    route, debug mode), and the list of models to serve.
    The \py{App} class supports three construction paths:
    from a Python import string
    (\py{StrApp("mymodule:app")}), from a dictionary
    (\py{DictApp.from\_dict(data)}), or from a live \flama
    instance (\py{FlamaApp(app)}).

  \item[Uvicorn]
    wraps all Uvicorn server options into a single data
    structure: bind address, port, number of workers, SSL
    configuration, event loop selection, protocol
    implementations, reload settings, logging configuration,
    and timeouts. Every field has a sensible default and can
    be overridden from the CLI, from environment variables, or
    from the JSON configuration file.

  \item[Config]
    combines an \py{App} and a \py{Uvicorn} instance and
    provides the \py{run()} method that starts the server.
    Config objects can be serialized to JSON
    (\py{config.dumps()}) and deserialized from JSON
    (\py{Config.loads(data)}) or from a file handle
    (\py{Config.load(fs)}).
\end{description}

\subsection{Application settings}
\label{sec:app-settings}

The dataclasses above describe how the CLI is told what to run.
They are distinct from \py{flama.config}, which is what an
application uses to read its \emph{own} settings: credentials,
feature flags, database coordinates, and anything else that
varies between environments. A \py{Config} object is bound to an
optional configuration file, in \texttt{ini}, \texttt{json},
\texttt{yaml}, or \texttt{toml} format, and resolves each key by
looking first at the environment, then at that file, and finally
at an explicit default; a key that is found nowhere and has no
default raises \py{KeyError} rather than silently yielding
\py{None}.

\begin{lstlisting}
import dataclasses
from flama.config import Config, Secret

config = Config("config.yaml", format="yaml")

DEBUG = config("DEBUG", cast=bool)
API_KEY = config("API_KEY", cast=Secret)
DB_HOST = config("DATABASE.host")
TIMEOUT = config("TIMEOUT", default=30, cast=int)

@dataclasses.dataclass
class FeatureFlags:
    enable_new_ui: bool
    max_daily_limit: int

FEATURES = config("FEATURE_FLAGS", cast=FeatureFlags)
\end{lstlisting}

The \py{cast} argument accepts any type or callable, and two
cases are worth singling out. A dotted key
(\py{"DATABASE.host"}) walks into nested structures in the
configuration file, so a hierarchical document can be read a
leaf at a time. Casting to a dataclass parses the value as JSON
when it arrives as a string, which is what an environment
variable always is, and then constructs the dataclass from the
matching keys, ignoring any others; this makes a structured
setting expressible either as a nested block in the file or as a
single JSON-valued environment variable, with the application
reading it the same way in both cases.

\py{Secret} wraps a value so that printing it, logging it, or
including it in a traceback shows \py{Secret('*****')} rather
than the value itself, which remains available through
\py{str()}. It is a guard against accidental disclosure in
diagnostics, not an encryption mechanism.

\subsection{Application generation}
\label{sec:app-generation}

When used with the \verb|serve| or \verb|start| commands, the
configuration system generates a temporary \flama application
from a Jinja2 template. The template creates a \flama instance
with the configured metadata and registers each model:

\begin{lstlisting}
from flama import Flama

app = Flama(
    debug=True,
    openapi={
        "info": {
            "title": "ML Platform",
            "version": "2.0.0",
            "description": "Production model serving",
        }
    },
    schema="/schema/",
    docs="/docs/"
)

models = [{"url": "/sentiment", "path": "sentiment.flm",
           "name": "sentiment"}]
for model in models:
    app.models.add_model(
        path=model["url"],
        model=model["path"],
        name=model["name"],
    )
\end{lstlisting}

The generated module is written to a temporary file and passed
to Uvicorn as an import path. This approach means that the
\verb|serve| and \verb|start| commands produce fully standard
\flama applications that can be inspected, debugged, and
extended as needed.

\subsection{Debug mode}
\label{sec:debug-mode}

When debug mode is enabled (\verb|--app-debug| or
\py{debug=True}), the framework activates the
\py{ServerErrorMiddleware} in its interactive mode. Unhandled
exceptions produce HTML error pages that include:

\begin{itemize}[leftmargin=2em, itemsep=2pt]
  \item The traceback, one entry per frame, each carrying the
    filename, the function, the line number, and ten lines of
    surrounding source.
  \item A marker on each frame recording whether it belongs to
    the application or to a vendored dependency, so the frames
    worth reading can be told from the ones that are only
    passing the exception along.
  \item The request that caused it: path, method, query
    parameters, headers, cookies, and the client's host and
    port.
\end{itemize}

The page is served only when the request's \texttt{Accept}
header asks for HTML; anything else gets a plain
\texttt{Internal Server Error}, which keeps API clients from
receiving a page of source code. The exception is re-raised
after the response is sent either way, so the server still logs
it and a test client can still catch it.

Debug mode has no place in production. Source, file paths, and
request headers are all exposed by it, and the framework does
not currently warn about this at startup, so the responsibility
for not shipping \py{debug=True} rests with the deployment
configuration.

\subsection{Deployment patterns}
\label{sec:deployment-patterns}

\subsubsection{Direct server}

The simplest deployment runs the \flama application directly
with Uvicorn:

\begin{lstlisting}[style=flamabash]
$ flama run myapp:app \
    --server-host 0.0.0.0 \
    --server-port 8000 \
    --server-workers 4
\end{lstlisting}

This is suitable for internal services, development servers, and
containerized deployments behind a load balancer.

\subsubsection{Container deployment}

A minimal Dockerfile for a \flama model server:

\begin{lstlisting}[style=flamabash]
FROM python:3.12-slim
WORKDIR /app
COPY requirements.txt .
RUN pip install --no-cache-dir -r requirements.txt
COPY models/ models/
COPY flama.json .
EXPOSE 8000
CMD ["flama", "start", "flama.json"]
\end{lstlisting}

The \verb|flama start| command reads the configuration from
\verb|flama.json|, which specifies the models, routes, and
server options. Environment variables can override any option
at runtime, making the same container image deployable in
different environments:

\begin{lstlisting}[style=flamabash]
$ docker run -e HOST=0.0.0.0 -e PORT=8080 \
    -v /models:/app/models myimage:latest
\end{lstlisting}

\subsubsection{Reverse proxy}

For production deployments, \flama applications are typically
placed behind a reverse proxy (Nginx, Traefik, or a cloud load
balancer) that handles TLS termination, rate limiting, and
static file serving. The \py{TrustedHostMiddleware} validates
the \texttt{Host} header against a whitelist, and the
\py{CORSMiddleware} manages cross-origin access policies.

\section{Lifespan management}
\label{sec:lifespan}

\flama implements the ASGI lifespan protocol, which provides
a structured way to run initialization code at server startup
and cleanup code at server shutdown. This replaces the ad-hoc
signal handlers and module-level initialization patterns common
in WSGI applications.

\flama provides three complementary mechanisms for managing
application lifecycle:

\begin{description}[leftmargin=2em, itemsep=4pt,
  font=\sffamily\bfseries]
  \item[Event handlers]
    are individual async callables registered for the
    \texttt{startup} or \texttt{shutdown} events. They can be
    provided at construction time or registered dynamically
    with decorators.

  \item[Lifespan context manager]
    is a callable that receives the application and returns an
    async context manager. The code before \py{yield} runs
    after all startup event handlers, and the code after
    \py{yield} runs before all shutdown event handlers.

  \item[Module hooks]
    are \py{on\_startup()} and \py{on\_shutdown()} methods on
    \py{Module} subclasses. They are registered automatically
    when the module is added to the application.
\end{description}

Event handlers can be registered at construction time:

\begin{lstlisting}
from flama import Flama

notifications = NotificationService()

async def connect_notifications():
    await notifications.connect()

async def close_notifications():
    await notifications.disconnect()

app = Flama(
    events={
        "startup": [connect_notifications],
        "shutdown": [close_notifications],
    }
)
\end{lstlisting}

Alternatively, the decorator syntax registers handlers after
construction. An application using the \py{SQLAlchemyModule}
typically uses a startup handler to create its schema, reaching
the engine through the module rather than through any
application-level state bag:

\begin{lstlisting}
app = Flama(modules=[SQLAlchemyModule(DATABASE_URL)])

@app.on_event("startup")
async def create_schema():
    async with app.sqlalchemy.engine.begin() as connection:
        await connection.run_sync(metadata.create_all)
\end{lstlisting}

For applications that prefer the context-manager pattern, the
\py{lifespan} parameter accepts a callable that returns an
async context manager:

\begin{lstlisting}
from contextlib import asynccontextmanager
from flama import Flama

@asynccontextmanager
async def lifespan(app):
    # Runs after startup event handlers
    model = load_heavy_model()
    setattr(app, "model", model)
    yield
    # Runs before shutdown event handlers
    delattr(app, "model")

app = Flama(lifespan=lifespan)
\end{lstlisting}

All startup event handlers run concurrently (via
\py{run\_task\_group}), then the lifespan context manager is
entered. On shutdown, the lifespan context manager exits first,
then all shutdown event handlers run concurrently. \flama also
propagates lifespan events to child mounted applications,
ensuring that sub-applications initialise and tear down
correctly.

Whatever a lifespan or a module sets up is reachable from the
application object, which is itself injectable, so a component
is all that is needed to hand it to request handlers as a typed
dependency:

\begin{lstlisting}
class DatabaseEngineComponent(Component):
    def resolve(self, app: Flama) -> AsyncEngine:
        return app.sqlalchemy.engine
\end{lstlisting}

\flama has no general-purpose \py{state} bag on the
application; state belongs either to a module, which owns it
behind its own name (Section~\ref{sec:module-system}), or to a
component, which produces it on demand. Both are reached
through the application instance, and both are typed.

\section{Testing}
\label{sec:testing}

\flama ships its own test client, \py{flama.client.Client}. It
subclasses \py{httpx.AsyncClient}, so the request API is the
familiar one, and wraps the application's lifespan: entering the
context manager runs the startup handlers and module hooks,
leaving it runs the shutdown ones. This matters more than
convenience. A \flama application refuses to serve requests
before its lifespan has run, so driving it through a bare
\py{httpx.ASGITransport}, which does not implement the lifespan
protocol, fails on the first request rather than exercising the
application:

\begin{lstlisting}
import pytest

from flama.client import Client
from myapp import app

@pytest.fixture(scope="function")
async def client():
    async with Client(app=app) as client:
        yield client

async def test_get_user(client):
    response = await client.get("/users/1/")
    assert response.status_code == 200
    assert response.json()["name"] == "Ada Lovelace"

async def test_create_user(client):
    response = await client.post(
        "/users/",
        json={"name": "Alan Turing",
              "email": "alan@example.com"},
    )
    assert response.status_code == 201
    assert response.json()["id"] is not None
\end{lstlisting}

Because resource routes are named, a test need not hard-code
URLs: \py{app.resolve\_url()} builds them from the resource and
operation, so a change of mount point does not ripple through
the suite.

\begin{lstlisting}
async def test_list_users(client):
    url = client.app.resolve_url("user:list").path
    response = await client.get(str(url), params={"page_size": 100})
    assert response.status_code == 200
\end{lstlisting}

\subsection{Testing ML model endpoints}
\label{sec:testing-ml}

Model endpoints can be tested by mounting the application with
a model file and sending prediction requests:

\begin{lstlisting}
import pytest

from flama.client import Client

@pytest.fixture(scope="function")
async def ml_client():
    async with Client(
        models=[("classifier", "/classifier/",
                 "tests/fixtures/classifier.flm")]
    ) as client:
        yield client

async def test_model_inspect(ml_client):
    response = await ml_client.get("/classifier/")
    assert response.status_code == 200
    assert response.json()["meta"]["framework"]["lib"] == "sklearn"

async def test_model_predict(ml_client):
    response = await ml_client.model_request(
        "classifier", "POST", "/predict/",
        json={"input": [[5.1, 3.5, 1.4, 0.2]]},
    )
    assert response.status_code == 200
    assert "output" in response.json()
\end{lstlisting}

The \py{models} argument is a shortcut that builds an
application and registers each \py{(name, url, path)} triple
through \py{add\_model()}, which saves constructing one by hand
when the models are all the test is about.
\py{model\_request()} then addresses a model by name, resolving
the rest of the URL from where it was mounted.

\subsection{Testing with Workers and repositories}
\label{sec:testing-ddd}

Applications that use the Worker and Repository patterns
(Section~\ref{sec:resources-ddd}) can be tested with a real
in-memory database or with mocked repositories:

\begin{lstlisting}
import pytest

from flama import Flama
from flama.client import Client
from flama.sqlalchemy import SQLAlchemyModule, metadata

DATABASE_URL = "sqlite+aiosqlite:///:memory:"

@pytest.fixture(scope="function")
async def client():
    app = Flama(modules=[SQLAlchemyModule(DATABASE_URL)])
    app.resources.add_resource("/users/", UserResource)

    async with Client(app=app) as client:
        engine = client.app.sqlalchemy.engine
        async with engine.begin() as connection:
            await connection.run_sync(metadata.create_all)
        yield client
\end{lstlisting}

The engine is taken from the \py{SQLAlchemyModule} once the
client has entered the application's lifespan, which is what
creates it; creating a second engine by hand would leave the
tests talking to a different database from the application. An
in-memory SQLite database gives each test run its own isolated
schema without an external server.

For suites that must not pay for schema creation per test, the
same structure works one level up: create the schema once in a
session-scoped fixture, then wrap each test in a transaction
that is rolled back on teardown. The module exposes
\py{open\_connection()}, \py{begin\_transaction()},
\py{end\_transaction()}, and \py{close\_connection()} for
exactly that, so a test can share the application's connection
and still leave no trace behind.
% ==========================================================================
% End part_4_operations.tex
% ==========================================================================

% ============================================================
% PART V: Ecosystem and outlook
% ============================================================
\flamapart{V}{Ecosystem and Outlook}
% ==========================================================================
% Begin part_5_ecosystem.tex
% ==========================================================================
% ============================================================
%  Part V — Ecosystem and outlook
%  Sections: Related Work, Feature Comparison, Conclusion
% ============================================================

\section{Related work}
\label{sec:related-work}

Python web frameworks and ML serving systems have evolved along
tracks that barely touch. Web frameworks took up asynchronous
execution and type-driven validation and largely stopped there,
while model serving platforms grew up separately to move trained
models into production, bringing packaging formats, batching,
and monitoring but almost none of the HTTP machinery. A third
track opened more recently with the LLM inference engines built
for generative workloads specifically. Each is worth taking in
turn, because what \flama is depends on what these are not.

\subsection{Web frameworks}
\label{sec:related-web}

\textbf{Flask} \citep{flask} has been the default answer for
small and medium Python web applications since 2010, and its
minimalism is deliberate: URL routing, template rendering, and a
request/response abstraction are provided, and everything else,
validation and serialization and authentication and database
access alike, is left to extensions. That works until
concurrency enters the picture. Being synchronous and
WSGI-based, Flask has no answer of its own for I/O-bound or
CPU-bound work beyond an external queue such as Celery or RQ,
which is a second piece of infrastructure to run and monitor.
Nothing in it is ML-specific.

\textbf{Django} \citep{django} takes the opposite position and
supplies almost everything: an ORM, an admin interface,
authentication, templates, with the Django REST Framework
\citep{djangorest} adding serialization, viewsets, and generated
API documentation on top. The cost is weight and a
synchronous ORM whose async support remains partial, which
tells against it for lightweight API services and real-time
workloads. On models it is as silent as Flask.

\textbf{Starlette} \citep{starlette} is a lightweight ASGI
toolkit supplying routing, request and response classes,
middleware, WebSocket support, and a test client. It is a
toolkit rather than a framework: it does not include validation,
dependency injection, or API documentation. \flama's 1.x series
was built on top of Starlette; the 2.0 rewrite replaced it with
a native HTTP, routing, and middleware core, part of it
Rust-accelerated (Section~\ref{sec:layered-design}), removing
the dependency.

\subsection{ML serving platforms}
\label{sec:related-ml}

\textbf{TensorFlow Serving} \citep{tfserving} is very good at
one thing. Given a SavedModel artifact it will version it, batch
requests against it, and expose it over gRPC or REST at high
throughput. What it will not do is anything else: the models
have to be TensorFlow's, the endpoints are the ones it
generates, and there is no route by which an inference result
might be joined to a database row before being returned.

\textbf{TorchServe} \citep{torchserve} occupies the same
position for PyTorch, with model archiving, worker management,
metrics, and A/B testing. Custom handler classes provide some
room to manoeuvre, but the shape of the deployment is fixed: a
standalone server that application logic has to be arranged
around rather than composed with.

Serving is only a part of what \textbf{MLflow} \citep{mlflow}
does, most of it being experiment tracking and a model registry.
A model can be put behind a REST API from there, though the
component that does it is a thin wrapper, and none of the
routing, validation, authentication, or database integration
that a web framework would bring comes with it.

Closest of all in ambition is \textbf{BentoML} \citep{bentoml},
which has a packaging format of its own (Bento), batching and
concurrency in its serving layer, and support across ML
frameworks. The divergence is in what that serving layer is
for. It is built for model inference specifically, so pluggable
schema validation, dependency injection, database-backed
resources, domain-driven design, and JWT authentication are
outside its remit, and an application needing both inference and
ordinary API endpoints ends up running BentoML alongside a web
framework rather than instead of one.

\subsection{LLM inference engines}
\label{sec:related-llm}

\textbf{vLLM} \citep{vllm} is a high-throughput inference engine
for large language models. It introduces PagedAttention for
efficient KV-cache management, continuous batching for maximizing
GPU utilization, and tensor parallelism for multi-GPU
deployments. An OpenAI-compatible API server comes with it,
which is often mistaken for the thing being a web framework. It
is not one. Schema validation, dependency injection, resource
generation, authentication, database access, and any endpoint
that is not an LLM endpoint all fall outside it, so an
application wanting inference and an ordinary API runs vLLM
beside itself as a separate service. \flama uses vLLM as a
backend for exactly this reason: the engine is worth having,
and the framework is a different problem.

\textbf{Ollama} \citep{ollama} aims lower and hits its target,
handling model download, quantization, and local serving behind
a REST API of its own design. The composition limits are the
same as vLLM's, and for the same structural reason.

\textbf{LiteLLM} \citep{litellm} is neither engine nor
framework but a translator, converting between the OpenAI,
Anthropic, Ollama, and other formats so that an application can
change provider without changing code. It runs no models
itself. Where \flama renders several dialects from one
in-process model, LiteLLM forwards to a provider that does the
rendering, which is a genuine alternative to multi-dialect
serving and costs a network hop per request.

\subsection{Positioning \flama}
\label{sec:positioning}

Three categories come out of the survey. There are web
frameworks that can be extended toward model serving (Flask,
Django), predictive model servers that can be extended a little
way toward the web (TensorFlow Serving, TorchServe, BentoML),
and LLM engines presenting a single protocol (vLLM, Ollama).
What they share is the direction of the gap: whichever one is
picked, the others' capabilities arrive later as external
tools, glue code, or an operational workaround.

\flama sits where none of them do, as a web framework whose
model serving is native rather than bolted on, for predictive
and generative models alike. The claim is narrower than it
sounds and rests on a single implementation fact: the routing,
validation, dependency injection, middleware, and
authentication that a conventional endpoint goes through are
not merely similar to what an inference endpoint goes through,
they are the same objects. A prediction route is a route. It
carries the same tags the authentication middleware reads, and
the paginator does not know or care whether the collection it
wraps came out of a table or a model.

The multi-dialect layer follows from the same arrangement. One
deployment answers the OpenAI SDK, the Anthropic SDK, and the
Ollama CLI at once, with no protocol-specific application code
and no proxy in between, because the dialects are renderers
over one canonical event stream rather than separate servers.

\section{Feature comparison}
\label{sec:comparison}

Table~\ref{tab:comparison} summarizes the feature sets of \flama
and six representative frameworks across three dimensions: web API
capabilities, predictive ML serving, and LLM inference capabilities.
Features are evaluated based on built-in support only; third-party
extensions and custom code are not counted.

\begin{table}[ht]
\centering
\small
\begin{tabularx}{\textwidth}{@{}l*{7}{>{\centering\arraybackslash}X}@{}}
\toprule
\textbf{Feature} &
\rotatebox{55}{\textbf{\flama{}}} &
\rotatebox{55}{\textbf{Flask}} &
\rotatebox{55}{\textbf{TF Serving}} &
\rotatebox{55}{\textbf{TorchServe}} &
\rotatebox{55}{\textbf{BentoML}} &
\rotatebox{55}{\textbf{vLLM}} &
\rotatebox{55}{\textbf{Ollama}} \\
\midrule
ASGI / async-native
  & \checkmark & -- & -- & -- & -- & -- & -- \\
WebSocket \& streaming
  & \checkmark & -- & -- & -- & -- & -- & -- \\
SSE / NDJSON responses
  & \checkmark & -- & -- & -- & -- & \checkmark & \checkmark \\
Type-driven validation
  & \checkmark & -- & -- & -- & -- & -- & -- \\
Pluggable schema libs
  & \checkmark & -- & -- & -- & -- & -- & -- \\
Component-based DI
  & \checkmark & -- & -- & -- & -- & -- & -- \\
CRUD resource gen.
  & \checkmark & -- & -- & -- & -- & -- & -- \\
DDD patterns
  & \checkmark & -- & -- & -- & -- & -- & -- \\
JWT authentication
  & \checkmark & -- & -- & -- & -- & -- & -- \\
OpenAPI generation
  & \checkmark & -- & -- & -- & -- & -- & -- \\
Multi-framework ML
  & \checkmark & -- & -- & -- & \checkmark & -- & -- \\
Model packaging format
  & \checkmark & -- & -- & -- & \checkmark & -- & -- \\
Codeless model serving
  & \checkmark & -- & \checkmark & \checkmark & \checkmark & \checkmark & \checkmark \\
Multi-backend LLM
  & \checkmark & -- & -- & -- & -- & -- & -- \\
Multi-dialect serving
  & \checkmark & -- & -- & -- & -- & -- & -- \\
MCP support
  & \checkmark & -- & -- & -- & -- & -- & -- \\
Model acquisition CLI
  & \checkmark & -- & -- & -- & -- & -- & \checkmark \\
Codebase migration
  & \checkmark & -- & -- & -- & -- & -- & -- \\
Unified API + ML + LLM
  & \checkmark & -- & -- & -- & -- & -- & -- \\
\bottomrule
\end{tabularx}
\caption{Feature comparison across frameworks, serving
platforms, and LLM inference engines. A checkmark indicates
built-in support without third-party extensions. \flama is the
only system that provides full web API capabilities, native
multi-framework ML model serving, multi-backend LLM inference
with simultaneous multi-dialect exposure, and MCP support
within a single unified architecture.}
\label{tab:comparison}
\end{table}

Several observations emerge from this comparison:

Read down the columns rather than across the rows and the
pattern is a clean split. Everything to the right of \flama is
strong in one band of the table and empty in the others, and
nothing in the survey is strong in two. That is the observation
the table exists to make.

Three qualifications are worth attaching to it. The first is
that a checkmark records presence, not quality: vLLM's
PagedAttention is a better piece of inference engineering than
anything \flama contributes, which is why \flama runs on top of
it rather than against it, and the row marked
\emph{multi-backend LLM} should be read as coverage rather than
as a claim about throughput. The second is that several of
these systems are not trying to fill the other bands, so an
empty cell is often a scope decision rather than an omission;
TensorFlow Serving is not a worse web framework than \flama, it
is not one. The third is that the table rewards breadth, and
breadth is only a virtue for an application that actually needs
it. A service doing nothing but batch inference over one
TensorFlow model is better served by TensorFlow Serving, and
the argument made here does not say otherwise.

What the comparison does support is narrower: for an
application that needs conventional endpoints and predictive
inference and generative serving at once, the alternatives
require composing two or three systems, and \flama requires
one.

\section{Conclusion and future work}
\label{sec:conclusion}

This paper has presented \flama, an open-source Python framework
that unifies web API development, predictive machine-learning
model serving, and large-language-model inference under a single
programming model. The framework rests on seven design
principles: async-first execution on the ASGI standard, type
annotations as the source of truth for validation and
documentation, pluggable schema libraries for data validation,
component-based dependency injection, convention over
configuration for resource generation, native ML model serving
for the four major predictive frameworks, and protocol-agnostic
LLM serving with multi-backend inference and multi-dialect wire
format support.

The FLM binary format provides a portable, compressed,
metadata-rich container for both predictive models and LLM
checkpoints, with two protocol versions addressing the distinct
storage requirements of each family. The three-level predictive
integration system (components, \py{add\_model()}, and model
resources) and the multi-dialect LLM serving architecture
(backends, transport, codec, dialects) allow developers to
choose the degree of customization that matches their
deployment requirements. The MCP module, now implementing the
current stateless protocol revision with Tasks, Elicitation, and
Apps extensions, connects \flama applications to the broader AI
tool ecosystem.

\flama is in production use for conventional REST APIs,
real-time inference services, and multi-dialect LLM endpoints.
Underneath sits the Rust core, seven modules covering route
resolution, path and host matching, JSON encoding, compression,
multipart parsing, cookie handling, and HTTP utilities, and the
effect of having it is visible in the numbers rather than only
in the design: growing a route table from ten entries to two
hundred moves the cost of a request by roughly 4\%, and ten
layers of middleware add about 0.6\% over none at all, both
measured under Callgrind in continuous integration. The point
of compiling those paths was to stop the framework's own
overhead from being the thing an application has to plan
around.

Several directions for future work are planned:

\begin{itemize}[leftmargin=2em, itemsep=4pt]
  \item \textbf{Model versioning and A/B testing.} Built-in
    support for serving multiple model versions concurrently and
    routing traffic between them based on configurable policies
    (percentage splits, header-based routing, canary
    deployments).

  \item \textbf{Adaptive batching.} Automatic request batching
    for GPU-bound inference to maximize throughput without
    requiring user configuration. Incoming prediction requests
    would be accumulated into micro-batches based on latency
    budgets and batch size targets.

  \item \textbf{GraphQL support.} Extending the schema and
    routing system to support GraphQL endpoints alongside REST,
    with the same type-driven validation and dependency
    injection infrastructure.

  \item \textbf{Observability integration.} Structured logging,
    distributed tracing (OpenTelemetry), and metrics export as
    built-in middleware, providing production-grade
    observability without third-party instrumentation.

  \item \textbf{Agentic workflows.} Native support for
    multi-step agent orchestration, combining LLM generation
    with tool use in structured loops, with built-in state
    management and conversation persistence.
\end{itemize}

\flama is released under the Apache~2.0 licence. The source code
is available at \url{https://github.com/vortico/flama}, with
documentation at \url{https://flama.dev} and package distribution
via PyPI (\texttt{pip install flama}).
% ==========================================================================
% End part_5_ecosystem.tex
% ==========================================================================

% ============================================================
% Bibliography
% ============================================================
\clearpage
\bibliographystyle{plainnat}
\bibliography{references}

\end{document}